\documentclass[11pt]{article}

\usepackage{amssymb,amsmath,amsfonts,amsthm}

 \usepackage[numbers,sort&compress]{natbib}
\usepackage{graphicx}

  \usepackage{lscape}
\usepackage{titletoc}

\usepackage{enumitem}

\usepackage{hyperref}
\hypersetup{
  colorlinks   =  true,
     linkcolor    = blue,
    citecolor    = blue,
    urlcolor	=blue,  
}

\def\be{\begin{equation}}
\def\ee{\end{equation}}
\def\bea{\begin{eqnarray}}
\def\eea{\end{eqnarray}}
\def\ba{\begin{align}}
\def\ea{\end{align}}

\def\>#1{{\bf #1}}
 \def\1{\'{\i}}
 
 \def\k{\omega}

\newcommand{\JJ}{{\hat x}}

  \newcommand{\astt}{{\perp}}
 \newcommand{\asttk}{{\perp,\k}}

 \newcommand{\sz}{\mathbb S_{z,\k} }
   
\newcommand{\bfI}{{\mathbf I}}
\newcommand\ka{\Lambda}

\newcommand\dsum{\oplus_S}

\newcommand\la{\lambda}

\newcommand\ca{\mathfrak{c}}

\newcommand\nh{\mathfrak{n}}

\newcommand\lo{\mathfrak{l}}

\newcommand{\kk}{ {\mathfrak {so}}_{\Lambda,c,\la} (5)}

\newcommand{\KK}{ {  {\rm SO}}_{\Lambda,c,\la} (5)}
 
\newcommand{\Hst}{ { H_{\rm st}}}

\newcommand{\Hline}{ { H_{\rm line}} }

\newcommand{\nn}{ {n} }

\newcommand{\xxi}{ {x} }
\newcommand{\A}{ {G_z} }
\newcommand{\UU}{ {z} }

\begin{document}
 
\
  \vskip1cm

\begin{center}
\noindent
 {\Large \bf  
Cayley--Klein Lie Bialgebras: Noncommutative Spaces,\\[6pt] Drinfel'd Doubles and
Kinematical Applications} 
\end{center}

\medskip 
\medskip 
\medskip

\begin{center}

{\sc   Ivan Gutierrez-Sagredo$^{1,2}$ and Francisco J.~Herranz$^1$}

\medskip

{$^1$Departamento de F\'isica, Universidad de Burgos, 
09001 Burgos, Spain}

{$^2$Departamento de Matem\'aticas y Computaci\'on, Universidad de Burgos, 
09001 Burgos, Spain}

 \medskip
 
e-mail: {\href{mailto:igsagredo@ubu.es}{igsagredo@ubu.es}, \href{mailto:fjherranz@ubu.es}{fjherranz@ubu.es}}

\end{center}

\medskip

\begin{abstract}
\noindent
The Cayley--Klein (CK) formalism is applied to the real algebra  $\mathfrak{so}(5)$ by making use of  four graded contraction   parameters describing in a unified setting   81 Lie algebras, which cover 
the (anti-)de Sitter,  Poincar\'e, Newtonian and Carrollian algebras. Starting with the  Drinfel'd--Jimbo real Lie bialgebra for $\mathfrak{so}(5)$ together  with its Drinfel'd double structure, we obtain the corresponding  CK bialgebra and the CK $r$-matrix coming from a  Drinfel'd double.   As a novelty, we construct  the (first-order) noncommutative  CK spaces of points, lines, 2-planes and 3-hyperplanes, studying their structural properties. 
By  requiring to deal with real structures, it comes out that there  exist 63  specific  real Lie bialgebras together with their sets of four  noncommutative spaces. Furthermore, we find   14 classical $r$-matrices  coming from Drinfel'd doubles, obtaining  new    results  for the de Sitter $\mathfrak{so}(4,1)$ and anti-de Sitter $\mathfrak{so}(3,2)$  and for some of their contractions.
 These   geometric results are exhaustively applied onto the (3+1)D kinematical algebras, not only considering the usual (3+1)D spacetime but also the 6D space of lines. We  establish different  assignations between the geometrical CK generators and the kinematical ones which convey     physical identifications for the CK contraction parameters in terms of the cosmological constant/curvature $\ka$ and speed of light $c$. We finally obtain four classes of kinematical $r$-matrices
 together with their noncommutative spacetimes and spaces of lines, comprising  all  $\kappa$-deformations as particular cases.  \end{abstract}
\medskip
\medskip

\noindent
PACS:  02.20.Uw; 02.20.Sv; 02.40.Gh; 04.60.--m 

\medskip
\noindent
MSC: 17B37; 17B62; 14M17; 81R60
\medskip

\noindent
KEYWORDS:  quantum groups; classical $r$-matrices; contractions;  symmetric homogeneous spaces;  Anti-de Sitter;  Carroll;  Newton--Hooke;  kappa-deformation;  noncommutative spacetimes;  noncommutative spaces of lines

 \newpage

\tableofcontents

\newpage


 \section{Introduction}
 
 The notion  of Cayley--Klein (CK) Lie algebras along with their corresponding Lie groups and symmetric homogeneous spaces date back to  early studies of projective metrics. In particular,   CK Lie groups  appear in a natural way   within
the context of the consideration by  Klein that most geometries are, in fact, subgeometries
of projective geometry and also in relation to Cayley's theory of projective metrics~\cite{Yaglom2,Ros,Yaglom}. 
However,  the complete
classification of   CK geometries, understood as geometries  endowed with a projective metric,  was not given by Klein himself. Meanwhile the 
two-dimensional (2D) ones were studied under the name of ``quadratic geometries" by
Poincar\'e, who followed a modern group theoretical approach. The  classification of CK geometries for arbitrary
dimension $N$ was finally achieved by Sommerville in 1909~\cite{Sommerville}, where   he  
showed that there exist exactly $3^N$ different CK geometries in dimension $N$, each of them
corresponding to a different choice of the kind of measure of distance between
points, lines, 2-planes,\dots, $(N-1)$-hyperplanes,   which can be either of elliptic, parabolic
or hyperbolic type~\cite{Yaglom}. Then CK   groups are just the motion groups of the CK geometries   acting as groups of isometries of the symmetric homogeneous CK spaces. In dimension $N$, such CK groups are semisimple pseudo-orthogonal groups ${\rm SO}(p,q)$  $(p+q=N+1)$
and some of their contractions such as, for instance, the inhomogeneous ${\rm ISO}(p',q')$  $(p'+q'=N)$. 

In order to set up the main ideas and the formalism that we shall follow along the whole paper, let us consider the   well  known nine 2D CK geometries~\cite{Yaglom,GromovTran,GromovCKq,CK2d,trigo,CKconf,McRae1,McRae2,Varna2D}. These emerge as the different 
possibilities for considering the  measure of distance between two points and the     measure of an angle between two lines, being   each of them either of  elliptic, parabolic or hyperbolic type. The CK groups are the simple real Lie groups $ {\rm SO}(3)$ and  $ {\rm SO}(2,1)$,    the non-simple inhomogeneous Euclidean $ {\rm ISO}(2)$ and Poincar\'e $ {\rm ISO}(1,1)$, and the twice inhomogeneous Galilean  $ {\rm IISO}(1)$    (in this notation $ {\rm ISO}(1)\equiv \mathbb R$).  The 2D CK geometries are constructed through    the coset spaces of the above 3D Lie groups with a precise 1D isotropy subgroup. Early, the usual procedure for describing  these geometries made use of hypercomplex numbers with two hypercomplex units $\iota_1$ and $\iota_2$~\cite{Ros,Yaglom,GromovTran,GromovCKq}. We recall that a hypercomplex number  is defined by $z:= x+\iota\, y$, where    $(x,y)$ are two real coordinates and  $\iota$ is a  hypercomplex unit  such that $\iota^2 \in\{ -1,+1,0\}$. Hence there  are   three possible  kind of hypercomplex numbers
according to   the specific unit $\iota$:
(1)  If $\iota^2=-1$, then $\iota$ is an {\em elliptical}   unit providing   the usual  complex numbers; (2)    If $\iota^2=+1$,   $\iota$ is a {\em hyperbolic}   unit  yielding the  so-called  split complex,  double or Clifford numbers; and (3)      if $\iota^2=0$,  $\iota$ is a {\em parabolic}   unit  leading to the dual or Study numbers. Alternatively, 2D CK geometries can also be studied in terms of two real graded contraction parameters $\k_1$ and $\k_2$, which  can take      positive, negative or zero values~\cite{CK2d,trigo,CKconf,McRae1,McRae2,Varna2D}. By taking into account the above   two approaches, we display the specific  2D CK geometries in Table~\ref{table0}, where they are named in their original geometric form~\cite{Yaglom}, as well as in their physical  (or kinematical) terminology (second and third rows).

 \begin{table}[htp] 
{\small
\caption{\small The nine two-dimensional Cayley--Klein
geometries as   homogeneous spaces  according to two  real graded contraction  parameters $(\k_1,\k_2)$ and to  two hypercomplex units $(\iota_1,\iota_2)$.}
\label{table0}
 \begin{center}
 \begin{tabular}{llll}
\hline  

\hline

\\[-0.25cm]
 &\multicolumn{3}{c}{{\bf Measure of distance}}\\[0.15cm]
\cline{2-4}
\\[-0.2cm] 
{\bf Measure}&Elliptic&Parabolic&Hyperbolic\\
{\bf of angle}&$\k_1>0 \quad \iota_1^2=-1$&$\k_1=0\quad \iota_1^2=0$&$\k_1<0\quad \iota_1^2=+1$\\[0.15cm]
\hline
\\[-0.2cm]
Elliptic &$\bullet$ Spherical&$\bullet$ Euclidean&$\bullet$ Hyperbolic\\[2pt]
 $\k_2>0\quad   \iota_2^2=-1$& $ {\rm SO}(3)/{\rm SO}(2)$&$   {\rm ISO}(2)/{\rm SO}(2)$ &$    {\rm SO}(2,1)/{\rm SO}(2)$ \\[0.25cm]

Parabolic&$\bullet$ Co-Euclidean &$\bullet$ Galilean&$\bullet$ Co-Minkowskian \\[1pt]
 $\k_2=0\quad  \iota_2^2=0$&or  Oscillating NH & &or  Expanding NH \\[2pt]
$ $ &$   {\rm ISO}(2)/ \mathbb R$ &$   {\rm IISO}(1)/ \mathbb R$ &$   {\rm ISO}(1,1)/ \mathbb R$\\[0.25cm]

Hyperbolic &$\bullet$ Co-Hyperbolic &$\bullet$  Minkowskian&$\bullet$ Doubly Hyperbolic  \\[1pt]
 $\k_2<0\quad  \iota_2^2=+1$&or  Anti-de Sitter& &or  De Sitter \\[2pt]
$ $&$   {\rm SO}(2,1)/ {\rm SO}(1,1)$  & $   {\rm ISO}(1,1)/ {\rm SO}(1,1)$   & $  {\rm SO}(2,1)/ {\rm SO}(1,1)$\\[6pt]
 \hline  

\hline

\end{tabular} 
\end{center}
}
 \end{table}

Consequently, the family of 2D CK geometries contains nine homogeneous spaces of constant curvature:  the three classical    Riemannian spaces   in the first row of Table~\ref{table0}; the three  
(Newtonian) spaces with a degenerate metric in  the second row; and the
 three   pseudo-Riemannian or Lorentzian spaces    in the third row.

 In the procedure  that makes use of the real parameters $(\k_1,\k_2)$~\cite{CK2d,trigo,CKconf,McRae1,McRae2,Varna2D}, 
  the generic CK Lie algebra is denoted by  $\mathfrak{so}_{\k_1,\k_2}(3)=\mbox{span}\{ J_{01},J_{02},J_{12}\}$, which corresponds to  a two-parametric family of Lie algebras  
  with commutation relations given by 
\be
[J_{12},J_{01}]=J_{02},\qquad [J_{12},J_{02}]=-\k_2 J_{01},\qquad [J_{01},J_{02}]=\k_1 J_{12} .
\label{adx}
\ee
The vanishing of a given parameter $\k_m$ (i.e., $\k_m\to 0 $)  is equivalent to apply an In\"on\"u--Wigner contraction~\cite{IW}.  The CK algebra has a single quadratic Casimir given by
\be
\mathcal{C}=\k _2 J_{01}^2+J_{02}^2+\k_1 J_{12}^2 .
\label{aex}
\ee

The 2D CK geometry is then defined as the following coset space between the  CK Lie group  ${\rm SO}_{\k_1,\k_2}(3)$  with Lie algebra $\mathfrak{so}_{\k_1,\k_2}(3)$  (\ref{adx}) and the isotropy subgroup of a point $H$, spanned by $J_{12}$~\cite{CKconf}:
\be
\mathbb{S}^2_{ [\k_1],\k_2}:={\rm SO}_{\k_1,\k_2}(3)/H,\qquad 
H={\rm SO}_{\k_2}(2)=\langle J_{12}\rangle  .
\label{afx}
\ee
Hence $J_{12}$ leaves a point $O$  invariant (the origin) so acting as the generator of rotations around $O$,   while $J_{01}$ and $J_{02}$ are the generators of translations that move $O$ along two basic directions on the space. 
The   CK geometry~\eqref{afx}  has a metric, provided by the Casimir (\ref{aex}), which is of constant (Gaussian) curvature equal to  $\k_1$ and  with metric signature   given by $\mbox{diag}(+1,\k_2)$ (see Table~\ref{table0}).
 
 Spacetimes of constant curvature  in (1+1) dimensions arise as particular cases of the CK geometry \eqref{afx} through different assignations between the geometrical generators $J_{ab}$ and the kinematical ones, which require appropriate relations between the CK parameters $(\k_1,\k_2)$ and   physical quantities. Explicitly, let $\{ P_0,P_1,K\}$ be the generators of  
 time translations, space translations and boost transformations. Under the particular identification 
 \be 
  J_{01} \equiv P_0,\qquad J_{02}\equiv  P_1 ,\qquad  J_{12}\equiv  K,  \qquad \k_1\equiv -\Lambda,\qquad \k_2\equiv - {c^{-2}},
  \label{agx}
 \ee
 where $\Lambda$ is the cosmological constant and $c$ is the speed of light, we find that the CK algebra (\ref{adx}) adopts the form
\be
[K,P_0]=P_1,\qquad [K,P_1]= \frac 1 {c^{2}}\, P_0,\qquad [P_0,P_1]=-\Lambda K.
\label{adx2}
\ee
Thus, under the   relations (\ref{agx}), the CK group ${\rm SO}_{\k_1,\k_2}(3)$  with Lie algebra (\ref{adx2}) becomes a kinematical group acting as the 
  group of isometries of six relevant (1+1)D spacetime models~\cite{BLL}   of constant curvature equal to  $-\Lambda$,  which are all of them contained in \eqref{afx}. These are the three Lorentzian spacetimes   with metric signature $\mbox{diag}(+1,-{c^{-2}})$,  mentioned   in the third row of Table~\ref{table0}, along with their non-relativistic limit $c\to \infty$ leading to the three Newtonian spacetimes 
  with degenerate metric   $\mbox{diag}(+1,0)$  in the second row of Table~\ref{table0}   (NH means Newton--Hooke). It should be noted that all of the kinematical algebras and spacetimes
   can also be described within the CK framework except for the static algebra~\cite{BLL}; at this dimension the latter is just the abelian algebra.

 In principle, the very same results can be obtained by making use of the formalism in terms of  hypercomplex numbers~\cite{Ros,Yaglom,GromovTran,GromovCKq} since, roughly speaking, one finds the relations $\iota^2\sim  {-\k}$. 
   Nevertheless,  besides to simply  deal with real numbers instead of hypercomplex ones,  the main differences   between both approaches 
   clearly appear   in the pure contracted case  corresponding to consider the parabolic unit  $\iota^2=0$ and to set  $\k=0$.  In particular,  the   contraction of exponentials of a Lie 
generator  $J$ could give rise to different results; for instance:
$$
\exp( {\iota^2}x J)\to 1 ,\quad\  \exp( {\iota}xJ)\to 1+\iota x J ,\quad\  \exp( {\k}xJ)\to 1 ,\quad\    \exp( {\sqrt{\k}}\,xJ )\to 1 ,
$$
where $x$ is a real number.  We stress that this kind of exponentials often appear in quantum groups and, moreover, 
  terms depending on some $ {\sqrt{\k_m}}$ will be omnipresent  throughout the paper (within  Lie   bialgebra structures),  so that
   both procedures  could be no longer equivalent in a quantum deformation framework.     Thus we shall make use of the graded contraction approach with   real parameters $\k_m$ in such manner that   a smooth and well-defined $\k_m\to 0$ limit of all the expressions that we shall present here  will be always feasible.

 In  addition, we stress that the same CK group $  {\rm SO}(2,1) $ appears three times in Table~\ref{table0}, and two of their CK spaces  are ``similar" (see~\cite{KisilBook, KisilProcs} for  a  very detailed  description of  these three geometries in terms of  hypercomplex numbers).  The structure of the three CK geometries involved,  hyperbolic and the (anti-)de Sitter ones, can be better understood by considering not only the usual CK space (\ref{afx}) shown in  Table~\ref{table0} but also their  2D homogeneous spaces of lines, as it was performed in~\cite{trigo,Varna2D}, and likewise for 
 the Euclidean and Poincar\'e groups which appear twice  in  Table~\ref{table0}.

 In arbitrary dimension $N$, the CK algebra  depends on $N$ real graded contraction parameters $\k_m$ $(m=1,\dots,N)$ and is   denoted by  $\mathfrak{so}_{\k_1,\dots,\k_N}(N+1)$.  This family      comprises $3^N$   semisimple and non-semsimple  real Lie algebras   (being some of them isomorphic)  which share common geometric and algebraic properties.  The signs  of the parameters  $\k_m\ne 0$  determine a specific real form $\mathfrak{so}(p,q)$ and when at least one $\k_m$ vanishes the CK algebra becomes a non-semisimple one.  
 
  In algebraic terms, a CK algebra can be defined as a 
  graded contracted Lie algebra from $\mathfrak{so}(N+1)$~\cite{CKMontigny} which keeps the same number of algebraically independent Casimir invariants as in the semisimple case, regardless of the values of $\k_m$~\cite{casimirs}.   This definition implies that all of the $3^N$ particular CK algebras   have the same rank (even for the most contracted case with all $\k_m=0$), so that they are also known as quasisimple orthogonal algebras~\cite{casimirs,extensions}. From this viewpoint they can be seen as the  ``closest" contracted algebras to the semisimple ones. In this respect, we remark that   the CK contraction sequence ensures to always obtain a non-trivial quadratic Casimir (like  \eqref{aex}) which, in turn, means that there always exists a non-trivial metric on the $N$D CK geometry, although degenerate in many cases.
This fact explains the absence of the static algebra in the CK family. Obviously,  if one goes beyond the CK Lie algebra contraction sequence, then one can get the  static algebra,  and   finally arrive at the abelian algebra~\cite{CKGraded}. 

From the CK  algebra, the corresponding CK Lie group $ {\rm SO}_{\k_1,\dots,\k_N}(N+1)$ can be constructed, and the $N$D CK geometry is defined as the coset space (see (\ref{afx}))
  \be
 \mathbb{S}^N_{ [\k_1],\k_2,\dots,\k_N}:= {\rm SO}_{\k_1,\dots,\k_N}(N+1) /{\rm SO}_{\k_2,\dots,\k_N}(N) , 
 \label{afxx}
\ee
which has a metric of constant (sectional) curvature equal to $\k_1$   with  signature   given by $\mbox{diag}(+1,\k_2,\dots,\k_N)$.

 In this paper we shall focus on the physically relevant dimension $N=4$   thus   covering the    (3+1)D spacetimes of constant curvature. Hence the CK algebra and group  will depend on a set of four  real graded contraction parameters $\k=(\k_1,\k_2,\k_3,\k_4)$  so comprising $3^4=81$ specific Lie algebras.  For each CK algebra/group   we shall consider four types of symmetric homogeneous spaces:   the usual 4D CK space of points (spacetimes) (\ref{afxx}) along with the 6D space of lines, 6D space of 2-planes and 4D space of 3-hyperplanes, being all of them of constant curvature and equal to $\k_1,\dots,\k_4$, respectively. Therefore in this paper by a CK geometry it will be understood the set of these four homogeneous spaces associated with a given Lie algebra in the CK family and not only the usual space of points (\ref{afxx}), which is the   one commonly considered in the literature.

  In this  geometrical setting, we initially  study CK Lie  bialgebras and their associated noncommutative spaces in order to further develop    their physical applications. Thus we, firstly, review the basics on quantum groups that will be used along the paper in Section~\ref{s1}. And, secondly,     we present a two-fold work in Sections~\ref{s2}--\ref{s5} with two main but interrelated parts, whose structure, objectives and results are as follows.

   \begin{enumerate} 
   
   \item  Starting with the  Drinfel'd--Jimbo Lie bialgebra for $\mathfrak{so}(5)$ and also considering its Drinfel'd double structure  in Section~\ref{s2}, we obtain the corresponding CK bialgebra along with the classical CK $r$-matrix coming from a  Drinfel'd double in Section~\ref{s3}. As a novelty, we construct, by means of quantum duality, the first-order (in the quantum coordinates) noncommutative  CK spaces of points, lines, 2-planes and 3-hyperplanes.     We    analyse their properties and    always require to deal with real structures. It comes out that, finally, there  are 63  specific  real Lie bialgebras together with their sets of four (first-order) noncommutative spaces, which are summarized in Tables~\ref{table1} and~\ref{table2}, respectively. Additionally, we find   14 classical $r$-matrices  coming from Drinfel'd double real structures: there are four cases (I)--(IV) for the simple algebras and 10 more cases for their contractions. In this way, we   obtain  new results  for  de Sitter $\mathfrak{so}(4,1)$ (case (II))  and anti-de Sitter $\mathfrak{so}(3,2)$ (case (IV))    Drinfel'd doubles and for some of their contractions, which are displayed in Table~\ref{table3}.

\item   The above geometric results are exhaustively applied onto the (3+1)D kinematical algebras~\cite{BLL}, not only considering the usual (3+1)D spacetime but also the 6D space of lines; the classical   picture  for each kinematical algebra  is presented in Section~\ref{s4} and outlined in Table~\ref{table4}. In Section~\ref{s5},  we establish different  assignations between the geometrical CK generators and the kinematical ones which convey   appropriate physical identifications for the CK contraction parameters $\k$ in terms of the cosmological constant/curvature $\ka$ and speed of light $c$. In this process we obtain four classes of kinematical $r$-matrices
and, for some algebras, also $r$-matrices coming from Drinfel'd doubles. These classes are    called A, B, C and D,   matching, in this order, with the above cases (I)--(IV). The resulting kinematical bialgebras   are given in Table~\ref{table5}, while their corresponding first-order noncommutative spacetimes and spaces of lines are shown in Table~\ref{table6}. We stress that the class C covers the kappa-deformations.

  \end{enumerate}

We would like to mention that although in this work we do not construct   the complete quantum kinematical algebras and their associated full noncommutative spacetimes and spaces of lines, we comment on related known results and open problems in Sections~\ref{s55} and \ref{s56}, respectively. To finish with,  several conclusions  and a more exhaustive list of open problems,    also  concerning the geometric CK setting, are drawn in Section~\ref{s6}.


\section{Fundamentals on Quantum Groups}
\label{s1}
 
In this Section we review the basic background on quantum groups necessary for the paper along with   updates and physical motivation related  to   the  main  results  here presented.  We shall focus on   quantum deformations of Lie algebras (with a Hopf algebra structure) along with their connection with Lie bialgebras, Poisson--Lie groups, Poisson  homogeneous spaces, noncommutative spaces and  Drinfel'd doubles. More details on these topics can be found in~\cite{ResTakFad89,JimboEd,Takhtajan,CP,majid,Abe}.


\subsection{Lie Bialgebras and Quantum Algebras}
\label{s11}
 
Let us consider an  $n$D  
 Lie algebra  $\mathfrak{g}=\mbox{span}\{ X_1,\dots,X_n\}$  
with   commutation relations given by
\begin{equation}
[X_i,X_j]=\sum_{k=1}^\nn c^k_{ij}X_k    .
\label{ta} 
\end{equation}
The Lie algebra  $\mathfrak{g}$ is endowed with a {\em Lie bialgebra} structure $(\mathfrak{g},\delta)$~\cite{Drinfeld1983hamiltonian} if there exists a  map $\delta:\mathfrak{g} \to \mathfrak{g} \wedge \mathfrak{g}$ called the {\em  cocommutator }verifying two conditions:
\smallskip

\noindent
 (i) $\delta$ is a 1-cocycle,  
\begin{equation}
\delta([X_i,X_j])=[\delta(X_i),\,  X_j\otimes 1+ 1\otimes X_j] + 
[ X_i\otimes 1+1\otimes X_i,\, \delta(X_j)] ,\quad\  \forall  X_i,X_j\in
\mathfrak{g}.
\label{tb} 
\end{equation}

\noindent
 (ii) The dual map $\delta^\ast:\mathfrak{g}^\ast\otimes \mathfrak{g}^\ast \to \mathfrak{g}^\ast$ is a Lie bracket on the dual Lie algebra $\mathfrak{g}^\ast$ of $\mathfrak{g}$.
\smallskip

Therefore any  cocommutator $\delta$ can be written in a skew-symmetric form  as
\begin{equation}
\delta(X_i)=\sum_{j,k=1}^\nn f^{jk}_i X_j \otimes X_k \, ,\qquad  f^{jk}_i=- f^{kj}_i,  
\label{tc}
\end{equation}
in such a manner that the antisymmetric factors $f^{jk}_i$  turn out to be the structure constants of the dual Lie algebra $\mathfrak{g}^\ast=\mbox{span}\{ \hat\xxi_1,\dots, \hat\xxi_n  \}$:
\begin{equation}
[\hat\xxi^j,\hat\xxi^k]=\sum_{i=1}^n  f^{jk}_i \hat\xxi^i   .
\label{td}
\end{equation}
The duality between the generators of $\mathfrak{g}$ and $\mathfrak{g}^\ast$  is determined by a canonical pairing given by the bilinear form
\be
\langle  \hat\xxi^i,X_j \rangle=\delta_j^i  \,,\qquad \forall i,j .
\label{tdd}
\ee
And the cocycle condition (\ref{tb})   leads to the following compatibility equations among the structure constants $c_{ij}^k$ (\ref{ta}) and $f^{jk}_i$  (\ref{td}):
\be
\sum_{k=1}^\nn f^{lm}_k c^k_{ij} =\sum_{k=1}^\nn\left( f^{lk}_i c^m_{kj}+f^{km}_i c^l_{kj}
+f^{lk}_j c^m_{ik} +f^{km}_j c^l_{ik} \right) ,\qquad \forall i,j,l,m .
\nonumber
\ee

For some Lie bialgebras the 1-cocycle $\delta$ is   coboundary~\cite{Drinfeld1983hamiltonian}, that is, it can be  obtained  from an element $r\in   \mathfrak{g} \otimes \mathfrak{g} $
in the form
\be
\delta(X_i)=[ X_i \otimes 1+1\otimes X_i ,\,  r],\qquad 
\forall X_i\in \mathfrak{g} .
\label{tf}
\ee
The element $r$ is   the so-called  {\em classical $r$-matrix} which can always be written in  a skew-symmetric form 
\be
r=\sum_{i,j=1}^n r^{ij} X_i \otimes X_j   \,,\qquad r^{ij} =-r^{ji} ,
\label{tg}
\ee
and    must
  be a solution of the modified classical Yang--Baxter equation  
\be
[X_i\otimes 1\otimes 1 + 1\otimes X_i\otimes 1 +
1\otimes 1\otimes X_i,[[r,r]]\, ]=0, \qquad \forall X_i\in \mathfrak{g},
\label{th}
\ee
where $[[r,r]]$ is the  Schouten bracket    defined by
\be
[[r,r]]:=[r_{12},r_{13}]+ [r_{12},r_{23}]+ [r_{13},r_{23}] ,
\label{ti}
\ee
  such that 
  \be
  r_{12}=\sum_{i,j=1}^n r^{ij} X_i \otimes X_j\otimes 1 \, , 
  \quad\ 
  r_{13}=\sum_{i,j=1}^n r^{ij} X_i \otimes 1\otimes X_j \, , 
  \quad\ 
  r_{23}= \sum_{i,j=1}^n r^{ij} 1 \otimes X_i\otimes X_j  \,.
\nonumber
  \ee
If  the  Schouten bracket (\ref{ti}) does not vanish for  an $r$-matrix written in the skew-symmetric form (\ref{tg}), then  the Lie algebra $\mathfrak{g}$ is endowed with a  {\em quasitriangular} (or standard) Lie bialgebra structure $(\mathfrak{g},\delta(r))$.    The vanishing of the Schouten bracket   corresponds to the classical Yang--Baxter equation
  \be[[r,r]]=0,
    \label{tk}
  \ee
   and $(\mathfrak{g},\delta(r))$ is called {\em triangular} (or nonstandard)  Lie bialgebra.
  
 We remark that the deformation parameter, that we shall denote $z$ throughout the paper and such that $q={\rm e}^z$, is already contained within $\delta$ (\ref{tc}) and $r$ (\ref{tg})  in the   factors $f^{jk}_i$    and $ r^{ij}$ as a global multiplicative constant. This, in turn, means that the non-deformed or ``classical" limit $z\to 0$ (i.e.~$q\to 1$) leads to a trivial coproduct $\delta =0$ with all $f^{jk}_i\equiv 0$ and classical $r$-matrix  $r=0$ with all $ r^{ij}\equiv 0$, being $\mathfrak{g}^\ast$      an abelian Lie algebra thus  with commutative generators $\hat x^i$~(\ref{td}).

A {\em quantum algebra} ${\mathcal U}_z(\mathfrak{g})$ is a Hopf algebra deformation of the universal enveloping algebra ${\mathcal U}(\mathfrak{g})$ of $\mathfrak{g}$ constructed as formal power series  $\mathbb{C}[[z]]$ in  a deformation indeterminate parameter $z$
and coefficients in ${\mathcal U}(\mathfrak{g})$, that is,  ${\mathcal U}_z(\mathfrak{g})={\mathcal U}(\mathfrak{g}) \, \hat \otimes  \,\mathbb{C}[[z]]$. 
 The Hopf algebra structure of ${\mathcal U}_z(\mathfrak{g})$  is determined by the coproduct $\Delta_z$, 
counit $\epsilon$ and antipode $\gamma$ mappings~\cite{CP,majid,Abe}.  In particular, the coproduct $\Delta_z: {\mathcal U}_z(\mathfrak{g})\to {\mathcal U}_z(\mathfrak{g})\otimes {\mathcal U}_z(\mathfrak{g})$  
must be   
an algebra homomorphism and fulfil 
the 
  coassociativity condition 
\be
({\rm Id}\otimes\Delta_z)\Delta_z  =(\Delta_z\otimes {\rm Id})\Delta_z  \, ,
\nonumber
\ee
where ${\rm Id}$ is the identity map, giving    rise to  a {\em coalgebra structure} $({\mathcal U}_z(\mathfrak{g}), \Delta_z)$.  Once $ \Delta_z$ is obtained,  the remaining maps, $\epsilon$ and   $\gamma$, can directly be deduced from the Hopf algebra axioms providing the complete Hopf algebra structure. Hence hereafter we shall only focus on the coalgebra structure of ${\mathcal U}_z(\mathfrak{g})$  assuming the existence of the corresponding  counit and antipode.

The remarkable point is that  any quantum   algebra ${\mathcal U}_z(\mathfrak{g})$ is determined at the first-order in  $z$ by  a Lie bialgebra $(\mathfrak{g},\delta)$.  Explicitly, if we write the coproduct $\Delta_z$ as a formal power series in $z$, the     cocommutator $\delta$ (\ref{tc})  is just the  skew-symmetric part of the first-order term $\Delta_1$ in $z$, namely
\bea
&& \Delta_z(X_i)=\Delta_0(X_i)+ \Delta_1(X_i)+o[z^2], \nonumber\\[2pt]  
&& \Delta_0(X_i)=X_i\otimes 1+ 1\otimes X_i   \,,\label{tm}\\[2pt]
&&  \delta(X_i)=\Delta_1(X_i)-\sigma\circ \Delta_1(X_i) ,
\nonumber
\eea
where $\sigma$ is the flip operator  $\sigma(X_i\otimes X_j)=X_j\otimes X_i$  and $\Delta_0$ is called the {\em primitive} (non-deformed) coproduct. Therefore 
each  Lie bialgebra $(\mathfrak{g},\delta)$ determines a    quantum deformation $({\mathcal U}_z(\mathfrak{g}), \Delta_z)$ and the equivalence classes (under automorphisms) of Lie bialgebra structures on   $\mathfrak{g}$ will provide       all its possible quantum algebras.

   We recall that for semisimple Lie algebras all their Lie bialgebra structures are coboundaries, so that all their possible quantum deformations are determined by classical $r$-matrices. The paradigmatic  type of them is provided by the   so-called Drinfel'd--Jimbo   deformations~\cite{Drinfeld1985Hopf,Jimbo,Drinfeld1987icm}; the corresponding Drinfel'd--Jimbo $r$-matrix for the compact real form $\mathfrak{so}(5)$~\cite{LBC} will be our starting point for the detailed study of the CK Lie bialgebras which will be performed in Sections~\ref{s2} and \ref{s3}, respectively.
    However, even  for semisimple Lie algebras the determination of all the Lie bialgebra structures through classical  $r$-matrices is a cumbersome task and, in fact, there are only classifications for the Lorentz algebra  $\mathfrak{so}(3,1)$~\cite{Zak94} and for the related real forms   $\mathfrak{so}(4)$ and  $\mathfrak{so}(2,2)$~\cite{BorowiecLukierskiTolstoy2016rmatrices,BorowiecLukierskiTolstoy2016rmatricesaddendum};      from a kinematical viewpoint these  classifications for $\mathfrak{so}(3,1)$ and  $\mathfrak{so}(2,2)$ correspond to   (2+1)D (anti-)de Sitter $r$-matrices~\cite{BHMJPCS12}. Therefore, in the (3+1)D case, which  is the one that we shall consider  throughout this paper, there are no such classifications for the simple algebras $\mathfrak{so}(p,q)$ with $p+q=5$, although we remark that there are some partial results. In particular, it was shown in~\cite{BrunoAdS} that there only exist two two-parametric classical $r$-matrices for 
    the (anti-)de Sitter algebras $\mathfrak{so}(4,1)$ and  $\mathfrak{so}(3,2)$ keeping primitive (undeformed) the time translation generator and a single rotation generator.  And the classification of their $r$-matrices which preserve a Lorentz   $\mathfrak{so}(3,1)$ sub-bialgebra has been, very recently,  obtained in~\cite{LorentzdS} starting with the former full (2+1)D classification~\cite{Zak94,BorowiecLukierskiTolstoy2016rmatrices}.

 Furthermore, all the Lie bialgebra structures for  inhomogeneous pseudo-orthogonal algebras $\mathfrak{iso}(p,q)$ with $p+q\ge 3$ are also coboundaries~\cite{Zakrzewski1997,PW1997}. Their classification for the (3+1)D Poincar\'e algebra $\mathfrak{iso}(3,1)$ was obtained in~\cite{Zakrzewski1995,PW1996,Zakrzewski1997}, while for the   (2+1)D Poincar\'e algebra $\mathfrak{iso}(2,1)$ and 3D Euclidean one $\mathfrak{iso}(3)$   it  was performed in~\cite{Stachura1998};  the latter   classifications  
 have also been  recovered in~\cite{Kowalski2020} by contracting the  (2+1)D (anti-)de Sitter  and $\mathfrak{so}(4)$ $r$-matrices given  in~\cite{BorowiecLukierskiTolstoy2016rmatrices,BorowiecLukierskiTolstoy2016rmatricesaddendum}.
 
 Concerning other kinematical algebras, we also recall that the obtention of 
      Lie bialgebras  mainly cover  low-dimensional cases  such as the (1+1)D Galilei algebra (isomorphic to the Heisenberg--Weyl algebra $\mathfrak{h}_3$)~\cite{Kupershmidt93,Hussin94,Parashar97,Kowalczyk97},  the 2D Euclidean algebra $\mathfrak{iso}(2)$~\cite{Sobczyk}, the (1+1)D centrally extended Galilei algebra~\cite{Opanowicz1998,Opanowicz2000,Galieli2000} and the   (1+1)D centrally extended Poincar\'e algebra in the light-cone basis  (isomorphic to the oscillator algebra $\mathfrak{h}_4$)~\cite{h496,h497}. With the exception of the latter,  all of them have    both coboundary and non-coboundary Lie bialgebra structures.    In general,  for solvable and nilpotent Lie algebras many of their Lie bialgebra structures are non-coboundaries; in this respect, see~\cite{gomez,BBMpl} and references therein for the classifcation of  3D Lie bialgebras. Finally, we point out that, very recently,  the classification of 4D indecomposable coboundary Lie bialgebras has been carried out in~\cite{Lucas}, which shows how the difficulties of this task grow when the dimensions of the Lie bialgebras increase.


\subsection{Quantum Groups and Noncommutative Spaces}
\label{s12}

Let us consider a quantum algebra  $({\mathcal U}_z(\mathfrak{g}), \Delta_z)$ with underlying   Lie bialgebra $(\mathfrak{g},\delta)$ and let $G$ be the Lie group with Lie algebra $\mathfrak{g}$.
 A  {\em quantum group} $(G_z,\Delta_{G_z})$ is a noncommutative algebra of functions on $G$ defined as the dual Hopf algebra  to the quantum algebra $({\mathcal U}_z(\mathfrak{g}), \Delta_z)$. Explicitly, let   $m_{\A}$ and $m_\UU$  be the noncommutative products in $\A$ and ${\mathcal U}_z(\mathfrak{g})$, respectively. The duality between the Hopf algebras  $(\A,m_\A,\Delta_\A)$ and
$({\mathcal U}_z(\mathfrak{g}), m_\UU,\Delta_z)$ is established by means of  a
canonical pairing
$\langle
\, , \, \rangle: \A\times {\mathcal U}_z(\mathfrak{g}) \rightarrow \mathbb R$   such that
\bea
&& \langle m_\A(f\otimes g), X \rangle=\langle f\otimes g ,
\Delta_z(X)\rangle,\label{pairing1}
\\[2pt]
&& \langle \Delta_\A(f) , X\otimes Y \rangle= \langle f,
m_\UU(X\otimes Y)\rangle,
\label{pairing2}
\eea
where $X,Y\in {\mathcal U}_z(\mathfrak{g})$, $f,g\in \A$, and 
$
\langle f\otimes g , X\otimes Y \rangle=\langle f , X
\rangle\,\langle g , Y\rangle.
$

The duality relation (\ref{pairing1}) implies that the noncommutative product $m_\A$  in the quantum group $\A$ is defined by the coproduct  $\Delta_z$ in the quantum algebra  ${\mathcal U}_z(\mathfrak{g})$ and, conversely,  the expression (\ref{pairing2}) implies that the coproduct $\Delta_{G_z}$ in $\A$ is given by the noncommutative product $m_z$ in ${\mathcal U}_z(\mathfrak{g})$. By taking into account that  the first-order term  in $z$  of the coproduct $\Delta_z$  is defined by the cocommutator $\delta$ (\ref{tm}), 
  a straightforward consequence of the above Hopf algebra duality is that   the  commutation relations for the quantum group $\A$ at the first-order in the quantum (noncommutative) coordinates $\hat x^i$ are  given by the dual map $\delta^\ast$ of $\delta$ (\ref{tc}), that is, with fundamental Lie brackets  (\ref{td}).

   Furthermore, each quantum group  $(G_z,\Delta_{G_z})$  can be associated with a {\em Poisson--Lie group} $(G,\Pi)$, with Poisson structure $\Pi$,  and  the latter with a unique Lie bialgebra structure $(\mathfrak{g},\delta)$. In particular, it is well known~\cite{Drinfeld1983hamiltonian} that Poisson--Lie structures on a connected and simply connected Lie group $G$ are in one-to-one correspondence with Lie bialgebra structures. Hence,  quantum groups  are quantizations of  Poisson--Lie  groups, that is,  quantizations of the Poisson--Hopf algebras of multiplicative Poisson structures on Lie groups~\cite{CP, majid, Drinfeld1987icm}. In the case of coboundary Lie bialgebras 
    $(\mathfrak{g},\delta(r))$, coming from a skew-symmetric classical $r$-matrix (\ref{tg}), the Poisson structure $\Pi$  of the Poisson--Lie group is   given by the so-called Sklyanin bracket \cite{CP,Drinfeld1987icm}
\be
\{f,g\}=\sum_{i,j=1}^n r^{ij} \bigl( \nabla^L_i f \nabla^L_j g - \nabla^R_i f \nabla^R_j g \bigr), \qquad f,g \in   C ^\infty(G),
\label{sklyanin}
\ee
where  $\nabla^L_i$ and $\nabla^R_i$ are left- and right-invariant vector fields on  $G$.

Next, a {\em Poisson homogeneous space}  of a Poisson--Lie group $(G,\Pi)$ is a Poisson manifold $(M,\pi)$ endowed with a transitive group action $\rhd: G\times M\to M$ which is a Poisson map with respect to the Poisson structure on the manifold $M$ and the product $\Pi \oplus \pi$ of the Poisson structures on $G$ and $M$.   In this paper we shall consider that the manifold $M$ is an   $\ell$D homogeneous space 
\be
M=G/{H}
\label{tp}
\ee of a  Lie group $G$ (the motion group of $M$)  with   isotropy subgroup $H$ whose   Lie algebras   are  $\mathfrak{g}$ and  $\mathfrak{h}$, respectively. Moreover, throughout this paper we will be interested in pointed Poisson homogeneous spaces, i.e. Poisson homogeneous spaces in which the origin is fixed, and we will not study how the Poisson structure is modified when this origin is changed. The Lie algebra $\mathfrak{g}$, understood as a vector space, can be written as the   sum of two vector subspaces
\be
{\mathfrak{g}}=  \mathfrak{h} \oplus \mathfrak{t} , \qquad  [\mathfrak{h} ,\mathfrak {h} ] \subset \mathfrak{h}  .
\nonumber
\ee
The generators of $\mathfrak{h}$ leave a point of $M$ invariant, which is taken as the origin $O$  of the space, so they play the role of rotations around $O$,  while the $\ell$ generators belonging to $\mathfrak{t}$ move $O$  along  $\ell$ basic directions,  behaving as translations on   $M$. The group parameters $(u^1,\dots,u^\ell)$  of the generators of $\mathfrak{t}$ lead to    $\ell$   coordinates of $M$ and they span the annihilator $\mathfrak{h}_\astt$ of  the vector subspace $\mathfrak{h}$ in  the dual Lie algebra ${\mathfrak{g}}^\ast$~\cite{Ciccoli2006}.

  In principle,  the Lie algebra $\mathfrak g$ of $G$    may admit several coboundary Lie bialgebra structures $(\mathfrak g,\delta(r))$, that is, different classical $r$-matrices.  Then a particular Poisson homogeneous space $(M,\pi)$ can be constructed by endowing the motion group $G$ with the Poisson--Lie structure $\Pi$~\eqref{sklyanin} for a given classical $r$-matrix, and the homogeneous space  $M$ (\ref{tp})
with a Poisson bracket $\pi$ 
that has to be compatible with the group action $\rhd: G\times M\to M$. Therefore, according to the possible classical  $r$-matrices of $\mathfrak g$ it follows  that the Lie group  $G$ may be endowed with several 
Poisson--Lie structures $\Pi$~\eqref{sklyanin}, each of them leading to a different Poisson homogeneous space~\cite{Drinfeld1993}.

A distinguished  type of Poisson homogeneous spaces are those  in which the Poisson bracket $\pi$ is obtained as the canonical projection on   $M$ with coordinates $(u^1,\dots,u^\ell)$  of the Poisson--Lie bracket $\Pi$~\cite{Ciccoli2006,BMN2017homogeneous} 
(see also~\cite{PHS,IvanTesis}). In terms of the underlying Lie bialgebra, this  requirement 
 corresponds  to imposing the so-called {\em coisotropy condition} for the cocommutator $\delta$ with respect to the isotropy subalgebra $\mathfrak h$ of $H$ given by~\cite{Ciccoli2006,BMN2017homogeneous} 
 \be
\delta(\mathfrak h) \subset \mathfrak h \wedge \mathfrak g.
\label{coisotropy}
\ee
A particular and very restrictive case of the above condition is that the subalgebra $\mathfrak h $ be a  sub-Lie bialgebra,
\be
\delta\left(\mathfrak{h}\right) \subset \mathfrak{h} \wedge \mathfrak{h} ,
\label{coisotropy2}
\ee
which implies that  the Poisson homogeneous space  is constructed through an isotropy subgroup $H$  which is a Poisson subgroup $(H,\Pi)$  of $(G,\Pi)$. Furthermore, since the quantum group  $(G_z,\Delta_{G_z})$  is the quantization   of the Poisson--Lie group $(G,\Pi)$, the    quantization of a coisotropic Poisson homogeneous space  $(M,\pi)$ fulfilling (\ref{coisotropy}) provides  a quantum homogeneous space $M_z$ with quantum coordinates $(\hat u^1,\dots, \hat u^\ell)$, onto which the quantum group $G_z$ co-acts covariantly \cite{Dijkhuizen1994}. The  coisotropy condition (\ref{coisotropy}) ensures that the commutation relations that define $M_z$ at the first-order in all the quantum coordinates close a Lie subalgebra which is just the   annihilator $\mathfrak{h}_\astt$ of $\mathfrak{h}$ on the dual Lie algebra $\mathfrak g^*$ and such relations determine a  Lie subalgebra of $\mathfrak{g}^\ast$ (\ref{td}).
 Generically, $M_z$ is called a {\em noncommutative space}.

Probably, the best known  and   most studied example of noncommutative spaces is the so-called $\kappa$-Minkowski spacetime coming from   the $\kappa$-Poincar\'e algebra~\cite{LNR1991realforms,LRNT1991,GKMMK1992,LNR1992fieldtheory,Maslanka1993,MR1994,Zakrzewski1994poincare}
where $\kappa$ is the quantum deformation parameter which is proportional to the Planck mass and here related to $z$ as $\kappa=z^{-1}$.  The quantum algebra 
${\mathcal U}_z(\mathfrak{g})$ and quantum group $G_z$ correspond to the $\kappa$-Poincar\'e algebra and $\kappa$-Poincar\'e group.  In this case, the underlying homogeneous space (\ref{tp})  is the flat (3+1)D Minkowski spacetime constructed as the coset space of the Poincar\'e group $G={\rm ISO}(3,1)$ with the Lorentz isotropy subgroup $H={\rm SO}(3,1)$:
\be
{\bf M}^{3+1}={\rm ISO}(3,1)/{\rm SO}(3,1). 
\label{trr}
\ee
Thus   the dimension is $\ell=4$ and the   coordinates $(  u^1,\dots,   u^4)$ are identified with   the time and spatial ones $(x^0,x^i)$ $(i=1,2,3)$. The $\kappa$-Poincar\'e classical $r$-matrix~\cite{Maslanka1993} provides a quasitriangular quantum deformation of the Poincar\'e algebra~\cite{LRNT1991,GKMMK1992,LNR1992fieldtheory} such that the Lorentz subalgebra $\mathfrak{h}=\mathfrak{so}(3,1)$ fulfils the coisotropy condition (\ref{coisotropy}) thus giving rise to the subalgebra $\mathfrak{h}_\astt$  whose generators are the quantum coordinates 
$(\hat x^0,\hat x^i)$ dual to the time and space translation generators.
 The complete quantization of $\mathfrak{h}_\astt$ is just the  $\kappa$-Minkowski spacetime ${\bf M}^{3+1}_\kappa$ which is defined by the commutation relations given by~\cite{Maslanka1993}:
\be
 [\hat x^0,\hat x^i] = -\frac{1}{\kappa} \, \hat x^i, \qquad [\hat x^i,\hat x^j] =0, 
\qquad
i,j=1,2,3 ,
\label{tss}
\ee
which are covariant under the $\kappa$-Poincar\'e quantum group~\cite{Zakrzewski1994poincare}. We remark that ${\bf M}^{3+1}_\kappa$  is a linear algebra which coincides exactly with the one obtained through the   Sklyanin bracket (\ref{sklyanin}) of the underlying classical  $r$-matrix~\cite{Maslanka1993} which provides the (linear) Poisson homogeneous spacetime. Therefore no higher-order terms  in the classical and quantum coordinates arise. By contrast, when the $\kappa$-deformation is applied to a 
curved manifold instead of (\ref{trr}), higher-order terms in the coordinates  appear in the   Poisson homogeneous spacetime, so that the corresponding quantization is not straightfoward at all,  as the recent constructions of the $\kappa$-noncommutative (anti-)de Sitter~\cite{kappaAdS},  Newtonian and Carrollian~\cite{kappaN}  spacetimes explicitly show.

It is worth stressing that the $\kappa$-Minkowski spacetime ${\bf M}^{3+1}_\kappa$ (\ref{tss})  has become the paradigmatic noncommutative space in the same form as Drinfel'd--Jimbo deformations are for quantum algebras;  in fact, we recall that the $\kappa$-Poincar\'e algebra was formerly obtained  as a real quantum algebra coming from a contraction of  the  
Drinfel'd--Jimbo deformation of $\mathfrak{sp}(4,\mathbb{C})$~\cite{LNR1991realforms, LRNT1991}. Among other issues and within the vast literature,  let us mention that 
$\kappa$-Minkowski space along with the  $\kappa$-Poincar\'e algebra  have been studied  in relation with noncommutative differential calculi~\cite{Sitarz1995plb,JMPS2015}, 
wave propagation on noncommutative spacetimes~\cite{AM2000waves}, deformed or doubly special relativity at the Planck scale~\cite{Amelino-Camelia:2002mn,Amelino-Camelia:2002vy,Kowalski-Glikman:2002we,Kowalskib,Lukierski:2002df,BP2010sigma},  noncommutative field theory~\cite{DJMTWW2003epjc, FKN2007plb, DJP2014sigma}, representation theory on Hilbert spaces~\cite{Agostini2007jmp,LMMP2018localization}, generalized $\kappa$-Minkowski spacetimes through twisted $\kappa$-Poincar\'e deformations~\cite{Daszkiewicz2008,BP2014sigma}, 
 deformed dispersion relations~\cite{BorowiecPachol2009jordanian,BGMP2010,ABP2017},   curved momentum spaces~\cite{KowalskiGlikman:2004tz,AmelinoCamelia:2011nt,GM2013relativekappa,BGGH2018momentum,MercatiMomentumMink},  relative locality phenomena~\cite{ALR2011speed},  star products~\cite{DS2013jncg},   deformed phase spaces~\cite{LSW2015hopfalgebroids},   noncommutative spaces of worldlines~\cite{Lines2014,BGH2019worldlinesplb} and light cones~\cite{MS2018plblightcone}   (in all cases see references therein).

 In Section~\ref{s53} we shall recover      the $\kappa$-Minkowski space ${\bf M}^{3+1}_\kappa$ (\ref{tss}) and the $\kappa$-Poincar\'e Lie bialgebra as a particular case of  ``time-like" deformations within the CK family of Lie bialgebras, a fact that is already well known~\cite{LBC,CK4d}.   However, as we shall show in Section~\ref{s33}, what is  a striking point  is that   a formally similar  structure to ${\bf M}^{3+1}_\kappa$  arises  as the first-order noncommutative  CK  space of points which is shared    by  63     CK bialgebras. Moreover the complete (in all orders in the quantum coordinates) noncommutative CK space of points is kept linear and shared by 27   CK bialgebras. Consequently,  a  linear noncommutative space  similar to  (\ref{tss})  is somewhat  ``ubiquitous" which, in turn,  suggests that additional ``structures" should be taken into account.
 In fact, this is one of the main aims of this paper  and  as a novel result we shall explicitly show in Section~\ref{s33} that the consideration of other noncommutative spaces beyond the space of points (kinematically, spacetimes)  associated with a given quantum algebra (namely, noncommutative spaces of lines, 2-planes and 3-hyperplanes) does allow one to distinguish mathematical and physical properties  
 between two quantum algebras with the same underlying  noncommutative space of points.


\subsection{Drinfel'd Double Structures}
\label{s13}

Let us assume that the dimension of the Lie algebra $\mathfrak{g}$ is even $n=2d$. In this case,   $\mathfrak{g}$ is a Lie algebra of  a  Drinfel'd double  Lie group~\cite{Drinfeld1987icm}   if there exists a basis $\{Y_1,\dots,Y_d,y^1,\dots,y^d \}$ of $\mathfrak g$ such that  the commutation relations (\ref{ta})  can be written as
\be
\begin{array}{l}
\displaystyle{ [Y_\alpha,Y_\beta]=\sum_{\gamma=1}^d C^\gamma_{\alpha \beta}Y_\gamma \,,   } \qquad 
\displaystyle{  
[y^\alpha,y^\beta]=\sum_{\gamma=1}^d F^{\alpha \beta}_\gamma y^\gamma }, \\[2pt] 
\displaystyle{  [y^\alpha ,Y_\beta] =\sum_{\gamma=1}^d\bigl( C^\alpha_{\beta \gamma}y^\gamma- F^{\alpha \gamma}_\beta Y_\gamma \bigr)}.
 \end{array}
\label{tr}
\ee
 Hence $\mathfrak{g}$ can be split into two Lie subalgebras   
 $$
 \mathfrak{g}_1=\mbox{span}\{Y_1,\dots,Y_d\},\qquad\mathfrak{g}_2=\mbox{span}\{y^1,\dots,y^d \}
 $$   
 with structure constants  $C^\gamma_{\alpha \beta}$ and $F^{\alpha \beta}_\gamma$, respectively. 
Both subalgebras are dual to each other,  $\mathfrak{g}_2^\ast= \mathfrak{g}_1$, by means of the duality   defined with respect to the nondegenerate symmetric bilinear form $\langle \, , \, \rangle : \mathfrak{g} \times \mathfrak{g} \rightarrow \mathbb{R}$ given by
\begin{align}
 \langle Y_\alpha,Y_\beta\rangle= 0,\qquad \langle y^\alpha,y^\beta\rangle=0, \qquad
\langle y^\alpha,Y_\beta\rangle= \delta^\alpha_\beta  \,,\qquad \forall \alpha,\beta  ,
\label{ts}
\end{align}
which is ``associative" or invariant in the sense that
\begin{align} 
\langle [X,Y],Z \rangle = \langle X,[Y,Z] \rangle,  \qquad \forall X,Y,Z \in \mathfrak{g}. 
\nonumber
\end{align}
The triple $(\mathfrak{g_1},  \mathfrak g_2= \mathfrak{g}^\ast_1,\mathfrak{g})$ is called a Manin triple and the Drinfel'd double  Lie group  is the unique connected and simply connected  Lie group $G$ with Lie algebra   $\mathfrak{g}$. 
Therefore the Lie algebra
$\mathfrak{g}$, verifying (\ref{tr}),  is  the double Lie algebra of $\mathfrak{g}_1$ and of its dual   algebra $\mathfrak{g}_1^\ast=\mathfrak{g}_2$.

By construction,  each Drinfel'd double structure for   $\mathfrak{g}$ has a canonical  classical
$r$-matrix  
\be
r_{\rm can}=\sum_{\alpha=1}^d {y^\alpha\otimes Y_\alpha} \, ,
\label{tu}
\ee
which is a solution of   the classical Yang--Baxter equation (\ref{tk}). Moreover,  the   universal enveloping algebra $\mathcal{U}(\mathfrak{g})$ of  $\mathfrak{g}$  has always a   quadratic Casimir element  given by
\begin{align}
C=\frac12\sum_{\alpha=1}^d{(y^\alpha\,Y_\alpha+Y_\alpha\,y^\alpha)} ,
\nonumber
\end{align}
which is directly related to the bilinear form~\eqref{ts}. The tensorized form of  $C$ reads
\be
\Omega= \frac12\sum_{\alpha=1}^d {(y^\alpha\otimes Y_\alpha+Y_\alpha\otimes y^\alpha)},
\label{tw}
\ee
which is   ad-invariant under the action of  $\mathfrak g$, that is  
\be [ X
\otimes 1+1\otimes X ,\Omega]=0 ,\qquad\forall X\in \mathfrak g.
\nonumber
\ee
The element  $\Omega$ leads to a skew-symmetric classical $r$-matrix for the Drinfel'd double Lie algebra   $\mathfrak{g}$ from the canonical one (\ref{tu}) in the form:
 \be 
r_D=r_{\rm can}-\Omega=\frac12\sum_{\alpha=1}^d   {y^\alpha \wedge Y_\alpha}  \,,
\label{ty}
\ee
which is a solution of the  modified classical Yang--Baxter equation (\ref{th}) (its Schouten bracket does not vanish now), so that $r_D$ defines a quasitriangular or standard quantum deformation of  $\mathfrak g$ with a coboundary Lie bialgebra $(\mathfrak g,\delta_D(r_D))$ determined by a cocommutator  $\delta_D$ through the relation (\ref{tf}).

 Concerning the   Drinfel'd--Jimbo quantum deformations of semisimple Lie algebras~\cite{Drinfeld1985Hopf,Jimbo,Drinfeld1987icm}, it is known that   they are closely related to quantum deformations of  Drinfel'd doubles, that is, quantum doubles, in such a manner that they are 
 ``almost" but not strictly speaking  quantum doubles~\cite{CP}.  Nevertheless,  it is remarkable that   proper Drinfel'd double structures for the four Cartan series of       semisimple Lie algebras on $\mathbb{C}$ have been obtained in~\cite{doubles1,doubles2} by enlarging the Lie algebras with an appropriate number of central extensions.

From a physical viewpoint, it is worth stressing that  Drinfel'd double structures are  naturally  related to  (2+1)D gravity, which is  a quite different theory from the full (3+1)D one~\cite{DJH1984,Carlip2003book}. In particular,  (2+1)D gravity  is a topological theory  which admits a description as a Chern--Simons theory with gauge group given by the group of isometries of the corresponding spacetime of constant curvature~\cite{AT1986,Witten1988}. The phase space structure of (2+1)D gravity is related to the moduli space of flat connections on a Riemann surface, the  symmetries of which are given by certain Poisson--Lie   groups \cite{AM1995moduli,FR1999moduli} such that the  Poisson structure on this space admits a   description in terms of coboundary Lie bialgebras associated with the gauge group.  Hence    quantum group symmetries arise as the quantum   counterparts of the (semiclassical) Poisson--Lie symmetries of the classical theory. The essential fact in the  (2+1)D gravity  framework is that 
 the relevant quantum group symmetries are those coming from   some classical $r$-matrices corresponding to 
    Drinfel'd double structures~\cite{MSquat,MS,MeusburgerN,BHM2010plb,BHM2013cqg,BHM2014plb,OS2018}, which    ensures that the Fock--Rosly condition~\cite{FR1999moduli} is fulfilled. The
     symmetric component of such
    admissible classical $r$-matrices, which  is just the  element $\Omega$ (\ref{tw}) when these are written in the symmetric form  (\ref{tu}),    must  be dual to the Ad-invariant symmetric bilinear form in the Chern--Simons action. As a consequence,   the $\kappa$-Poincar\'e and $\kappa$-(anti-)de Sitter symmetries are    not compatible~\cite{MSquat,MS} with the Chern--Simons formulation of (2+1)D gravity. Furthermore, the Chern--Simons approach to non-relativistic (2+1)D quantum gravity  has also   been developed in~\cite{SchroersGalilei,SchroersNH} by making use of a 
    two-fold central extension of the Galilei~\cite{LeblondGalilei}   and Newton--Hooke algebras, and their  full quantum  deformation has been obtained in~\cite{NaranjoNH}. 
   Additionally,    Drinfel'd doubles also play  a prominent role in state sum or spin foam models for (2+1)D gravity as shown in~\cite{balskir,turvire}   in the context of the Turaev--Viro model and invariant.
 
We recall that   the classifications of non-isomorphic  4D and 6D real  Drinfel'd double structures were   carried out in~\cite{HS2002tdual2D} and~\cite{SH2002}, respectively, while their Hopf algebra quantizations  were constructed in~\cite{BCO2005quantizationDD}. From these results and also from~\cite{gomez}, there were obtained   
 the classifications of Drinfel'd double structures for the  (2+1)D (anti-)de Sitter algebras in~\cite{BHM2013cqg},  (2+1)D Poincar\'e algebra
 and    centrally extended (1+1)D  Poincar\'e algebra in~\cite{PoinDD}, and  3D Euclidean algebra in~\cite{EuclDD}.
By contrast,  results concerning Drinfel'd double structures  in  the (3+1)D case are very scarce,  only covering the real $\mathfrak{so}(5)$  and 
  anti-de Sitter  $\mathfrak{so}(3,2)$ algebras given in~\cite{BHN2015towards31}. In this respect, we advance that   in Section~\ref{s34}  we shall obtain two new classical $r$-matrices coming from   Drinfel'd doubles: one for the de Sitter    $\mathfrak{so}(4,1)$ and  another for  the anti-de Sitter    $\mathfrak{so}(3,2)$.


\section{The Drinfel'd--Jimbo Lie Bialgebra for $\mathfrak{so}(5)$}
\label{s2}

Let us consider the   real orthogonal Lie algebra  $\mathfrak{so}(5)$  with  generators
$\{J_{ab}\}$   $(a<b;\,  a,b=0,1,\dots,4)$
  fulfilling the Lie brackets   
\be
[J_{ab}, J_{ac}] = J_{bc} , \qquad
[J_{ab}, J_{bc}] = -J_{ac}  ,\qquad
[J_{ac}, J_{bc}] = J_{ab} ,  \qquad a<b<c,
  \label{ab}
\ee
and such that   those commutators involving four different indices  are equal to zero.
 The universal enveloping algebra of the Lie algebra  $\mathfrak{so}(5)$  is endowed with two  (second- and fourth-order) Casimir operators~\cite{Gelfand,casimirs}.  The quadratic  one, coming from the Killing--Cartan form, is given by  
\be
\mathcal{C}=\!\sum_{0\le a<b\le 4} \! J_{ab}^2\,  .
\label{acc}
\ee

A fine grading group ${\mathbb {Z}}_2^{\otimes 4}$   of $\mathfrak{so}(5)$  is spanned 
by the   four commuting involutive automorphisms $\Theta^{(m)}$ $(m=1,\dots,4)$ of  
(\ref{ab}) defined by~\cite{CKMontigny, CKGraded}:
\be
\Theta^{(m)}(J_{ab}):=
\left\{
 \begin{array}{rl}
J_{ab},&\mbox{if either $m \le a $ or $b<m$};\\[2pt]
-J_{ab},&\mbox{if   $a<m \le b$} .
\end{array}\right.
\label{inv}
\ee
Each involution $\Theta^{(m)}$ provides a Cartan  decomposition  of $\mathfrak{so}(5)$ in invariant   and
anti-invariant subspaces  denoted
$\mathfrak{h}^{(m)}$ and $\mathfrak{t}^{(m)}$  respectively:
\be
{\mathfrak{so}}(5)= \mathfrak{h}^{(m)}\oplus \mathfrak{t}^{(m)},\qquad m=1,\dots, 4 .
\label{db}
\ee
These  subspaces verify that
\be
[\mathfrak{h}^{(m)},\mathfrak {h}^{(m)}] \subset \mathfrak{h}^{(m)}  ,\qquad 
[ \mathfrak{h}^{(m)}, \mathfrak{t}^{(m)}] \subset \mathfrak{t}^{(m)} , \qquad 
[ \mathfrak {t}^{(m)}, \mathfrak{t}^{(m)}] \subset \mathfrak{h}^{(m)}   ,
\label{hp}
\ee
    where $\mathfrak{h}^{(m)}$  is a Lie subalgebra such that
    \be
\mathfrak{h}^{(1)}={\mathfrak{so}}(4) ,\quad\ \mathfrak{h}^{(2)}={\mathfrak{so}}(2)\oplus
{\mathfrak{so}}(3),\quad\ \mathfrak{h}^{(3)}={\mathfrak{so}}(3)\oplus
{\mathfrak{so}}(2),\quad\ \mathfrak{h}^{(4)}={\mathfrak{so}}(4) ,
\label{dc}
\ee
while the vector subspace $\mathfrak{t}^{(m)}$ is not   a subalgebra.

A faithful matrix representation of     $\mathfrak{so}(5)$, $\rho:\mathfrak{so}(5)  \rightarrow \text{End} (\mathbb{R}^5)$, is given by 
\be
\rho(J_{ab})=  - e_{ab}+e_{ba}, \quad
\label{dd}
\ee
where $e_{ab}$ is the $5\times 5$ matrix with a single non-zero  entry 1 at row $a$
and column $b$ $(a,b=0,1,\dots,4)$, fulfilling  the orthogonal matrix condition
\begin{equation}
\rho(J_{ab})^T \bfI\,+\bfI \, \rho(J_{ab})=0,\qquad \bfI={\rm diag}(1,1,1,1,1)  ,
\label{de}
\end{equation}
where $\rho(J_{ab})^T$ is the transpose matrix of $\rho(J_{ab})$.

\subsection{Symmetric Homogeneous Spaces}
\label{s21}

According to each automorphism  $\Theta^{(m)}$ (\ref{inv}) and its associated    Cartan  decomposition (\ref{db})  we construct      four  symmetric homogeneous  spaces, of the type (\ref{tp}), as the coset spaces~\cite{gil, HS1997phasespaces,deform} 
\be
\mathbb S^{(m)}= {\rm SO}(5)/H^{(m)}, \qquad m=1,\dots,4,
\label{dff}
\ee
where $H^{(m)}$ is the isotropy    subgroup with Lie algebra $\mathfrak{h}^{(m)}$ in (\ref{db})  (see (\ref{dc})). We briefly describe their structure.

\begin{enumerate}

\item {\em 4D space of points}. We write  the ten generators of $\mathfrak{so}(5)$  in array form and the decomposition (\ref{db}) for  $\Theta^{(1)}$ gives
\be
\noindent
\begin{tabular}{|cccc}
$J_{01} $&$ J_{02} $& $  J_{03} $&$  J_{04} $\\[1pt] 
\cline{1-4}
 \multicolumn{1}{c}{\,} &$  J_{12} $& $  J_{13} $&$  J_{14} $\\
 \multicolumn{1}{c}{\,} & & $  J_{23} $&$  J_{24} $\\
 \multicolumn{1}{c}{\,} &  & &$  J_{34} $ 
\end{tabular}
\label{dia}
\ee
where the four generators in the rectangle span the subspace $\mathfrak{t}^{(1)}$.
Hence we get the coset space
\be
\mathbb S^{(1)}={\rm SO}(5)/{\rm SO}(4) ,\qquad H^{(1)}\equiv {\rm SO}(4)=\langle  J_{12},J_{13},J_{14},J_{23},J_{24},J_{34}\rangle ,
\nonumber
\ee
which is identified with the symmetric homogeneous  space of points.  The  subgroup $H^{(1)}$ is the isotropy (or stabilizer) group of a    point which is taken as the origin in $\mathbb S^{(1)}$   so that its generators play the role of rotations on $\mathbb S^{(1)}$ (leaving the origin invariant), 
while the generators of $\mathfrak{t}^{(1)}$
  play the role of translations on $\mathbb S^{(1)}$ (so moving the origin along four basic directions).
  
\item {\em 6D space of lines}. The decomposition (\ref{db}) for  $\Theta^{(2)}$ can be represented as
\be
\noindent
\begin{tabular}{c|ccc}
$J_{01} $&$ J_{02} $& $  J_{03} $&$  J_{04} $\\ 
  &$  J_{12} $& $  J_{13} $&$  J_{14} $\\[1pt] 
\cline{2-4}
 \multicolumn{1}{c}{\,} & & $  J_{23} $&$  J_{24} $\\
 \multicolumn{1}{c}{\,} &  & &$  J_{34} $ 
\end{tabular}
\label{diab}
\ee
where now the six generators in the rectangle span the subspace $\mathfrak{t}^{(2)}$.
We find the coset space
\be
\mathbb S^{(2)}={\rm SO}(5)/\bigl( {\rm SO}(2) \otimes {\rm SO}(3) \bigr) , \quad\  H^{(2)}\equiv {\rm SO}(2) \otimes {\rm SO}(3) =\langle  J_{01}     \rangle \otimes \langle  J_{23},J_{24},J_{34}\rangle ,
\nonumber
\ee
which is interpreted as the symmetric homogeneous  space of lines.  The subgroup $H^{(2)}$ is the isotropy group of a line (the origin in  $\mathbb S^{(2)}$), while the six generators in $\mathfrak{t}^{(2)}$ play the role of translations on $\mathbb S^{(2)}$  (moving the origin-line).

\item {\em 6D space of 2-planes}. The decomposition (\ref{db}) for  $\Theta^{(3)}$  is displayed as 
\be
\noindent
\begin{tabular}{cc|cc}
$J_{01} $&$ J_{02} $& $  J_{03} $&$  J_{04} $\\ 
  &$  J_{12} $& $  J_{13} $&$  J_{14} $\\
 \multicolumn{1}{c}{\,}  & & $  J_{23} $&$  J_{24} $\\[1pt] 
\cline{3-4}
 \multicolumn{2}{c}{\,}   & &$  J_{34} $ 
\end{tabular}
\label{diac}
\ee
where   the six generators in the rectangle span the subspace $\mathfrak{t}^{(3)}$.
The coset space reads
\be
\mathbb S^{(3)}={\rm SO}(5)/\bigl( {\rm SO}(3) \otimes {\rm SO}(2) \bigr) , \quad\  H^{(3)}\equiv {\rm SO}(3) \otimes {\rm SO}(2) =\langle  J_{01} ,J_{02} ,J_{12}     \rangle \otimes \langle J_{34}\rangle ,
\nonumber
\ee
which corresponds to the symmetric homogeneous  space of 2-planes.  The six generators in $\mathfrak{t}^{(3)}$ play the role of translations on $\mathbb S^{(3)}$, while $H^{(3)}$ is the isotropy group of a 2-plane. 

\item {\em 4D space of 3-hyperplanes}. Finally, the decomposition (\ref{db}) for  $\Theta^{(4)}$ yields
\be
\noindent
\begin{tabular}{ccc|c}
$J_{01} $&$ J_{02} $& $  J_{03} $&$  J_{04} $\\ 
  &$  J_{12} $& $  J_{13} $&$  J_{14} $\\
 \multicolumn{1}{c}{\,}  & & $  J_{23} $&$  J_{24} $\\[1pt] 
 \multicolumn{2}{c}{\,}   & &$  J_{34} $ \\
 \cline{4-4}
\end{tabular}
\label{diad}
\ee
where   the four generators in the rectangle span the subspace $\mathfrak{t}^{(4)}$.
The coset space is given by
\be
\mathbb S^{(4)}={\rm SO}(5)/{\rm SO}(4) , \qquad H^{(4)}\equiv {\rm SO}(4)=\langle  J_{01},J_{02},J_{03},J_{12},J_{13},J_{23}\rangle ,
\nonumber
\ee
which is interpreted as the symmetric homogeneous  space of 3-hyperplanes.  The four generators in $\mathfrak{t}^{(4)}$ play the role of translations on $\mathbb S^{(4)}$, while $H^{(4)}$ is the isotropy group of a 3-hyperplane. 

 \end{enumerate}

Some remarks are in order. Firstly,  the four spaces (\ref{dff}) are of positive constant curvature in the sense that their  sectional curvature $K$ is equal to $+1$ and they are endowed with a Riemmanian metric (so with positive definite signature).  
Secondly, the 4D spaces of points and   3-hyperplanes are of rank 1, that is, there is a single invariant under the action of ${\rm SO}(5)$ for a pair of points (the ordinary distance) or 3-hyperplanes.  
The  6D spaces of lines and   2-planes are of rank 2~\cite{HS1997phasespaces,deform} so that there are two independent invariants under the action of ${\rm SO}(5)$ for a pair of lines (an angle and the distance between two lines) or 2-planes (see~\cite{Jordan} for the Euclidean case).  
And thirdly,  there is a relevant automorphism of $\mathfrak{so}(5)$ defined by
\be
\mathcal{D}(J_{ab}):=-J_{4-b\, 4-a}   \,,
\label{od}
\ee  
that  in the array display of the generators  is visualized as
\be
\noindent
\begin{tabular}{cccc}
$J_{01} $&$ J_{02} $& $  J_{03} $&\fbox{$  J_{04}$}  \\ 
  &$  J_{12} $&\fbox{ $  J_{13} $}&$  J_{14} $ \\ 
 & & $  J_{23} $&$  J_{24} $ \\ 
&  & &$  J_{34} $ 
\end{tabular}
\qquad   \stackrel{\mathcal{D}}{\longleftrightarrow}  \quad 
\begin{tabular}{cccc}
$-J_{34} $&$- J_{24} $& $  -J_{14} $&\fbox{$ - J_{04} $} \\ 
  &$  -J_{23} $&\fbox{ $ -J_{13} $}&$  -J_{03} $ \\ 
 & & $ - J_{12} $&$ - J_{02} $ \\ 
&  & &$ - J_{01} $ 
\end{tabular}
\label{d0}
\ee
leaving  $J_{04}$ and $J_{13}$ invariant (up to the minus sign).
Thus the map $\mathcal{D}$ interchanges the spaces of points and 3-hyperplanes, and the spaces of lines and 2-planes:
\be
\mathbb S^{(1)}\,  \stackrel{\mathcal{D}}{\longleftrightarrow}\ \mathbb S^{(4)} ,\qquad 
\mathbb S^{(2)} \, \stackrel{\mathcal{D}}{\longleftrightarrow}\ \mathbb S^{(3)} .
\label{d01}
\ee
Note that $\mathcal{D}^2={\rm Id}$. The map $\mathcal{D}$ will be called 
 {\em polarity},  since for $\mathfrak{so}(3)$  reduces to the well known duality in projective geometry   interchanging the 2D space of points with the 2D space of lines (see~\cite{trigo} and references therein) which, only  at this dimension, are both of rank 1. Note that this map is sometimes called ordinary duality~\cite{trigo}, although in this paper we will always call it polarity in order to avoid confusion with the completely unrelated notion of quantum duality.

\subsection{Lie Bialgebra}
\label{s22}

Let us consider the   Drinfel'd--Jimbo quantum deformation of the real compact form $\mathfrak{g}=\mathfrak{so}(5)$~\cite{Drinfeld1985Hopf,Jimbo,Drinfeld1987icm} which in the basis (\ref{ab})  is generated by the following classical $r$-matrix~\cite{LBC}
\be
r_{04,13}=z(J_{14}\wedge J_{01} + J_{24}\wedge J_{02}
+ J_{34}\wedge J_{03} + J_{23}\wedge J_{12}) .
\label{ec}
\ee
Recall that $z$ is the quantum deformation parameter (such that $q={\rm e}^z$) and, hereafter,  it will be assumed that $z$  is an {\em    indeterminate real parameter}. We remark that $r_{04,13}$ is a   solution of the modified classical Yang--Baxter equation (\ref{th}) so that this underlies a quasitriangular Hopf algebra structure.   Therefore the corresponding cocommutator is    coboundary~\cite{Drinfeld1983hamiltonian},  $\delta=\delta(r)$, so that this is obtained from (\ref{ec}) through the relation (\ref{tf}),   yielding
\bea
&& \delta(J_{04})=0,\qquad  \delta(J_{13})=0,\nonumber\\[2pt]
&& \delta(J_{12})=z J_{12}\wedge
J_{13}, \qquad
\delta(J_{23})=z J_{23}\wedge J_{13},\nonumber\\[2pt]
&& \delta(J_{01})=z(
J_{01}\wedge J_{04}+
J_{24}\wedge J_{12}+
J_{34}\wedge J_{13}+
J_{02}\wedge J_{23}), \nonumber\\[2pt]
&& \delta(J_{02})=z(
J_{02}\wedge J_{04}+
J_{12}\wedge J_{14}+
J_{34}\wedge J_{23}+
J_{23}\wedge J_{01}+
J_{12}\wedge J_{03}), \nonumber\\[2pt]
&& \delta(J_{03})=z(
J_{03}\wedge J_{04}+
J_{13}\wedge J_{14}+
J_{23}\wedge J_{24}+
J_{02}\wedge J_{12}), \label{coco} \\[2pt]
&& \delta(J_{14})=z(
J_{14}\wedge J_{04}+
J_{13}\wedge J_{03}+
J_{12}\wedge J_{02}+
J_{24}\wedge J_{23}), \nonumber\\[2pt]
&& \delta(J_{24})=z(
J_{24}\wedge J_{04}+
J_{23}\wedge J_{03}+
J_{01}\wedge J_{12}+
J_{12}\wedge J_{34}+
J_{23}\wedge J_{14}), \nonumber\\[2pt]
&& \delta(J_{34})=z(
J_{34}\wedge J_{04}+
J_{02}\wedge J_{23}+
J_{01}\wedge J_{13}+
J_{24}\wedge J_{12}). \nonumber
\eea
The resulting real Lie bialgebra $(\mathfrak{so}(5),\delta(r_{04,13}))$ is so determined by the commutation rules (\ref{ab}) and cocommutator (\ref{coco}). The indices in $r_{04,13}$ (\ref{ec}) indicate the primitive generators $J_{04}$ and $J_{13}$. The 
first  primitive generator   $J_{04}$ is the ``main"  one  in the sense that, once the CK scheme of contractions be introduced and further applied to kinematical algebras,  it will  provide dimensions of the deformation parameter $z$ since the product $zJ_{04}$ must be dimensionless~\cite{CK2d,CK3d}; this fact will be studied in detail in Section~\ref{s5}.
 The last term  of $r_{04,13}$  is a classical $r$-matrix $r_{13}=z J_{23}\wedge J_{12}$
giving rise  to the  Drinfel'd--Jimbo Lie bialgebra $(\mathfrak{so}(3),\delta(r_{13}))$, with $\mathfrak{so}(3)= \mbox{span}\{ J_{12},J_{13},J_{23}\}$  and primitive generator $J_{13}$,  
which is a sub-Lie bialgebra of
$(\mathfrak{so}(5),\delta(r_{04,13}))$; thus $J_{13}$ plays the role of a
``secondary" primitive generator  in  $r_{04,13}$~\cite{LBC}.

Now we analyse how to implement the ${\mathbb {Z}}_2^{\otimes 4}$-grading of $\mathfrak{so}(5)$ into $(\mathfrak{so}(5),\delta(r_{04,13}))$.  This requires to generalize the   action of the automorphisms $\Theta^{(m)}: \mathfrak{so}(5)\to \mathfrak{so}(5)$~\eqref{inv} onto the cocommutator  $\delta: \mathfrak{so}(5)\to \mathfrak{so}(5)\otimes\mathfrak{so}(5)$ and also to consider 
a possible action on the  quantum deformation parameter~\cite{CK2d,CK3d}. Recall that $\delta$   (\ref{coco}) is the  skew-symmetric part of the first-order term  in $z$ (\ref{tm})  of the full  coproduct  $\Delta$ of the  real quantum   algebra  ${\mathcal U}_z(\mathfrak{so}(5))={\mathcal U}(\mathfrak{so}(5)) \, \hat \otimes  \,\mathbb{R}[[z]]$ such that, as mentioned above,  $z$  is an   indeterminate parameter.
 Since   $z$ is linked to the ``main" primitive generator $J_{04}$ through the product $zJ_{04}$,   both elements  must be transformed in the same way. By taking into account that $J_{04}\to -J_{04}$ under the four maps $\Theta^{(m)}$, then $z\to -z$ as well.  Hence we define four $z$-maps   $(m=1,\dots,4)$ 
\be
\begin{array}{l}
\Theta^{(m)}_z\bigl(\delta(J_{ab})\bigr):= \delta\bigl(\Theta^{(m)} (J_{ab})  \bigr) ,\\[2pt]
\Theta^{(m)}_z(z J_{ab}\otimes J_{cd} ):= \bigl( -z \,\Theta^{(m)} (J_{ab})\otimes \Theta^{(m)} (J_{cd})  \bigr)  ,
\end{array}
\label{invz}
\ee 
where  $\Theta^{(m)}$ is given in (\ref{inv}).  Notice that the second relation in (\ref{invz}) can directly be  applied to the $r$-matrix (\ref{ec}), consistently with  the relation (\ref{tf}), and extended to higher-order tensor product spaces.   When the $z$-maps (\ref{invz}) are applied either to $\delta$  (\ref{coco})  or  to the $r$-matrix (\ref{ec})  one finds that only $\Theta^{(2)}_z$ and $\Theta^{(3)}_z$ remain as involutive automorphisms of $(\mathfrak{so}(5),\delta(r_{04,13}))$, meanwhile  $\Theta^{(1)}_z$ and $\Theta^{(4)}_z$ are no longer  involutions. This is a consequence of the presence of the term   $r_{13}=zJ_{23}\wedge J_{12}$ in $r_{04,13}$ which does not appear neither in the Drinfel'd--Jimbo quantum deformation of 
 $\mathfrak{so}(3)$~\cite{CK2d} nor in  $\mathfrak{so}(4)$~\cite{CK3d}; for these latter deformations the whole  initial ${\mathbb {Z}}_2^{\otimes 2}$- and  ${\mathbb {Z}}_2^{\otimes 3}$-grading  is kept, respectively, but for $(\mathfrak{so}(5),\delta(r_{04,13}))$ there only remains a  ${\mathbb {Z}}_2^{\otimes 2}$-grading spanned by  $\Theta^{(2)}_z$ and $\Theta^{(3)}_z$.

Likewise, the polarity (\ref{od}) can   be implemented into  $(\mathfrak{so}(5),\delta(r_{04,13}))$ by also considering an action on the deformation parameter defined by~\cite{CK2d,CK3d}
\be
\begin{array}{l}
\mathcal{D}_z\bigl(\delta(J_{ab})\bigr):= \delta\bigl(\mathcal{D}(J_{ab})  \bigr) ,\\[2pt] 
\mathcal{D}_z(z J_{ab}\otimes J_{cd}  ):= \bigl( -z \, \mathcal{D} (J_{ab})\otimes \mathcal{D} (J_{cd}) \bigr)  ,
\end{array}
\label{odz}
\ee 
with $ \mathcal{D}$ given in  (\ref{od}), so that   $\mathcal{D}_z^2={\rm Id}$.
It can be checked that  the $r$-matrix (\ref{ec}) remains invariant under this ``$z$-polarity" (and, obviously, the cocommutator  (\ref{coco}) as well). Note that both primitive generators $J_{04}$ and $J_{13}$ are  kept unchanged under  $ \mathcal{D} $ (up to the minus sign) as shown in (\ref{d0}).

From (\ref{coco})  it is straightforward to prove that this deformation fulfils the coisotropy condition (\ref{coisotropy})~\cite{Ciccoli2006,BMN2017homogeneous} for the 
 four isotropy subalgebras  $\mathfrak{h}^{(m)}$ (\ref{dc}):
 \be
\delta( \mathfrak{h}^{(m)})\subset \mathfrak{h}^{(m)}\wedge   \mathfrak{so}(5) , \qquad m=1,\dots,4.
\label{coiso}
\ee
  Thus each of them would provide a Poisson homogeneous space.
Notice, however, that  none of them   leads to  a Poisson subgroup   since the condition (\ref{coisotropy2}) is not satisfied:
 \be
\delta( \mathfrak{h}^{(m)})\not\subset \mathfrak{h}^{(m)}\wedge    \mathfrak{h}^{(m)}, \qquad m=1,\dots,4.
\nonumber
\ee

\subsection{Dual Lie Algebra and Noncommutative Spaces}
\label{s23}

According to Section~\ref{s11},  we denote by $\JJ^{ab}$ $(a<b;\,  a,b=0,1,\dots,4)$ the generators in $\mathfrak g^\ast=\mathfrak{so}(5)^\ast$  dual to  $J_{ab}$ in 
$\mathfrak g =\mathfrak{so}(5)$    with canonical    pairing  defined by (\ref{tdd}):
\be
\langle \JJ^{ab},J_{cd}\rangle= \delta _{c}^{a}\delta _{d}^{b} .
\label{pairing}
\ee
From the cocommutator (\ref{coco}), read as (\ref{tc}), we obtain the commutation relations of the dual  Lie algebra $\mathfrak{so}(5)^\ast$ (\ref{td}), which are given by
\be
\begin{array}{lll}
[\JJ^{01}, \JJ^{02}] =0 , &\quad
[\JJ^{01}, \JJ^{12}] =z  \JJ^{24}  ,&\quad
[\JJ^{02}, \JJ^{12}] =z( \JJ^{03} -  \JJ^{14} ),\\[2pt]
[\JJ^{01}, \JJ^{03}] =0 ,&\quad
[\JJ^{01}, \JJ^{13}] = z  \JJ^{34} , &\quad
[\JJ^{03}, \JJ^{13}] =-z   \JJ^{14} ,\\[2pt]
[\JJ^{01}, \JJ^{04}] =z  \JJ^{01} , &\quad
[\JJ^{01}, \JJ^{14}] =0 , &\quad
[\JJ^{04}, \JJ^{14}] =-z  \JJ^{14} ,\\[2pt]
[\JJ^{02}, \JJ^{03}] =0 ,&\quad
[\JJ^{02}, \JJ^{23}] = z(\JJ^{01} +\JJ^{34}) ,&\quad
[\JJ^{03}, \JJ^{23}] =-z  \JJ^{24} ,\\[2pt]
[\JJ^{02}, \JJ^{04}] = z \JJ^{02}  ,&\quad
[\JJ^{02}, \JJ^{24}] = 0 , &\quad
[\JJ^{04}, \JJ^{24}] = -z \JJ^{24}, \\[2pt]
[\JJ^{03}, \JJ^{04}] =z  \JJ^{03} , &\quad
[\JJ^{03}, \JJ^{34}] =0 ,&\quad
[\JJ^{04}, \JJ^{34}] = -z \JJ^{34} ,\\[2pt]
[\JJ^{12}, \JJ^{13}] =z  \JJ^{12}  ,&\quad
[\JJ^{12}, \JJ^{23}] = 0 , &\quad
[\JJ^{13}, \JJ^{23}] = -z \JJ^{23} ,\\[2pt]
[\JJ^{12}, \JJ^{14}] = z \JJ^{02} , &\quad
[\JJ^{12}, \JJ^{24}] = -z(\JJ^{01} +\JJ^{34}) ,&\quad
[\JJ^{14}, \JJ^{24}] =0, \\[2pt]
[\JJ^{13}, \JJ^{14}] =z  \JJ^{03}  ,&\quad
[\JJ^{13}, \JJ^{34}] =-z \JJ^{01} ,&\quad
[\JJ^{14}, \JJ^{34}] =0,\\[2pt]
[\JJ^{23}, \JJ^{24}] =z( \JJ^{03} -  \JJ^{14} ), &\quad
[\JJ^{23}, \JJ^{34}] = -z \JJ^{02}  ,&\quad
[\JJ^{24}, \JJ^{34}] =0 ,
\end{array}
\label{aab}
\ee
\be
\begin{array}{lll}
[\JJ^{01}, \JJ^{23}] =  -z \JJ^{02} , &\quad
[\JJ^{01}, \JJ^{24}] = 0, &\quad\ 
[\JJ^{01}, \JJ^{34}] = 0, \\[2pt]
[\JJ^{02}, \JJ^{13}] = 0 ,&\quad
[\JJ^{02}, \JJ^{14}] = 0 ,&\quad\ 
[\JJ^{02}, \JJ^{34}] = 0, \\[2pt]
[\JJ^{03}, \JJ^{12}] =  -z \JJ^{02} , &\quad
[\JJ^{03}, \JJ^{14}] = 0, &\quad\ 
[\JJ^{03}, \JJ^{24}] = 0, \\[2pt]
[\JJ^{04}, \JJ^{12}] = 0 ,&\quad
[\JJ^{04}, \JJ^{13}] = 0 ,&\quad\ 
[\JJ^{04}, \JJ^{23}] = 0, \\[2pt]
[\JJ^{12}, \JJ^{34}] =   z \JJ^{24} , &\quad
[\JJ^{13}, \JJ^{24}] = 0, &\quad\ 
[\JJ^{14}, \JJ^{23}] = -  z \JJ^{24} .
\end{array}
\label{aacx}
\ee
In contrast to the commutation rules of $\mathfrak{so}(5)$ (\ref{ab}), the Lie brackets of $\mathfrak{so}(5)^\ast$ involving four different indices  (\ref{aacx}) are no longer equal to zero.

Next we express the dual Lie algebra $\mathfrak{so}(5)^\ast$ as the sum of two vector spaces 
\be
{\mathfrak{so}}(5)^\ast= \mathfrak{h}^{(m)}_\astt\oplus \mathfrak{t}^{(m)}_\astt ,\qquad m=1,\dots, 4,
\label{fc}
\ee
where $\mathfrak{h}^{(m)}_\astt$ and $\mathfrak{t}^{(m)}_\astt$ are, in this order, the annihilators in $\mathfrak{so}(5)^*$ of the vector subspaces $\mathfrak{h}^{(m)}$ and $\mathfrak{t}^{(m)}$ of $\mathfrak{so}(5)$ given in (\ref{db})  and verifying (\ref{hp}). From the results presented in Section~\ref{s12}, it turns out that each $\mathfrak{h}^{(m)}_\astt$    leads to a  linear noncommutative   space which is      the first-order in the quantum coordinates of the full noncommutative space associated with the homogeneous space (\ref{dff}). We shall denote such a {\em first-order noncommutative space} by
$\mathbb S^{(m)}_z\equiv \mathfrak{h}^{(m)}_\astt $.

Following~\cite{PHS, IvanTesis} we analyse the relations
$\bigl[\mathfrak{h}^{(m)}_\astt,\mathfrak {h}^{(m)}_\astt \bigr] $, $\bigl[ \mathfrak{h}^{(m)}_\astt, \mathfrak{t}^{(m)}_\astt \bigr] $  and $\bigl[ \mathfrak {t}^{(m)}_\astt, \mathfrak{t}^{(m)}_\astt \bigr] $ for each $m$. Such structures do  depend on the chosen quantum deformation (here determined by (\ref{ec})) and they   are directly deduced from (\ref{aab}) and (\ref{aacx}):
\begin{enumerate} 

\item    {\em Noncommutative   space of points}   $\mathbb S^{(1)}_z\equiv \mathfrak{h}^{(1)}_\astt=\langle \JJ^{01},\JJ^{02},\JJ^{03},\JJ^{04}\rangle$:
\be
\bigl[ \mathfrak {h}^{(1)}_\astt, \mathfrak{h}^{(1)}_\astt \bigr] \subset \mathfrak{h}^{(1)}_\astt ,\qquad 
\bigl[ \mathfrak{h}^{(1)}_\astt, \mathfrak{t}^{(1)}_\astt \bigr] \subset \mathfrak{t}^{(1)}_\astt +\mathfrak{h}^{(1)}_\astt   , \qquad 
\bigl[\mathfrak{t}^{(1)}_\astt,\mathfrak {t}^{(1)}_\astt \bigr] \subset \mathfrak{h}^{(1)}_\astt  +  \mathfrak{t}^{(1)}_\astt  .
\label{fd}
\ee

\item  {\em Noncommutative space of lines}   $\mathbb S^{(2)}_z\equiv \mathfrak{h}^{(2)}_\astt=\langle \JJ^{02},\JJ^{03},\JJ^{04},\JJ^{12},\JJ^{13},\JJ^{14}\rangle$:
\be
\bigl[ \mathfrak {h}^{(2)}_\astt, \mathfrak{h}^{(2)}_\astt \bigr] \subset \mathfrak{h}^{(2)}_\astt ,\qquad 
\bigl[ \mathfrak{h}^{(2)}_\astt, \mathfrak{t}^{(2)}_\astt \bigr] \subset \mathfrak{t}^{(2)}_\astt   , \qquad 
\bigl[\mathfrak{t}^{(2)}_\astt,\mathfrak {t}^{(2)}_\astt \bigr] \subset \mathfrak{h}^{(2)}_\astt    .
\label{fe}
\ee

\item {\em Noncommutative space of 2-planes}    $\mathbb S^{(3)}_z\equiv \mathfrak{h}^{(3)}_\astt=\langle \JJ^{03},\JJ^{04},\JJ^{13},\JJ^{14},\JJ^{23},\JJ^{24}\rangle$:
\be
\bigl[ \mathfrak {h}^{(3)}_\astt, \mathfrak{h}^{(3)}_\astt \bigr] \subset \mathfrak{h}^{(3)}_\astt ,\qquad 
\bigl[ \mathfrak{h}^{(3)}_\astt, \mathfrak{t}^{(3)}_\astt \bigr] \subset \mathfrak{t}^{(3)}_\astt   , \qquad 
\bigl[\mathfrak{t}^{(3)}_\astt,\mathfrak {t}^{(3)}_\astt \bigr] \subset \mathfrak{h}^{(3)}_\astt     .
\label{ff}
\ee

\item {\em Noncommutative space of 3-hyperplanes} with  $\mathbb S^{(4)}_z\equiv \mathfrak{h}^{(4)}_\astt=\langle \JJ^{04},\JJ^{14},\JJ^{24},\JJ^{34}\rangle$:
\be
\bigl[ \mathfrak {h}^{(4)}_\astt, \mathfrak{h}^{(4)}_\astt \bigr] \subset \mathfrak{h}^{(4)}_\astt ,\qquad 
\bigl[ \mathfrak{h}^{(4)}_\astt, \mathfrak{t}^{(4)}_\astt \bigr] \subset \mathfrak{t}^{(4)}_\astt +\mathfrak{h}^{(4)}_\astt   , \qquad 
\bigl[\mathfrak{t}^{(4)}_\astt,\mathfrak {t}^{(4)}_\astt \bigr] \subset \mathfrak{h}^{(4)}_\astt  +  \mathfrak{t}^{(4)}_\astt   .
\label{fg}
\ee
\end{enumerate}
Consequenlty,  the four first-order noncommutative spaces $\mathbb S^{(m)}_z$  close on a Lie subalgebra of  $\mathfrak{so}(5)^\ast$
\be
\bigl[ \mathfrak {h}^{(m)}_\astt, \mathfrak{h}^{(m)}_\astt] \subset \mathfrak{h}^{(m)}_\astt ,
\nonumber
\ee
as it should be, since this     is a direct consequence of the coisotropy condition (\ref{coiso})~\cite{Ciccoli2006,BMN2017homogeneous,PHS, IvanTesis}. Furthermore, the noncommutative spaces of lines $\mathbb S^{(2)}_z$ and 2-planes $\mathbb S^{(3)}_z$ are both reductive and symmetric
\be
\bigl[ \mathfrak{h}^{(l)}_\astt, \mathfrak{t}^{(l)}_\astt] \subset \mathfrak{t}^{(l)}_\astt ,\qquad 
 \bigl[\mathfrak{t}^{(l)}_\astt,\mathfrak {t}^{(l)}_\astt \bigr] \subset \mathfrak{h}^{(l)}_\astt   ,\qquad l=2,3 .
\label{symmetryc}
\ee

The pairing (\ref{pairing}) allows us to define the following  maps in $\mathfrak{so}(5)^\ast$ from (\ref{inv})  by    considering again an action on $z$:
\be
\begin{array}{l}
 \theta^{(m)}(\JJ^{ab}):=
\left\{
 \begin{array}{rl}
\JJ^{ab},&\mbox{if either $m \le a $ or $b<m$};\\[2pt] 
-\JJ^{ab},&\mbox{if   $a<m \le b$} .
\end{array}\right.  \\[12pt]   
  \theta^{(m)}_z(\JJ^{ab},z):= \bigl(  \theta^{(m)} (\JJ^{ab}),-z \bigr) .
\label{invzb}
\end{array}
\ee 
Similarly to  (\ref{invz}), the action of $\theta^{(m)}_z$ can be extended to the tensor product space   $\mathfrak{so}(5)^\ast\otimes  \,\mathfrak{so}(5)^\ast$,  although we shall not make use of it in this paper.
 It  can be checked that only $\theta^{(2)}_z$ and $\theta^{(3)}_z$ are involutive automorphisms of the commutation relations (\ref{aab}) and (\ref{aacx}) in agreement with the symmetric property of $\mathbb S^{(2)}_z$ and   $\mathbb S^{(3)}_z$  (\ref{symmetryc}).

Moreover, a   dual polarity $ {d}_z$   can also be defined in  $\mathfrak{so}(5)^\ast$ in the form
\be
 {d}(\JJ^{ab}):=-\JJ^{4-b\, 4-a},\qquad 
 {d}_z(\JJ^{ab},z):=\bigl( {d}(\JJ^{ab}),-z\bigr) ,\qquad {d}_z^2={\rm Id},
\label{odzb}
\ee  
such that the map ${d}$ is dual  to  $\mathcal{D}$ (\ref{od}) through the pairing (\ref{pairing}) and  $ {d}_z$  is an automorphism of  $\mathfrak{so}(5)^\ast$  which, as expected, interchanges the noncommutative spaces  of points  and 3-hyperplanes, and the noncommutative spaces of lines  and 2-planes:
\be
\mathbb S_z^{(1)}\,  \stackrel{    {d}_z }{\longleftrightarrow}\ \mathbb S_z^{(4)} ,\qquad 
\mathbb S_z^{(2)} \, \stackrel{   {d}_z  }{\longleftrightarrow}\ \mathbb S_z^{(3)} ,
\label{odzc}
\ee
to be compared with (\ref{d01}).

 
\subsection{Drinfel'd Double Structure}
\label{s24}

In~\cite{BHN2015towards31} it was shown that 
the real Lie algebra $\mathfrak{so}(5)$ has  a classical $r$-matrix coming from   a Drinfel'd double structure for the classical complex Lie algebra $\mathfrak{c}_2$.    We now  
 review the main results according to the  notation introduced in Section~\ref{s13}.

Let us consider the  complex Lie algebra  $\mathfrak{c}_2$ in a Chevalley basis with generators $\{ h_l, e_{\pm l}\}$ ($l=1,2$) 
 fulfilling the Lie brackets  given by
  \bea 
&& [h_1,e_{\pm 1}] = \pm e_{\pm 1},     \qquad 
   [h_1,e_{\pm 2}] = \mp e_{\pm 2}, \qquad \ \ 
   [e_{+1},e_{-1}] = h_1, \nonumber\\[2pt]
&& [h_2,e_{\pm 1}] = \mp e_{\pm 1},   \qquad 
   [h_2,e_{\pm 2}] = \pm 2 e_{\pm 2}, \qquad 
   [e_{+2},e_{-2}] = h_2,    \nonumber  \\[2pt]
&& [h_1,h_2] = [e_{-1},e_{+2}] = [e_{+1},e_{-2}] = 0 .\nonumber
\eea 
We define   four new
generators $e_{\pm 3},\, e_{\pm 4}$   as
\be 
[e_{+1},e_{+2}]:=e_{+3},    \qquad
[e_{-2},e_{-1}]:=e_{-3}, \qquad 
[e_{+1},e_{+3}]:=e_{+4},    \qquad
[e_{-3},e_{-1}]:=e_{-4},  
\nonumber
\ee 
such that the Serre relations read
\be
[e_{+1},e_{+4}]=[e_{+2},e_{+3}]=0,\qquad [e_{-1},e_{-4}]=[e_{-2},e_{-3}]=0.
\nonumber
\ee
Then   the 10 generators $\{ h_l,e_{\pm m}\}$   with $l=1,2$ and $m=1,\dots, 4$  span 
 the Lie algebra  $\mathfrak{c}_2$ in the Cartan--Weyl basis. As a shorthand notation we denote 
 $e_m\equiv e_{+m} $ and $f_m \equiv e_{-m} $ so that the full commutation rules of $\mathfrak{c}_2$  read
\begin{align}
[h_1,h_2]&=0 , & [h_1,e_1]&=e_1 , & [h_1,f_1]&=-f_1 , \nonumber \\
[h_1,e_2]&=-e_2 , & [h_1,f_2]&=f_2 , & [h_1,e_3]&=0 , \nonumber \\
[h_1,f_3]&=0 , & [h_1,e_4]&=e_4 , & [h_1,f_4]&=-f_4 , \nonumber \\
[h_2,e_1]&=-e_1 , & [h_2,f_1]&=f_1 , & [h_2,e_2]&=2 e_2 , \nonumber \\
[h_2,f_2]&=-2 f_2 , & [h_2,e_3]&=e_3 , & [h_2,f_3]&=-f_3 , \nonumber \\
[h_2,e_4]&=0 , & [h_2,f_4]&=0 , & [e_1,f_1]&=h_1 , \nonumber \\
[e_1,e_2]&=e_3 , & [e_1,f_2]&=0 , & [e_1,e_3]&=e_4 , \nonumber \\
 [e_1,f_3]&=-f_2 , & [e_1,e_4]&=0 , & [e_1,f_4]&=-f_3 ,\label{nnd} \\
[f_1,e_2]&=0 , & [f_1,f_2]&=-f_3 , & [f_1,e_3]&=e_2 , \nonumber \\
[f_1,f_3]&=-f_4 , & [f_1,e_4]&=e_3 , & [f_1,f_4]&=0 , \nonumber \\
[e_2,f_2]&=h_2 , & [e_2,e_3]&=0 , & [e_2,f_3]&=f_1 , \nonumber \\
[e_2,e_4]&=0 , & [e_2,f_4]&=0 , & [f_2,e_3]&=-e_1 , \nonumber \\
[f_2,f_3]&=0 , & [f_2,e_4]&=0 , & [f_2,f_4]&=0 , \nonumber \\
[e_3,f_3]&=h_1+h_2 , & [e_3,e_4]&=0 , & [e_3,f_4]&=f_1 , \nonumber \\
[f_3,e_4]&=-e_1 , & [f_3,f_4]&=0 , & [e_4,f_4]&=2h_1+h_2 . \nonumber
\end{align}
To unveil the  Drinfel'd double structure for  $\mathfrak{c}_2$, we consider  the   linear combination of the two generators $ h_1$ and  $ h_2$   belonging to the Cartan subalgebra given by~\cite{BHN2015towards31}:
\be
 e_0 :=\tfrac{1}{\sqrt{2}}\big((1+\mathrm{i})h_1+\mathrm{i}h_2\big) , \qquad  f_0:=\tfrac{1}{\sqrt{2}}\big((1-\mathrm{i})h_1-\mathrm{i}h_2\big) .
\label{nne}
\ee
Finally, the identification 
\be
Y_a \equiv  e_{a} ,\qquad  y^a \equiv f_a    ,\qquad a=0,\dots, 4  ,
\nonumber
\ee
allows us to express the commutation relations (\ref{nnd}) with the new Cartan generators (\ref{nne}) in the required form  (\ref{tr}),  thus obtaining a Drinfel'd double structure  for  $\mathfrak{c}_2$ with two 5D subalgebras  $\mathfrak{g}_1=\mbox{span}\{Y_a \equiv  e_{a} \}$ and $\mathfrak{g}_2=\mbox{span}\{ y^a \equiv f_a \}$  $(a=0,\dots, 4)$, 
which are  dual to each other by means of the canonical pairing (\ref{ts}).  

 \noindent
 $\bullet$ Lie subalgebra  $\mathfrak{g}_1=\mbox{span}\{  e_{0} ,\dots, e_4\}$:
  \be
\begin{array}{lll}
\multicolumn{3}{l}{  [e_0,e_1]=\frac 1{ \sqrt{2} }\, e_1,  \qquad [e_0,e_2]= -\frac 1{\sqrt{2}}  (1-{\rm i} )e_2 , }\\[6pt]
\multicolumn{3}{l}{  [e_0,e_3]=\frac {\rm i}{ \sqrt{2} }\, e_3,  \qquad [e_0,e_4]= \frac 1{\sqrt{2}} (1+{\rm i} )e_4 , }\\[4pt]
[ e_1,e_2] = e_3, &\quad   
[ e_1,e_3] = e_4, &\quad   
[ e_1,e_4] = 0,\nonumber \\[2pt]
[ e_2,e_3] = 0, &\quad   
[ e_2,e_4] = 0, &\quad   
[ e_3,e_4] = 0. 
\end{array}
\ee

 \noindent
 $\bullet$ Lie subalgebra  $\mathfrak{g}_2=\mathfrak{g}^\ast_1=\mbox{span}\{  f_{0} ,\dots, f_4\}$:
  \be
\begin{array}{lll}
\multicolumn{3}{l}{  [f_0,f_1]=-\frac 1{ \sqrt{2} }\, f_1,  \qquad\, [f_0,f_2]= \frac 1{\sqrt{2}}  (1+{\rm i} )f_2 , }\\[6pt]
\multicolumn{3}{l}{  [f_0,f_3]=\frac {\rm i}{ \sqrt{2} }\, f_3,  \qquad\quad    [f_0,f_4]=- \frac 1{\sqrt{2}} (1-{\rm i} )f_4 , }\\[4pt]
[ f_1,f_2] = -f_3, &\quad   
[ f_1,f_3] =- f_4, &\quad   
[ f_1,f_4] = 0,\nonumber \\[2pt]
[ f_2,f_3] = 0, &\quad   
[ f_2,f_4] = 0, &\quad   
[ f_3,f_4] = 0. 
\end{array}
\ee

 \noindent
 $\bullet$ Crossed relations $[e_a,f_b]$:
   \be
 \begin{array}{llll}
  [e_0,f_0]=0,   & [e_1,f_1]= \frac 1{\sqrt{2}}  ( e_0+f_0) ,   &\multicolumn{2}{l}{\ [e_3,f_3]=-\frac {\rm i}{\sqrt{2}}  ( e_0-f_0) ,   }\\[6pt]
\multicolumn{2}{l}{  [e_2,f_2]=-\frac {1}{ \sqrt{2} } \bigl((1+{\rm i} )e_0+(1-{\rm i} )f_0 \bigr),  }  &\multicolumn{2}{l}{ \ [e_4,f_4]=\frac {1}{ \sqrt{2} } \bigl((1-{\rm i}) e_0+(1+{\rm i} )f_0 \bigr) , }\\[6pt]
[f_1, e_0] =\frac 1{ \sqrt{2} }\,f_1, &   \  
[f_2, e_0] =  \frac 1{ \sqrt{2} } ({\rm i} -1) f_2, &    \   
 [f_3,e_0] =\frac {\rm i}{ \sqrt{2} }\,f_3, &    \   
 [f_4, e_0] =  \frac 1{ \sqrt{2} } (1+{\rm i} ) f_4,  \\[6pt] 
 [e_1, f_0] =-\frac{ 1}{ \sqrt{2} }\,e_1, &     \  
[ e_2,f_0] =  \frac 1{ \sqrt{2} } (1+{\rm i} ) e_2, &     \  
 [e_3,f_0] =\frac {\rm i}{ \sqrt{2} }\,e_3, &     \  
 [ e_4,f_0] =  \frac 1{ \sqrt{2} } ({\rm i} -1) e_4,  \\[4pt] 
  [  e_1,f_2] =0, &     \  
  [  e_2,f_1] = 0 , &     \  
  [  e_3,f_1] =  -  e_2, &     \  
  [  e_4,f_1] =  -  e_3,  \\[2pt] 
[  e_1,f_3] =-f_2, &     \  
  [  e_2,f_3] = f_1 , &     \  
  [  e_3,f_2] =   e_1, &     \  
  [  e_4,f_2] = 0,  \\[2pt] 
[  e_1,f_4] =-f_3, &     \  
  [  e_2,f_4] =0 , &     \  
  [  e_3,f_4] =   f_1, &     \  
  [  e_4,f_3] = e_1.  
\end{array}
\nonumber
\ee

 From these results the real Lie algebra  $\mathfrak{so}(5)\sim \mathfrak{c}_2$    is obtained in the basis   with generators $\{J_{ab}\}$ obeying the commutation rules (\ref{ab}) through the following    change of basis~\cite{BHN2015towards31}:
\begin{align}
e_0&=-\tfrac 1{\sqrt{2} } (  J_{04} - \mathrm{i} J_{13} )   , & f_0&= \tfrac 1{\sqrt{2} } ( J_{04} + \mathrm{i} J_{13} )   , \nonumber \\[2pt]
e_1&=\tfrac{1}{\sqrt{2}} ( J_{23} +\mathrm{i} J_{12} ) , & f_1&=-\tfrac{1}{\sqrt{2}} (J_{23} - \mathrm{i} J_{12}) ,\nonumber \\[2pt]
 e_2&=\tfrac{1}{2} \bigl(J_{01}-J_{34}  -\mathrm{i} (J_{03}+J_{14}) \bigr) , & f_2&=-\tfrac{1}{2} \bigl(J_{01}-J_{34} +\mathrm{i} (J_{03}+ J_{14} )\bigr) , 
\nonumber \\[2pt]
e_3&=\tfrac{1}{\sqrt{2}} ( J_{24}+\mathrm{i} J_{02}) , & f_3&=-\tfrac{1}{\sqrt{2}} ( J_{24} -\mathrm{i} J_{02} ) ,\nonumber \\[2pt]
e_4&=\tfrac{1}{2} \bigl( J_{01} +J_{34}  +\mathrm{i} ( J_{03}- J_{14}  ) \bigr) , & f_4&=-\tfrac{1}{2} \bigr(J_{01}+J_{34}-\mathrm{i} (J_{03}- J_{14} )\bigl) ,\nonumber 
\end{align}
whose inverse reads
\begin{align}
J_{01}&= \tfrac 12   (e_2 - f_2 +e_4 - f_4)    , & J_{13}  &= -\tfrac {\rm i}{\sqrt 2} (e_0 + f_0)   , \nonumber \\[2pt]
 J_{02}  &=   -\tfrac{\rm i}{\sqrt{2}} (e_3 + f_3)  , & J_{14}   &=  \tfrac {\rm i}2   (e_2 + f_2 +e_4 + f_4)  , \nonumber  \\[2pt]
J_{03}&=   \tfrac {\rm i} 2  (e_2 + f_2 -e_4 - f_4)  , &J_{23} &=    \tfrac 1{\sqrt 2} (e_1 -f_1)  , \nonumber   \\[2pt]
J_{04} &= -\tfrac {1} {\sqrt 2} (e_0 - f_0)  , &  J_{24}  &= \tfrac 1{\sqrt{2}} (e_3 -  f_3)  ,\nonumber \\[2pt]
 J_{12} &=  - \tfrac {\rm i}{\sqrt 2}  (e_1 + f_1) , &J_{34}  &=    -\tfrac 12  (e_2 - f_2 -e_4 + f_4)   . \nonumber 
\end{align}
  The canonical pairing (\ref{ts}) now reads
\be
  \langle J_{ab},J_{cd} \rangle=-\delta_{ac} \delta_{bd} .
  \nonumber
  \ee
Then the canonical classical
$r$-matrix (\ref{tu}) turns out to be
\bea
  r_{\rm can}=\sum_{a=0}^4 {f_a\otimes e_a}=\frac{1}{2}\mathrm{i}\left(J_{01}\wedge J_{14}+J_{02}\wedge J_{24}+J_{03}\wedge J_{34}+J_{12}\wedge J_{23}+J_{04}\wedge J_{13}\right)+\Omega ,
  \nonumber
\eea
where 
\be
  \Omega =-\frac 12 \sum_{0\le a<b\le 4} J_{ab}\otimes J_{ab},
  \nonumber
\ee 
 is the ad-invariant  element (\ref{tw}) corresponding to the tensorized expression of the Casimir $\mathcal{C}$ (\ref{acc}). Hence the skew-symmetric form for $ r_{\rm can}$ is obtained by substracting $\Omega$, as in (\ref{ty}).  We    introduce explicitly the quantum deformation parameter $z$   multiplying  this result    by $2 \mathrm{i}z$ as $ \tilde r_D=   2 \mathrm{i}z (r_{\rm can}- \Omega )$,    obtaining the real $r$-matrix
\begin{equation}
\tilde r_D=z(J_{14}\wedge J_{01} + J_{24}\wedge J_{02}
+ J_{34}\wedge J_{03} + J_{23}\wedge J_{12}  +J _{13}\wedge J_{04} )  ,
\label{rdd}
\end{equation}
which  in terms of the Drinfel'd--Jimbo classical $r$-matrix  (\ref{ec})  considered for $\mathfrak{so}(5)$ reads
\be
\tilde r_D =r_{04,13}+ z J _{13}\wedge J_{04} .
\nonumber
\ee
Hence a Reshetikhin twist with the commuting primitive generators must be added to  (\ref{ec})  in order to obtain a classical $r$-matrix coming from a Drinfel'd double structure. Recall that it is possible to consider a generalized two-parametric $r$-matrix~\cite{BHMN2017kappa3+1}
\be
r_{z,\vartheta} = z(J_{14}\wedge J_{01} + J_{24}\wedge J_{02}
+ J_{34}\wedge J_{03} + J_{23}\wedge J_{12} ) +\vartheta J _{13}\wedge J_{04} ,
\label{rdde}
\ee
showing the effects of the twist with quantum deformation parameter $\vartheta$ on the former deformation determined by $r_{04,13}$ and   properly recovering the 
        Drinfel'd double  $r$-matrix  whenever $\vartheta=z$. We  remark that both $\tilde r_D$ and $r_{z,\vartheta} $ are quasitriangular classical $r$-matrices (like $r_{04,13}$), so that they are solutions of the modified classical Yang--Baxter equation  (\ref{th}), while the twist itself determines  a triangular $r$-matrix with vanishing Schouten bracket \eqref{tk}. In this sense, $\tilde r_D$ and $r_{z,\vartheta} $ can be regarded as ``hybrid" classical $r$-matrices~\cite{Dobrev2001,BHM2010plb}.


\section{The Drinfel'd--Jimbo Lie Bialgebra for  the Cayley--Klein  Algebra~$\mathfrak{so}_\k(5)$}
\label{s3}

The  ${\mathbb {Z}}_2^{\otimes 4}$-grading of  $\mathfrak{so}(5)$ generated by the four automorphisms $\Theta^{(m)}$ (\ref{inv})
 enables one to obtain a particular set of   contracted real Lie algebras~\cite{CKMontigny,CKGraded} through the graded contraction formalism~\cite{Montigny1,Moody}. These are the so-called orthogonal Cayley--Klein (CK) algebras or quasisimple orthogonal algebras~\cite{casimirs,deform,LBC,extensions} (see~\cite{GromovSO} for their description in terms of hypercomplex units). We collectively denote them by $\mathfrak{so}_\k(5)$     as this  family of contracted algebras   depends explicitly on {\em four} real graded contraction parameters $\k=(\k_1,\k_2,\k_3,\k_4)$.  
Alternatively,   each contraction parameter $\k_m$ $(m=1,2,3,4)$ can be introduced in the initial commutation rules of $\mathfrak{so}(5)$ (\ref{ab}) 
    by means of the following mapping provided by the involution $\Theta^{(m)}$ (\ref{inv}):
 \be
\phi^{(m)}(J_{ab}):=
\left\{
 \begin{array}{rl}
J_{ab},&\mbox{if either $m \le a $ or $b<m$};\\[2pt] 
\sqrt{\k_m} \, J_{ab},&\mbox{if   $a<m \le b$} .
\end{array}\right.
\nonumber
\ee
The composition of the four (commuting) mappings gives~\cite{LBC}
\be
\Phi (J_{ab}):=\phi^{(1)}\circ \phi^{(2)}\circ \phi^{(3)}\circ \phi^{(4)}(J_{ab})=\sqrt{\k_{ab} }\, J_{ab} ,
\label{maa}
\ee
 where   the  contraction parameter with two indices $\k_{ab}$ is defined by
  \be
 \k_{ab}:=\prod_{s=a+1}^b\k_s ,\qquad 0\le a<b \le 4.
 \label{mab}
 \ee
 Next we apply   the map (\ref{maa}) with all the factors $\sqrt{\k_{ab} } \ne 0$ onto the commutation rules  of 
 $\mathfrak{so}(5)$ (\ref{ab}) obtaining the Lie brackets corresponding to the CK family $\mathfrak{so}_\k(5)$  which are given by
 \be
[J_{ab}, J_{ac}] =\k_{ab} J_{bc} , \qquad
[J_{ab}, J_{bc}] = -J_{ac} , \qquad
[J_{ac}, J_{bc}] =\k_{bc} J_{ab}   ,\qquad a<b<c,
  \label{me}
\ee
without sum over repeated indices and with all the remaining brackets being equal to zero. This is 
    just the same result coming from a particular solution of the ${\mathbb {Z}}_2^{\otimes 4}$-graded contraction equations for $\mathfrak{so}(5)$~\cite{CKMontigny} (see~\cite{CKGraded} for the general solution). Explicitly, the non-vanishing commutation relations of $\mathfrak{so}_\k(5)$  read
\be
\begin{array}{lll}
[J_{01}, J_{02}] = \k_1 J_{12}  , &\qquad
[J_{01}, J_{12}] = -J_{02}  , &\qquad
[J_{02}, J_{12}] =  \k_2 J_{01} , \\[2pt]
[J_{01}, J_{03}] =  \k_1 J_{13}  , &\qquad
[J_{01}, J_{13}] = -J_{03},   &\qquad
[J_{03}, J_{13}] =  \k_2  \k_3 J_{01} , \\[2pt]
[J_{01}, J_{04}] = \k_1 J_{14}  , &\qquad
[J_{01}, J_{14}] = -J_{04}  , &\qquad
[J_{04}, J_{14}] = \k_2 \k_3 \k_4 J_{01},  \\[2pt]
[J_{02}, J_{03}] = \k_1\k_2 J_{23} ,  &\qquad
[J_{02}, J_{23}] = -J_{03} ,  &\qquad
[J_{03}, J_{23}] = \k_3 J_{02} , \\[2pt]
[J_{02}, J_{04}] = \k_1 \k_2J_{24} ,  &\qquad
[J_{02}, J_{24}] = -J_{04} ,  &\qquad
[J_{04}, J_{24}] = \k_3 \k_4 J_{02},  \\[2pt]
[J_{03}, J_{04}] = \k_1\k_2\k_3J_{34}  , &\qquad
[J_{03}, J_{34}] = -J_{04}  , &\qquad
[J_{04}, J_{34}] = \k_4 J_{03} , \\[2pt]
[J_{12}, J_{13}] = \k_2J_{23} , &\qquad
[J_{12}, J_{23}] = -J_{13} ,  &\qquad
[J_{13}, J_{23}] = \k_3 J_{12} , \\[2pt]
[J_{12}, J_{14}] = \k_2 J_{24} , &\qquad
[J_{12}, J_{24}] = -J_{14}  , &\qquad
[J_{14}, J_{24}] = \k_3 \k_4 J_{12} , \\[2pt]
[J_{13}, J_{14}] = \k_2 \k_3 J_{34}  , &\qquad
[J_{13}, J_{34}] = -J_{14} ,  &\qquad
[J_{14}, J_{34}] = \k_4 J_{13},  \\[2pt]
[J_{23}, J_{24}] = \k_3 J_{34}  , &\qquad
[J_{23}, J_{34}] = -J_{24} ,  &\qquad
[J_{24}, J_{34}] = \k_4 J_{23}   .
\end{array}
\label{mf}
\ee

 We remark  that although the factor $\sqrt{\k_{ab} }\ne 0$ in the map (\ref{maa}) can be an imaginary number, enabling to change the real form of the algebra, the resulting commutation relations (\ref{me})  of $\mathfrak{so}_\k(5)$ only comprise real Lie algebras. Moreover, the zero value for $\k_{ab}$ is consistently allowed  in~(\ref{me}),  which is equivalent to apply an In\"on\"u--Wigner contraction~\cite{LBC,IW}, leading to a more abelian (contracted) Lie algebra.
Consequently,  each   graded contraction parameter $\k_m$   can take a positive, negative or zero value in  \eqref{me}, and when $\k_m\ne 0$, it can be reduced to $\pm 1$ through a scaling of the Lie generators. Hence $\mathfrak{so}_\k(5)$ contains $3^4=81$ specific real Lie algebras, being some of them   isomorphic. 

Moreover, 
 the CK algebra $\mathfrak{so}_\k(5)$ (\ref{me}) is always endowed with    two non-trivial Casimirs regardless of the values of $\k$.   One of them is the quadratic Casimir coming from the Killing--Cartan form  which is given by~\cite{casimirs}
 \bea
&&\mathcal{C}= \k_2\k_3\k_4 J_{01}^2  +\k_3\k_4 J_{02}^2 
+\k_4J_{03}^2  +J_{04}^2 
  +\k_1\k_3\k_4J_{12}^2  +\k_1\k_4J_{13}^2 \nonumber\\[2pt] 
  &  &\qquad\qquad   +\k_1J_{14}^2+\k_1\k_2\k_4 J_{23}^2  +\k_1\k_2  J_{24}^2  
 + \k_1\k_2\k_3 J_{34}^2 ,
\eea
  to be compare with (\ref{acc}). Observe that in the  most contracted case, with all $\k_m=0$,
 $\mathcal{C}=J_{04}^2 $. The second Casimir is a fourth-order one that can be found explicitly in~\cite{casimirs} and this   is related to the Pauli--Lubanski operator. Also in the most contracted case the fourth-order Casimir does not vanish. In this respect, we recall that the CK Lie algebras are the only graded contracted algebras from $\mathfrak{so}(N+1)$~\cite{CKMontigny,casimirs} that preserve the rank of the semisimple algebra, understood as the number of algebraically independent Casimirs, which at this dimension  is equal to two.

To unveil the structure of CK family $\mathfrak{so}_\k(5)$, let us recall that the  vector   representation of  the CK algebra in  terms of $5\times 5$ real matrices, $\rho:\mathfrak{so}_\k(5)  \rightarrow \text{End} (\mathbb{R}^5)$, is given by~\cite{casimirs,extensions} 
\be
\rho(J_{ab})=  - \k_{ab} e_{ab}+e_{ba}, \quad
\label{mh}
\ee
 which fulfils that
\be
\begin{array}{l}
 \rho(J_{ab})^T \bfI_\k +\bfI_\k \, \rho(J_{ab})=0,\\[4pt]
\bfI_\k={\rm diag}(1,\k_{01},\k_{02} ,\k_{03},\k_{04})\\[3pt]
\quad\ ={\rm diag}(1,\k_1,\k_1\k_2,\k_1\k_2\k_3,\k_1\k_2\k_3\k_4)  ,
\end{array}
\label{mi}
\ee
to be compared with (\ref{dd})  and  (\ref{de}). The value of the parameter $\k_{ab}$ (\ref{mab}) determines the Lie subalgebra generated by   $J_{ab}$  (\ref{mh}), denoted $\mathfrak{so}_{\k_{ab}}(2)$, i.e.,      $\mathfrak{so}(2)$ for $\k_{ab}>0$, $\mathfrak{so}(1,1)$ for $\k_{ab}<0$, and $\mathfrak{iso}(1)\equiv \mathbb R$ for the pure contracted case with $\k_{ab}=0$.

According to the values of $\k=(\k_1,\k_2,\k_3,\k_4)$, we mention the most relevant  mathematical and physical  Lie algebras contained within $\mathfrak{so}_\k(5)$~\cite{casimirs,extensions}:
\begin{itemize}

\item   If all
$\k_m\ne 0$,  $\mathfrak{so}_\k(5)$ is a
pseudo-orthogonal algebra $\mathfrak{so}(p,q)$ ($p+q=5$) where
  $(p,q)$ are the number of positive and negative terms  in the invariant   quadratic form with
   matrix $ \bfI_\k$ (\ref{mi}). Obviously, for all $\k_m> 0$ we recover $\mathfrak{so}(5)$, otherwise we find either   $\mathfrak{so}(3,2)$ (isomorphic to the (3+1)D anti-de Sitter algebra)    or $\mathfrak{so}(4,1)$  (isomorphic to the (3+1)D  de Sitter algebra or to   the 4D hyperbolic one).

\item    When only $\k_1=0$ we find the    inhomogeneous  pseudo-orthogonal algebras 
with   semidirect sum  structure
\be
\mathfrak{so}_{0,\k_2,\k_3,\k_4}(5)
\equiv \mathbb{R}^4\dsum  \mathfrak{so}_{\k_2,\k_3,\k_4}(4)\equiv \mathfrak{iso}(p,q),  \quad 
p+q=4 , 
\label{ia} 
\ee
where the abelian subalgebra  $\mathbb{R}^4$ is spanned by $\langle
J_{01} , J_{02},J_{03},J_{04}\rangle$ and 
$ \mathfrak{so}_{\k_2,\k_3,\k_4}(4)$ is a pseudo-orthogonal  algebra,     preserving the
quadratic form with $4\times 4$ matrix   
$$
{\rm diag} (1,\k_{12},\k_{13},\k_{14})={\rm diag} (1,\k_{2},\k_{2}\k_3,\k_{2}\k_3\k_4) ,
$$ that   acts on  $\mathbb{R}^4$ through the vector representation (\ref{mh})  (see (\ref{dia})).   Hence, the 4D Euclidean  $\mathfrak{iso}(4)$,  the (3+1)D Poincar\'e   $\mathfrak{iso}(3,1)$ and $\mathfrak{iso}(2,2)$ algebras
belong to this class.

\item    If only $\k_4=0$ we again obtain     inhomogeneous  pseudo-orthogonal algebras 
with   semidirect sum  structure 
\be
\mathfrak{so}_{\k_1,\k_2,\k_3,0}(5)
\equiv  \mathbb{R}'^4 \dsum  \mathfrak{so}_{\k_1,\k_2,\k_3}(4)\equiv \mathfrak{i'so}(p,q),  \quad 
p+q=4 , 
\label{iaa}
\ee
where now the abelian subalgebra  $\mathbb{R}'^4=\langle
J_{04} , J_{14},J_{24},J_{34}\rangle$ and 
$ \mathfrak{so}_{\k_1,\k_2,\k_3}(4)$, that preserves the
quadratic form with $4\times 4$ matrix   
$$
{\rm diag} (1,\k_{01},\k_{02},\k_{03})= {\rm diag} (1,\k_{1},\k_{1}\k_{2},\k_{1}\k_{2}\k_{3}),
$$    acts on  $\mathbb{R}'^4$ through the contragredient of the vector representation  (\ref{mh}) (see \eqref{diad}).  These algebras are isomorphic to the previous ones with structure (\ref{ia}), e.g., $\mathfrak{iso}(4)\simeq  \mathfrak{i'so}(4)$.

\item  For $\k_1=\k_2=0$ we get   a ``twice-inhomogeneous"
pseudo-orthogonal algebra
\be
\mathfrak{so}_{0,0,\k_3,\k_4}(5) \equiv  \mathbb{R}^4\dsum \! \left(  \mathbb{R}^3\dsum  
\mathfrak{so}_{\k_3,\k_4}(3)\right)\equiv \mathfrak{ iiso}(p,q),\quad  p+q=3,
\label{ib}
\ee
where    $ \mathbb{R}^4=\langle
J_{01} , J_{02},J_{03},J_{04}\rangle$, $ \mathbb{R}^3=\langle
J_{12} , J_{13},J_{14}\rangle$  and 
$ \mathfrak{so}_{ \k_3,\k_4}(3)=\langle J_{23},J_{24},J_{34}\rangle$ is a pseudo-orthogonal  algebra that  preserves the
quadratic form with $3\times 3$ matrix   
$$
{\rm diag} (1,\k_{23},\k_{24} ) = {\rm diag} (1,\k_{3},\k_{3} \k_4).
$$  Here   we find the (3+1)D Galilean algebra 
   $\mathfrak{ iiso}(3)$ as well as  $\mathfrak{ iiso}(2,1)$.

\item If   $\k_1=\k_4=0$ we obtain that
\be
\mathfrak{so}_{0,\k_2,\k_3,0}(5) \equiv \mathbb{R}^4\dsum  \!  \left( \mathbb{R}'^3\dsum  
\mathfrak{so}_{\k_2,\k_3}(3)\right)\equiv \mathfrak{ ii'so}(p,q),\quad  p+q=3,
\label{ibb}
\ee
where $\mathbb{R}^4=\langle
J_{01} , J_{02},J_{03},J_{04}\rangle$, $\mathbb{R}'^3=\langle
J_{14} , J_{24},J_{34}\rangle$ are abelian subalgebras and $ \mathfrak{so}_{ \k_2,\k_3}(3)=\langle J_{12},J_{13},J_{23}\rangle$,   preserving     
$$
{\rm diag} (1,\k_{12},\k_{13} )=  {\rm diag} (1,\k_{2},\k_{2} \k_3),
$$
  such that    $ \mathfrak{so}_{ \k_2,\k_3}(3)$ acts on $\mathbb{R}'^3$ 
through the contragredient of the vector representation, while $ \mathbb{R}'^3\dsum  
\mathfrak{so}_{\k_2,\k_3}(3)$ acts on   $\mathbb{R}^4$ through the vector representation. 
Alternatively, the structure (\ref{ibb}) can also be expressed as
\be
\mathfrak{so}_{0,\k_2,\k_3,0}(5) \equiv \mathbb{R}'^4\dsum  \!  \left(\mathbb{R}^3\dsum  
\mathfrak{so}_{\k_2,\k_3}(3)\right)\equiv \mathfrak{ i'iso}(p,q),\quad  p+q=3,
\label{ibbc}
\ee
where $\mathbb{R}'^4=\langle
 J_{04},J_{14} , J_{24},J_{34}\rangle$ and  $\mathbb{R}^3=\langle
  J_{01},J_{02},J_{03}\rangle$; note that $ \mathbb{R}'^4 \simeq\mathbb{R}^4$  and  $ \mathbb{R}'^3 \simeq\mathbb{R}^3$ via $\mathcal{D}$ (\ref{od}).
As particular algebras we get the (3+1)D Carroll algebra  $\mathfrak{ ii'so}(3)\simeq \mathfrak{ i'iso}(3)$ formerly introduced in~\cite{LL,BLL} (see also~\cite{carroll,Gomis:2014,Bergshoeff:2017btm,Figueroa-OFarrill2018,Daszkiewicz2019,SnyderG} and references therein) and  $\mathfrak{ ii'so}(2,1)\simeq \mathfrak{ i'iso}(2,1)$.

\item  When $\k_2=0$,     these
contracted algebras are of Newton--Hooke type~\cite{BLL}  (see also~\cite{carroll,Figueroa-OFarrill2018,SnyderG,Aldrovandi,expansiones})  with structure~\cite{WB}
\be
\mathfrak{so}_{\k_1,0,\k_3,\k_4}(5) \equiv \mathbb{R}^6 \dsum  \! \bigl(\mathfrak{so}_{\k_1 }(2)\oplus
\mathfrak{so}_{\k_{3}, \k_{4}}(3) \bigr)\equiv\mathfrak{ i}_6\bigl(\mathfrak{so}_{\k_1 }(2)\oplus
\mathfrak{so}_{\k_{3}, \k_{4}}(3) \bigr) , 
\label{ic}
\ee
 where  $\mathbb{R}^6=\langle
J_{02} , J_{03},J_{04},J_{12},J_{13},J_{14}\rangle$ is  an abelian subalgebra and the direct sum is between  the subalgebras $ \mathfrak{so}_{\k_1}(2)=\langle J_{01} \rangle$ and $ \mathfrak{so}_{ \k_3,\k_4}(3)=\langle J_{23},J_{24},J_{34}\rangle$.

\item   The fully contracted case in the CK family corresponds to
setting the four    $\k_m=0$. This is the so-called flag algebra 
\be
\mathfrak{so}_{0,0,0,0}(5) \equiv    \mathbb{R}^4\dsum  \!  \bigl(    \mathbb{R}^3\dsum  \!   \bigl( \mathbb{R}^2\dsum  \!   \mathbb{R}  \bigr)  \bigr) \equiv \mathfrak{iiiiso}(1) ,
\label{id}
\ee
where 
$\mathbb{R}^4=\langle
J_{01} , J_{02},J_{03},J_{04}\rangle$, $\mathbb{R}^3=\langle
J_{12} , J_{13},J_{14} \rangle$, $\mathbb{R}^2=\langle
J_{23} , J_{24}  \rangle$ and $\mathbb{R}=\langle
 J_{34}  \rangle \equiv \mathfrak{iso}(1)$.

\end{itemize}

Therefore  the kinematical algebras associated with different models of (3+1)D spacetimes of constant
curvature~\cite{BLL,BN}   belong to the  CK family $\mathfrak{so}_\k(5)$~\cite {CKGraded,MontignyKinematical}.

It is worth stressing that
the polarity $\mathcal{D}$ (\ref{od}) also remains as an automorphism of the whole family of CK algebras   in such a manner that this map interchanges isomorphic Lie algebras within the family  in the form
\be
 \mathfrak{so}_{\k_1,\k_2,\k_3,\k_4}(5) \    \stackrel{\mathcal{D}}{\longleftrightarrow}\   \mathfrak{so}_{\k_4,\k_3,\k_2,\k_1}(5),
  \label{ms}
\ee 
so interchanging the contraction parameters $\k_1\leftrightarrow \k_4$ and $\k_2\leftrightarrow \k_3$. Consequently,   the 
CK algebras with $\k_4=0$ (\ref{iaa}) are  related,   through  $\mathcal{D}$, to  those with $\k_1=0$ \eqref{ia} and  so  they are isomorphic; those with  $\k_4=\k_3=0$ are twice-inhomogeneous 
algebras and  isomorphic  to  the ones with $\k_1=\k_2=0$ (\ref{ib}); those with $\k_3=0$ are also  Newton--Hooke type algebras isomorphic to   (\ref{ic}); and the (single) flag algebra (\ref{id})  remains unchanged under $\mathcal{D}$.

 We also recall that 
all the CK algebras in $\mathfrak{so}_\k(5)$ (even the flag algebra) have the same number of functionally independent Casimirs~\cite{casimirs}. At this dimension, there are two (second- and fourth-order) Casimir invariants, exactly equal to   the rank of the simple algebra $\mathfrak{so}(5)$; for this reason they are also called quasisimple orthogonal  algebras.

\subsection{Symmetric Homogeneous Cayley--Klein  Spaces}
\label{s31}

Since, by construction,  the  ${\mathbb {Z}}_2^{\otimes 4}$-grading is preserved for the CK algebra $\mathfrak{so}_\k(5)$, the same 
Cartan  decompositions (\ref{db}) in invariant  $\mathfrak{h}_\k^{(m)}$  and
anti-invariant   $\mathfrak{t}_\k^{(m)}$ subspaces  under  $ \Theta^{(m)}$ (\ref{inv}) also hold  $(m=1,2,3,4)$
\be
{\mathfrak{so}}_\k(5)= \mathfrak{h}_\k^{(m)}\oplus \mathfrak{t}_\k^{(m)}.
\label{mp}
\ee
But now from  (\ref{me}), we can express the   relations (\ref{hp}) by taking into account the contraction parameter $\k_m$:

\be
[\mathfrak{h}_\k^{(m)},\mathfrak {h}_\k^{(m)}] \subset \mathfrak{h}_\k^{(m)}  ,\qquad 
[ \mathfrak{h}_\k^{(m)}, \mathfrak{t}_\k^{(m)}] \subset \mathfrak{t}_\k^{(m)} , \qquad 
[ \mathfrak {t}_\k^{(m)}, \mathfrak{t}_\k^{(m)}] \subset \k_m \, \mathfrak{h}_\k^{(m)}   .
\label{mq}
\ee
This, in turn, means that again for any value of $\k_m$, $\mathfrak{h}_\k^{(m)}$ is always a 
  Lie subalgebra but     the subspace $\mathfrak{t}^{(m)}$ becomes an abelian subalgebra when $\k_m=0$.
  
  Next, as in Section~\ref{s21} we construct the  homogeneous CK spaces as the coset spaces~\cite{gil,HS1997phasespaces,deform}
  \be
  \mathbb S_\k^{(m)}={\rm SO}_\k(5)/ H_\k^{(m)},
  \label{mqq}
  \ee
  where ${\rm SO}_\k(5)$ is the CK Lie group with Lie algebra $\mathfrak{so}_\k(5)$ and $H_\k^{(m)}$ is the isotropy subgroup of ${\rm SO}_\k(5)$ with Lie algebra $\mathfrak{h}_\k^{(m)}$. We recall that,   usually, a CK geometry (\ref{afxx}) is identified  with   the space of points $ \mathbb S_\k^{(1)}$, without taking  into account other spaces.  Along this paper,  a CK geometry will be understood as  the full set of  the four homogeneous spaces (\ref{mqq}).

  The four spaces $  \mathbb S_\k^{(m)}$ (\ref{mqq}) are symmetric and reductive spaces of constant sectional curvature $K$ equal to the graded contraction parameter $\k_m$. These are (see (\ref{dia})--(\ref{diad})):
    \begin{enumerate} 
   
   \item   {\em 4D CK space of points}:
\be
 \mathbb S_\k^{(1)}={\rm SO}_\k(5)/  {\rm SO}_{\k_2,\k_3,\k_4}(4),\qquad K=\k_1.
 \label{space1}
\ee

\item {\em 6D CK space of lines}:
\be
 \mathbb S_\k^{(2)}={\rm SO}_\k(5)/ \bigl(   {\rm SO}_{\k_1}(2)\otimes   {\rm SO}_{\k_3,\k_4}(3)   \bigr),\qquad K=\k_2.
  \label{space2}
\ee

\item {\em 6D CK space of 2-planes}:
\be
 \mathbb S_\k^{(3)}={\rm SO}_\k(5)/ \bigl(   {\rm SO}_{\k_1,\k_2}(3)\otimes   {\rm SO}_{ \k_4}(2)   \bigr),\qquad K=\k_3.
  \label{space3}
\ee

\item {\em 4D CK space of 3-hyperplanes}:
\be
 \mathbb S_\k^{(4)}={\rm SO}_\k(5)/   {\rm SO}_{\k_1,\k_2,\k_3}(4),\qquad K=\k_4.
  \label{space4}
 \ee
 \end{enumerate}

We stress that, strictly speaking, only the rank-one spaces  $ \mathbb S_\k^{(1)}$ and $ \mathbb S_\k^{(4)}$ are of constant curvature in the sense that {\em all}   their sectional curvatures   are equal to   
  $\k_1$ and $\k_4$, respectively. However, the rank-two spaces    $ \mathbb S_\k^{(2)}$ and $ \mathbb S_\k^{(3)}$ are not, in general, of  constant curvature in the above sense,  but they are
as close to constant  curvature as a rank-two space would allow~\cite{HS1997phasespaces}. In particular,     the
sectional curvature $K$ of the    space of lines  $ \mathbb S_\k^{(2)}$  along any 2-plane direction spanned by any two tangent vectors
 $(J_{0i}, J_{0j})$,  $(J_{1i}, J_{1j})$  and $(J_{0i},J_{1i})$ $(i,j=2,3,4)$ is constant  and equal to
$\k_2$, but the remaining sectional curvatures  could be different but proportional  to $\k_2$. When  $\k_2=0$, $ \mathbb S_\k^{(2)}$ is a proper flat space with $K=0$.  And similarly for $ \mathbb S_\k^{(3)}$.

By taking into account the above comments, we can say, roughly speaking,   that the coefficients $\k=(\k_1,\k_2,\k_3,\k_4)$ that label the CK family $\mathfrak{so}_\k(5)$ are just the constant curvatures of the four aforementioned spaces.  Therefore two isomorphic algebras  in the family $\mathfrak{so}_\k(5)$ lead to two different sets of four homogeneous spaces through their corresponding Lie groups, and such sets of spaces are those which determine each specific {\em CK geometry} amongst the 81 ones.
In this respect, we remark that   the polarity $\mathcal{D}$ (\ref{od}), that   relates isomorphic CK algebras in the form (\ref{ms}), also 
interchanges the  homogeneous CK spaces as in (\ref{d01}):
\be
   \mathbb S_\k^{(1)} \    \stackrel{\mathcal{D}}{\longleftrightarrow}\   \mathbb S_\k^{(4)} ,
  \qquad  \mathbb S_\k^{(2)} \    \stackrel{\mathcal{D}}{\longleftrightarrow}\   \mathbb S_\k^{(3)}  .
\nonumber
\ee 
For instance, the 4D Euclidean algebra $\mathfrak{iso}(4)$ corresponding to take $\k=(0,+,+,+)$ yields a flat space of points $\mathbb S^{(1)}={\rm ISO}(4)/  {\rm SO}(4)$  (\ref{space1})   but a  positively  curved space of 3-hyperplanes $\mathbb S^{(4)}={\rm ISO}(4)/  {\rm ISO}(3)$  (\ref{space4}). Conversely,    the isomorphic   algebra  $\mathfrak{i'so}(4)\simeq \mathfrak{iso}(4)$  arising for $\k=(+,+,+,0)$, via $\mathcal{D}$, gives rise to 
a positively curved space of points $\mathbb S^{(1)}={\rm ISO}(4)/  {\rm ISO}(3)$    but a   flat space of 3-hyperplanes $\mathbb S^{(4)}={\rm ISO}(4)/  {\rm SO}(4)$.

It is worth remarking that it is possible to construct other 4D and 6D symmetric homogeneous spaces from the CK group ${\rm SO}_\k(5)$ which, depending on each particular CK geometry, could be different from the  four above ones (\ref{mqq}).  In particular, any  composition of the automorphisms   $ \Theta^{(m)}$ (\ref{inv}), which form a basis for the  ${\mathbb {Z}}_2^{\otimes 4}$-grading, gives rise to another  automorphism which provides another Cartan decomposition, like (\ref{mp}), and from it  the corresponding coset space can be constructed. For instance, the composition 
$ \Theta^{(1)} \Theta^{(4)}$ leads to the 6D symmetric homogeneous space
\be
{\rm SO}_\k(5)/H_\k' \,,\qquad H'_\k\equiv {\rm SO}_{\k_{04}}(2) \otimes {\rm SO}_{\k_2,\k_3} (3)=\langle  J_{04}     \rangle \otimes \langle  J_{12},J_{13},J_{23}\rangle 
\label{other}
\ee
(to be compared with (\ref{diab})),
which can be interpreted as another 6D CK space of lines.  This fact can clearly be  appreciated in the Lorentzian spacetimes where there exist time-like and space-like lines. In the single case of  $\mathfrak{so}(5)$ with $\k=(+,+,+,+)$ all of such possible CK spaces are equivalent to the four 
spaces (\ref{dff}). Furthermore, it is also possible to obtain generalizations of     the polarity $\mathcal{D}$ (\ref{od}) relating such other homogeneous spaces belonging to different CK geometries with the same (isomorphic) CK algebra. In the 2D case
shown in Table~\ref{table0}, the prefix ``Co-"  in the name of some CK geometries reminds the action of $\mathcal{D}$ that here interchanges $\k_1\leftrightarrow \k_2$, so keeping the three geometries in the diagonal unchanged. The full description of all the 2D CK spaces and the  generalizations of     the polarity  $\mathcal{D}$  can be found in~\cite{CK2d,Varna2D}.

\subsection{Cayley--Klein Lie Bialgebra}
\label{s32}

Let us start with the classical $r$-matrix $r_{04,13}$ (\ref{ec}) for $\mathfrak{so}(5)$. The Lie bialgebra contraction procedure introduced in~\cite{LBC} shows that it is not only necessary to apply the contraction map $\Phi$  (\ref{maa}) to the Lie generators of $\mathfrak{so}(5)$  in order to obtain a classical $r$-matrix for the CK algebra  $\mathfrak{so}_\k(5)$, but additionally a possible transformation of the quantum deformation parameter $z$ must be considered. We recall that the idea to transform  the deformation parameter in contractions of quantum groups was formerly   introduced   in~\cite{CGST1991heisemberg, CGST1992}. In our case, the coboundary Lie  bialgebra contraction that ensures a well defined  limit $\k_m\to 0$ (for any $m$) of both the classical $r$-matrix and the cocommutator   for $\mathfrak{so}(5)$    is given by  the transformation~\cite{LBC}
\be
\Psi(z)=\frac z{\sqrt{\k_{04}}}=\frac z{\sqrt{\k_{1}\k_{2}\k_{3}\k_{4}}} \, .
\label{na}
\ee
Then we apply the composition of the maps  (\ref{maa}) and  (\ref{na}) to $r_{04,13}$ in the form
\be
r=\bigl( \Phi^{-1}\otimes \Phi^{-1} \bigr)\circ  \Psi^{-1}(r_{04,13}) ,
\nonumber
\ee
obtaining that
\be
r=z\bigl(J_{14}\wedge J_{01} + J_{24}\wedge J_{02}
+ J_{34}\wedge J_{03} + \sqrt{\k_1\k_4}\, J_{23}\wedge J_{12} \bigr) .
\label{nb}
\ee
Its Schouten bracket (\ref{ti}) turns out to be
\bea 
&&\!\!\!\!\!\!\!\!\!\!\!\!\!\!\!\!
[[r,r]]=z^2\bigl( J_{01}\wedge J_{04}\wedge J_{14}+J_{02}\wedge J_{04}\wedge J_{24} + J_{03}\wedge J_{04}\wedge J_{34}+   \k_1\k_4  J_{12}\wedge J_{13}\wedge J_{23}\nonumber\\[2pt]
&&\qquad \quad + \k_4 (\k_3  J_{01}\wedge J_{02}\wedge J_{12}+J_{01}\wedge J_{03}\wedge J_{13} +J_{02}\wedge J_{03}\wedge J_{23}  ) \nonumber\\[2pt]
&&\qquad \quad + \k_1 (J_{12}\wedge J_{14}\wedge J_{24}+ J_{13}\wedge J_{14}\wedge J_{34}+\k_2 J_{23}\wedge J_{24}\wedge J_{34}) 
\bigr) .
\label{nby}
\eea
It can be checked that $r$ (\ref{nb})   is a solution of the  modified classical Yang--Baxter equation  (\ref{th})  for any Lie algebra within the CK family $\mathfrak{so}_\k(5)$ (so for any value of  $\k=(\k_1,\k_2,\k_3,\k_4)$). The corresponding cocommutator can either be  obtained from the Lie bialgebra contraction of (\ref{coco}) or through the relation (\ref{tf}) with (\ref{nb}) giving rise to the  CK Lie bialgebra $(\mathfrak{so}_{\k}(5),\delta(r))$; namely
  \be
  \begin{array}{l}
 \delta(J_{04})=0,\qquad  \delta(J_{13})=0, \\[4pt]
 \delta(J_{12})=z \sqrt{\k_1\k_4}\, J_{12}\wedge
J_{13}, \qquad
\delta(J_{23})=z \sqrt{\k_1\k_4}\, J_{23}\wedge J_{13}, \\[4pt]
  \delta(J_{01})=z\bigl(
J_{01}\wedge J_{04}+
\k_1 J_{24}\wedge J_{12}+
\k_1 J_{34}\wedge J_{13}+
 \sqrt{\k_1\k_4}\, J_{02}\wedge J_{23} \bigr),\\[4pt]
  \delta(J_{02})=z\bigl(
J_{02}\wedge J_{04}+
\k_1J_{12}\wedge J_{14}+
\k_1\k_2J_{34}\wedge J_{23} \\[4pt]
\qquad\qquad\qquad  +
\k_2  \sqrt{\k_1\k_4}\, J_{23}\wedge J_{01}+
 \sqrt{\k_1\k_4} \, J_{12}\wedge J_{03} \bigr), \\[4pt]
  \delta(J_{03})=z\bigl(
J_{03}\wedge J_{04}+
\k_1 J_{13}\wedge J_{14}+
\k_1 \k_2 J_{23}\wedge J_{24}+
\k_3  \sqrt{\k_1\k_4}\, J_{02}\wedge J_{12} \bigr), \\[4pt]
  \delta(J_{14})=z\bigl(
J_{14}\wedge J_{04}+
\k_4 J_{13}\wedge J_{03}+
\k_3 \k_4 J_{12}\wedge J_{02}+
\k_2  \sqrt{\k_1\k_4}\,  J_{24}\wedge J_{23} \bigr),
\\[4pt]
  \delta(J_{24})=z\bigl(
J_{24}\wedge J_{04}+
\k_4 J_{23}\wedge J_{03}+
\k_3\k_4 J_{01}\wedge J_{12} \\[4pt]
\qquad\qquad\qquad  +
\k_3   \sqrt{\k_1\k_4}\, J_{12}\wedge J_{34}+
 \sqrt{\k_1\k_4}\, J_{23}\wedge J_{14} \bigr), \\[4pt]
  \delta(J_{34})=z\bigl(
J_{34}\wedge J_{04}+
\k_4 J_{02}\wedge J_{23}+
\k_4 J_{01}\wedge J_{13}+
 \sqrt{\k_1\k_4}\,  J_{24}\wedge J_{12} \bigr).
 \end{array}
 \label{cocok}
\ee

Now some remarks are in order.
\begin{itemize} 
\item  The last term  in the CK $r$-matrix  (\ref{nb}) is also an $r$-matrix $r_{13}=z \sqrt{\k_1\k_4}\,J_{23}\wedge J_{12}$
  generating   the CK Lie bialgebra $(\mathfrak{so}_{\k_2,\k_3}(3),\delta(r_{13}))$ with generators   $\langle
J_{12},J_{13},J_{23}\rangle$ and primitive generator $J_{13}$,  
which is a sub-Lie bialgebra of
$(\mathfrak{so}_\k(5),\delta(r ))$. Notice that, as in (\ref{coco}),    $J_{04}$ is the ``main" primitive generator such that the product $zJ_{04}$ is dimensionless
while   $J_{13}$  is a
``secondary" primitive generator. 

  \item
The same $z$-polarity $\mathcal{D}_z$ (\ref{odz}) 
is an automorphism of the whole  family of CK bialgebras relating the   cocommutators (\ref{cocok})   as
\be
( \mathfrak{so}_{\k_1,\k_2,\k_3,\k_4}(5),\delta(r)) \    \stackrel{\mathcal{D}_z}{\longleftrightarrow}\   (\mathfrak{so}_{\k_4,\k_3,\k_2,\k_1}(5) ,\delta(r)) ,
\label{nc}
\ee
so interchanging   $\k_1\leftrightarrow \k_4$ and $\k_2\leftrightarrow \k_3$ as in  (\ref{ms}). Note that the classical  $r$-matrix \eqref{nb} and its Schouten bracket (\ref{nby})  remain unchanged    under (\ref{odz}) and (\ref{nc}). 

\item As far as    the     $z$-maps $\Theta^{(m)}_z$ (\ref{invz}) is concerned, it can directly be checked from the expression of the  $r$-matrix (\ref{nb})  that 
 $\Theta^{(2)}_z$ and $\Theta^{(3)}_z$ are again   involutive automorphisms of $(\mathfrak{so}_\k(5),\delta(r ))$ for any value of the contraction parameters $\k$.
 Moreover,  both $\Theta^{(1)}_z$ and $\Theta^{(4)}_z$ become involutions whenever, at least,  either  $\k_1=0$ or $\k_4=0$, that is, when the last term of $r$ (\ref{nb})  vanishes.  Therefore 
  a complete  ${\mathbb {Z}}_2^{\otimes 4}$-grading spanned by  the four $\Theta^{(m)}_z$ (\ref{invz})  is   kept for the 45 Lie bialgebras with  contraction parameters $\k=(0,\k_2,\k_3,\k_4)$ and 
 $\k=(\k_1,\k_2,\k_3,0)$, which cover inhomogeneous algebras and their further contractions. The quantum algebras for the  first  set of 27 bialgebras with $\k=(0,\k_2,\k_3,\k_4)$ were fully constructed in~\cite{CK4d}, while the second set is related to the first one by means of $\mathcal{D}_z$ (\ref{nc}); in these results it can be appreciated that the term $\exp(z J_{04}/2)$  always appears in the  deformed coproduct $\Delta_z$, for any value of $\k=(0,\k_2,\k_3,\k_4)$, showing that $z J_{04}$ is dimensionless and that, in this sense, $J_{04}$ is the principal primitive generator.

\item The    coisotropy condition (\ref{coisotropy}) is always satisfied by the four CK subalgebras  $\mathfrak{h}_\k^{(m)}$ 
 \be
\delta( \mathfrak{h}_\k^{(m)})\subset \mathfrak{h}_\k^{(m)}\wedge\mathfrak{so}_\k(5) ,  \qquad m=1,\dots,4 ,
\label{coisok}
\ee
but   none of them fulfils  the  Poisson subgroup condition for any value of the contraction  parameters (even for the flag algebra with all $\k_m=0$)
 \be
\delta( \mathfrak{h}_\k^{(m)})\not\subset \mathfrak{h}_\k^{(m)}\wedge    \mathfrak{h}_\k^{(m)} .
\nonumber
\ee

\end{itemize}

So far we have obtained, in a unified setting,   a  family of coboundary Lie bialgebra structures $(\mathfrak{so}_\k(5),\delta(r ))$, with quasitriangular classical $r$-matrix (\ref{nb})  and  cocommutator  $\delta$ (\ref{cocok}),   which covers  81 particular  Lie bialgebras with the aforementioned properties. However, it is worth stressing that for some values of the contraction parameters   the CK cocommutator could   involve imaginary quantities due to term $\sqrt{\k_1\k_4}$ in the CK $r$-matrix, although the CK algebras are always real ones. If 
we require to have a {\em real} Lie bialgebra then 
\be
\k_1\k_4\ge 0, 
\label{constraint}
\ee
which excludes the 18 cases with the following values for $(\k_1,\k_2,\k_3,\k_4)$:
\be
\mbox{Complex Lie bialgebras:}\quad (+,\k_2,\k_3,-) \quad  \mbox{and}\quad  (-,\k_2,\k_3,+) ,\quad \forall \k_2,\k_3 .
\nonumber
\ee
Hence there remain 63  real Lie bialgebras which are explicitly presented in Table~\ref{table1} according to the sign or zero value of the contraction parameters and following the notation (\ref{ia})--(\ref{id}).

\begin{table}[htp] 
{
\footnotesize
\caption{\small The 63 Cayley--Klein  algebras with commutation relations (\ref{mf}) endowed with a real Lie bialgebra structure $( \mathfrak{so}_\k(5), \delta(r))$, determined by the  $r$-matrix (\ref{nb}),  according to    the sign of the graded contraction parameters $\k=(\k_1,\k_2,\k_3,\k_4)$ such that $\k_1\k_4\ge 0$.}
\label{table1}
 \begin{center}
 \begin{tabular}{l l }
 
\hline  

\hline

\\[-0.20cm]
\multicolumn{2}{l}{$\bullet$ Simple Lie algebras $\mathfrak{so}(p,q)$}\\[0.2cm]
  $\mathfrak{so}(5)$  & $\k = (+,+,+,+) $  \\[4pt]
 $\mathfrak{so}(4,1)$  & $\k =(+,-,-,+)$   \\[4pt]
  $\mathfrak{so}(3,2)$   & $\k =(+,+,-,+),(+,-,+,+),(-,+,+,-),  (-,+,-,-),(-,-,+,-),(-,-,-,-)$    \\[6pt]
\hline
\\[-0.2cm]
\multicolumn{2}{l}{$\bullet$ Inhomogeneous Lie  algebras    $\mathfrak{iso}(p,q)\simeq  \mathfrak{i'so}(p,q)$}\\[0.2cm]
 $\mathfrak{iso}(4)$  & $\k = (0,+,+,+), (+,+,+,0) $  \\[4pt]
  $\mathfrak{iso}(3,1)$  & $\k = (0,+,+,-) ,(0,-,+,+),(0,+,-,-),(0,-,-,+), $  \\[4pt]
   & $\qquad\,  (-,+,+,0) ,(+,+,-,0),(-,-,+,0),(+,-,-,0) $  \\[4pt]
 $\mathfrak{iso}(2,2)$  & $\k =(0,+,-,+),(0,-,+,-),(0,-,-,-), ( +,-,+,0),(-,+,-,0),(-,-,-,0)$     \\[6pt]
\hline
\\[-0.2cm]
\multicolumn{2}{l}{$\bullet$ Newton--Hooke type  algebras    $\mathfrak{ i}_6 \bigl(\mathfrak{so}(p,q)\oplus
\mathfrak{so }(p',q') \bigr)$}\\[0.2cm]
 \multicolumn{2}{l}{   $\mathfrak{ i}_6  \bigl(\mathfrak{so}(2)\oplus
\mathfrak{so }(3) \bigr)$  \qquad \quad $\k = (+,0,+,+), (+,+,0,+) $}  \\[4pt]
 \multicolumn{2}{l}{  $\mathfrak{ i}_6 \bigl(\mathfrak{so}(2)\oplus
\mathfrak{so }(2,1) \bigr)$  \qquad $\k = (+,0,-,+), (+,-,0,+) $ } \\[4pt]
 \multicolumn{2}{l}{   $\mathfrak{ i}_6  \bigl(\mathfrak{so}(1,1)\oplus
\mathfrak{so }(2,1) \bigr)$  \quad $\k = (-,0,+,-),(-,0,-,-),  (-,+,0,-) ,  (-,-,0,-)$  }   \\[6pt]
\hline
\\[-0.2cm]
\multicolumn{2}{l}{$\bullet$ Twice inhomogeneous Lie  algebras    $\mathfrak{iiso}(p,q)\simeq  \mathfrak{i'i'so}(p,q)$}\\[0.2cm]
  $\mathfrak{iiso}(3)$  & $\k = (0,0,+,+), (+,+,0,0) $  \\[4pt]
  $\mathfrak{iiso}(2,1)$  & $\k = (0,0,+,-) ,(0,0,-,+),(0,0,-,-), (-,+,0,0),(+,-,0,0),(-,-,0,0)$   \\[6pt]
\hline
\\[-0.2cm] 
\multicolumn{2}{l}{$\bullet$ Carroll type  algebras    $\mathfrak{ii'so}(p,q)\simeq  \mathfrak{i'iso}(p,q)$}\\[0.2cm]
  $\mathfrak{ii'so}(3)$  & $\k = (0,+,+,0) $  \\[4pt]
  $\mathfrak{ii'so}(2,1)$  & $\k = (0,+,-,0) ,(0,-,+,0),(0,-,-,0) $   \\[6pt]
\hline
\\[-0.2cm]
\multicolumn{2}{l}{$\bullet$    $\mathfrak{ i}_6 \bigl(\mathfrak{so}(p,q)\oplus
\mathfrak{iso }(p',q') \bigr)$}\\[0.2cm]
 \multicolumn{2}{l}{    $\mathfrak{ i}_6 \bigl(\mathfrak{so}(2)\oplus
\mathfrak{iso }(2) \bigr)$  \qquad\quad $\k = (+,0,0,+)  $ } \\[4pt]
 \multicolumn{2}{l}{  $\mathfrak{ i}_6 \bigl(\mathfrak{so}(1,1)\oplus
\mathfrak{iso }(1,1) \bigr)$  \quad $\k = (-,0,0,-)  $  }   \\[6pt]
\hline
\\[-0.2cm]
\multicolumn{2}{l}{$\bullet$   Inhomogeneous Newton--Hooke type  algebras    $\mathfrak{ i}_4\bigl(  \mathfrak{ i}_4\bigl(\mathfrak{so}(p,q)\oplus
\mathfrak{so }(p',q') \bigr) \bigr)$}\\[0.2cm]
 \multicolumn{2}{l}{   $ \mathfrak{ i}_4\bigl(  \mathfrak{ i}_4\bigl(\mathfrak{so}(2)\oplus
\mathfrak{so }(2) \bigr) \bigr) $  \qquad\quad  $\k = (0,+,0,+) ,  (+,0,+,0)$  }\\[4pt]
\multicolumn{2}{l}{   $ \mathfrak{ i}_4\bigl(  \mathfrak{ i}_4 \bigl(\mathfrak{so}(2)\oplus
\mathfrak{so }(1,1) \bigr) \bigr) $  \qquad  $\k = (0,+,0,-) ,(0,-,0,+) ,  (-,0,+,0),(+,0,-,0) $  }  \\[4pt]
\multicolumn{2}{l}{   $ \mathfrak{ i}_4\bigl(  \mathfrak{ i}_4 \bigl(\mathfrak{so}(1,1)\oplus
\mathfrak{so }(1,1) \bigr) \bigr) $  \quad  $\k = (0,-,0,-) ,  (-,0,-,0)$  }
   \\[6pt]
\hline
\\[-0.2cm]
\multicolumn{2}{l}{$\bullet$ Thrice inhomogeneous Lie  algebras    $\mathfrak{iiiso}(p,q)\simeq  \mathfrak{i'i' i'so}(p,q)$}\\[0.2cm]
 $\mathfrak{iiiso}(2)$  & $\k = (0,0,0,+), (+,0,0,0) $  \\[4pt]
  $\mathfrak{iiiso}(1,1)$  & $\k = (0,0,0,-) ,(0,0,0,-) $   \\[6pt]
\hline
\\[-0.2cm] 
\multicolumn{2}{l}{$\bullet$  Inhomogeneous Carroll type  algebras    $\mathfrak{iii'so}(p,q)\simeq  \mathfrak{i'i'iso}(p,q)$}\\[0.2cm]
 $\mathfrak{iii'so}(2)$  & $\k = (0,0,+,0), (0,+,0,0) $  \\[4pt]
  $\mathfrak{iii'so}(1,1)$  & $\k = (0,0,-,0), (0,-,0,0)$   \\[6pt]
\hline
\\[-0.2cm] 
\multicolumn{2}{l}{$\bullet$ Flag algebra     $\mathfrak{iiiiso}(1)\simeq  \mathfrak{i'i'i'i'so}(1)$}\\[0.2cm]
   $\mathfrak{iiiiso}(1)$  & $\k = (0,0,0,0) $   \\[6pt]
\hline  

\hline

\end{tabular}
\end{center}
}
 \end{table}
 

 \newpage
\subsection{Dual Cayley--Klein   Algebra and Noncommutative Cayley--Klein Spaces}
\label{s33}

 We consider the generators  $\JJ^{ab}$  in  $ \mathfrak{so}_\k(5)^\ast$  dual to  $J_{ab}$ in 
$ \mathfrak{so}_\k(5)$    with     pairing  (\ref{pairing}) and compute the commutation rules of $ \mathfrak{so}_\k(5)^\ast$  (\ref{td}) from the CK    cocommutator (\ref{cocok})  getting  
 \be
\begin{array}{lll}
   [\JJ^{01}, \JJ^{02}] =0 ,    & 
[\JJ^{01}, \JJ^{12}] =z\k_3\k_4  \JJ^{24}  , &  \  \ 
[\JJ^{02}, \JJ^{12}] =z\k_3 (\sqrt{\k_1\k_4}\, \JJ^{03} - \k_4  \JJ^{14}  ),  \\[4pt]
[\JJ^{01}, \JJ^{03}] =0 ,& 
[\JJ^{01}, \JJ^{13}] = z\k_4   \JJ^{34} , &  \  \ 
[\JJ^{03}, \JJ^{13}] =-z\k_4   \JJ^{14} ,\\[4pt]
[\JJ^{01}, \JJ^{04}] =z \JJ^{01} , & 
[\JJ^{01}, \JJ^{14}] =0 , &  \  \ 
[\JJ^{04}, \JJ^{14}] =-z \JJ^{14} ,\\[4pt]
[\JJ^{02}, \JJ^{03}] =0 , &\multicolumn{2}{l}{   
[\JJ^{02}, \JJ^{23}] = z (\sqrt{\k_1\k_4}\, \JJ^{01} +\k_4 \JJ^{34}  ) , \quad\ 
[\JJ^{03}, \JJ^{23}] =-z\k_4  \JJ^{24} ,}\\[4pt]
[\JJ^{02}, \JJ^{04}] = z\JJ^{02}  ,&   
[\JJ^{02}, \JJ^{24}] = 0 , &  \  \ 
[\JJ^{04}, \JJ^{24}] = -z\JJ^{24}, \\[4pt]
[\JJ^{03}, \JJ^{04}] =z \JJ^{03} , &    
[\JJ^{03}, \JJ^{34}] =0 ,&  \  \ 
[\JJ^{04}, \JJ^{34}] = -z\JJ^{34} ,\\[4pt]
\multicolumn{3}{l}{  [\JJ^{12}, \JJ^{13}] =z\sqrt{\k_1\k_4}\, \JJ^{12}  , \qquad
[\JJ^{12}, \JJ^{23}] = 0 ,  \qquad\!\!
[\JJ^{13}, \JJ^{23}] = -z\sqrt{\k_1\k_4}\, \JJ^{23} , }\\[4pt]
 [\JJ^{12}, \JJ^{14}] = z\k_1  \JJ^{02} ,  &\multicolumn{2}{l}{ \   
[\JJ^{12}, \JJ^{24}] = -z (\k_1  \JJ^{01} +\sqrt{\k_1\k_4}\,  \JJ^{34}  ) , \quad
[\JJ^{14}, \JJ^{24}] =0,} \\[4pt]
[\JJ^{13}, \JJ^{14}] =z \k_1 \JJ^{03}  ,&   \ 
[\JJ^{13}, \JJ^{34}] =-z\k_1 \JJ^{01} ,&\ \ 
[\JJ^{14}, \JJ^{34}] =0,\\[4pt]
 \multicolumn{3}{l}{ [\JJ^{23}, \JJ^{24}] =z\k_2\bigl( \k_1 \JJ^{03} - \sqrt{\k_1\k_4}\,  \JJ^{14} \bigr), \quad 
[\JJ^{23}, \JJ^{34}] = -z\k_1\k_2 \JJ^{02}  ,   \quad 
[\JJ^{24}, \JJ^{34}] =0 ,}
\end{array}
\label{oa}
\ee
  \be
\begin{array}{lll}
[\JJ^{01}, \JJ^{23}] =  -z\k_2\sqrt{\k_1\k_4}\,   \JJ^{02} , &\quad 
[\JJ^{01}, \JJ^{24}] = 0, &\quad\  \,
[\JJ^{01}, \JJ^{34}] = 0, \\[4pt]
[\JJ^{02}, \JJ^{13}] = 0 ,&\quad
[\JJ^{02}, \JJ^{14}] = 0 ,&\quad\   \,
[\JJ^{02}, \JJ^{34}] = 0, \\[4pt]
[\JJ^{03}, \JJ^{12}] =  -z\sqrt{\k_1\k_4}\,  \JJ^{02} , &\quad
[\JJ^{03}, \JJ^{14}] = 0, &\quad\   \,
[\JJ^{03}, \JJ^{24}] = 0, \\[4pt]
[\JJ^{04}, \JJ^{12}] = 0 ,&\quad
[\JJ^{04}, \JJ^{13}] = 0 ,&\quad\   \,
[\JJ^{04}, \JJ^{23}] = 0, \\[4pt]
[\JJ^{12}, \JJ^{34}] =   z\k_3\sqrt{\k_1\k_4}\,  \JJ^{24} , &\quad
[\JJ^{13}, \JJ^{24}] = 0, &\quad\   \,
[\JJ^{14}, \JJ^{23}] = -  z\sqrt{\k_1\k_4}\,  \JJ^{24} .
\end{array}
\label{ob}
\ee

 Alternatively, the same result comes out by applying a contraction map directly to the commutation relations of the dual algebra $\mathfrak{so}(5)^\ast$ (\ref{aab}) and (\ref{aacx}). By taking into account the pairing (\ref{pairing}) and the maps $\Phi$  (\ref{maa}) and $\Psi$  (\ref{na}), the full contraction map for  $\mathfrak{so}(5)^\ast$ turns out to be
\be
\Phi\circ \Psi(\JJ^{ab},z):=  \bigl( \Phi(\JJ^{ab}),\Psi(z) \bigr)=  \biggl( \frac{\JJ^{ab}}{\sqrt{\k_{ab}}},\frac{z}{ \sqrt{\k_{04}}} \biggr) .
\nonumber
\ee
We remark that the commutation relations (\ref{oa}) and (\ref{ob}) define a real dual CK algebra  $\mathfrak{so}(5)^\ast$ under the constraint (\ref{constraint}), thus covering the 63 cases given in Table~\ref{table1}. Note also that all the commutators (\ref{ob}) vanish for either $\k_1=0$ or $\k_4=0$, corresponding to the dual algebra of inhomogeneous algebras and their contractions.

Similarly to (\ref{fc}),  we express  the dual CK algebra $\mathfrak{so}_\k(5)^\ast$ as the sum of two vector spaces 
\be
{\mathfrak{so}}_\k(5)^\ast= \mathfrak{h}^{(m)}_\asttk\oplus \mathfrak{t}^{(m)}_\asttk \, ,\qquad m=1,\dots, 4,
\nonumber
\ee
where $\mathfrak{h}^{(m)}_\asttk$ and $\mathfrak{t}^{(m)}_\asttk$ are the annihilators of the vector subspaces $\mathfrak{h}_\k^{(m)}$ and $\mathfrak{t}_\k^{(m)}$ introduced in (\ref{mp})  and fulfilling (\ref{mq}). As we already performed in  Section~\ref{s23} for ${\mathfrak{so}}(5)^\ast$, we   define the {\em first-order noncommutative CK spaces}     by
\be
\sz^{(m)} := \mathfrak{h}^{(m)}_\asttk \, ,\qquad  m=1,\dots, 4 ;
\label{oobb}
\ee
see (\ref{fd})--(\ref{fg}).
Each  linear noncommutative   space $\sz^{(m)} $ is    the first-order in the quantum coordinates of the complete noncommutative space associated with the homogeneous CK space $\mathbb S^{(m)}_{\k}$ (\ref{mqq}).   We display in Table~\ref{table2}   the defining commutation relations for the four  noncommutative CK spaces along with the Lie brackets among 
 $\mathfrak{h}^{(m)}_\asttk$ and $\mathfrak{t}^{(m)}_\asttk$.

 
\begin{table}[t!] 
{
\small
\caption{\small The   first-order noncommutative Cayley--Klein  spaces  $\sz^{(m)}$ (\ref{oobb})  and the relations among $\mathfrak{h}^{(m)}_\asttk$ and $\mathfrak{t}^{(m)}_\asttk$. Real commutation relations are ensured whenever $\k_1\k_4\ge 0$ covering the 63  cases shown in Table~\ref{table1}.}
\label{table2}
 \begin{center}
 
\begin{tabular}{l l l}
\hline  

\hline

\\[-0.2cm]
\multicolumn{3}{l}{$\bullet$   Noncommutative CK  space of points    $\mathbb S^{(1)}_{z,\k}\equiv \mathfrak{h}^{(1)}_\asttk=\langle \JJ^{01},\JJ^{02},\JJ^{03},\JJ^{04}\rangle$}\\[0.3cm]
\multicolumn{3}{l}{$ [\JJ^{0i}, \JJ^{04}] = z \JJ^{0i}\qquad   [\JJ^{0i}, \JJ^{0j}]=0\qquad i,j=1,2,3$}\\[0.3cm]
\multicolumn{3}{l}{ $\bigl[ \mathfrak {h}^{(1)}_\asttk, \mathfrak{h}^{(1)}_\asttk] \subset \mathfrak{h}^{(1)}_\asttk   \quad\  \ \,
 \bigl[ \mathfrak{h}^{(1)}_\asttk, \mathfrak{t}^{(1)}_\asttk] \subset \mathfrak{t}^{(1)}_\asttk +\k_1\k_4\mathfrak{h}^{(1)}_\asttk    \quad  \ \,
 \bigl[\mathfrak{t}^{(1)}_\asttk,\mathfrak {t}^{(1)}_\asttk \bigr] \subset \k_1\mathfrak{h}^{(1)}_\asttk  + \k_1\k_4 \mathfrak{t}^{(1)}_\asttk$}    \\[6pt]
\hline
\\[-0.2cm]
 \multicolumn{3}{l}{$\bullet$   Noncommutative  CK space of lines    $\mathbb S^{(2)}_{z,\k}\equiv \mathfrak{h}^{(2)}_\asttk=\langle \JJ^{02},\JJ^{03},\JJ^{04},\JJ^{12},\JJ^{13},\JJ^{14}\rangle$}\\[0.3cm]
$ [\JJ^{02}, \JJ^{03}] =0$& $[\JJ^{02}, \JJ^{04}] = z\JJ^{02}  $&  $[\JJ^{03}, \JJ^{04}] =z \JJ^{03} $\\[0.2cm]
$ [\JJ^{12}, \JJ^{13}] =z\sqrt{\k_1\k_4}\, \JJ^{12}   $& $ [\JJ^{12}, \JJ^{14}] = z\k_1  \JJ^{02}  $&  $[\JJ^{13}, \JJ^{14}] =z \k_1 \JJ^{03}  $\\[0.2cm]
\multicolumn{3}{l}{$ [\JJ^{02}, \JJ^{12}] =z\k_3\bigl(\sqrt{\k_1\k_4}\, \JJ^{03} - \k_4  \JJ^{14} \bigr)\qquad  [\JJ^{03}, \JJ^{12}] =  -z\sqrt{\k_1\k_4}\,  \JJ^{02} \qquad [\JJ^{04}, \JJ^{12}] = 0$}\\[0.2cm]
$ [\JJ^{02}, \JJ^{13}] = 0  $& $[\JJ^{03}, \JJ^{13}] =-z\k_4   \JJ^{14} $&  $[\JJ^{04}, \JJ^{13}] = 0$\\[0.2cm]
$[\JJ^{02}, \JJ^{14}] = 0   $& $[\JJ^{03}, \JJ^{14}] = 0 $&  $[\JJ^{04}, \JJ^{14}] =-z \JJ^{14} $\\[0.3cm]
$\bigl[ \mathfrak {h}^{(2)}_\asttk, \mathfrak{h}^{(2)}_\asttk] \subset \mathfrak{h}^{(2)}_\asttk $ &   
$\bigl[ \mathfrak{h}^{(2)}_\asttk, \mathfrak{t}^{(2)}_\asttk] \subset \mathfrak{t}^{(2)}_\asttk     $ &   
$\bigl[\mathfrak{t}^{(2)}_\asttk,\mathfrak {t}^{(2)}_\asttk \bigr] \subset \k_1 \k_2\mathfrak{h}^{(2)}_\asttk  $    \\[6pt]
\hline
\\[-0.2cm]
\multicolumn{3}{l}{$\bullet$   Noncommutative  CK space of 2-planes    $\mathbb S^{(3)}_{z,\k}\equiv \mathfrak{h}^{(3)}_\asttk=\langle \JJ^{03},\JJ^{04},\JJ^{13},\JJ^{14},\JJ^{23},\JJ^{24}\rangle$}\\[0.3cm]
$[\JJ^{03}, \JJ^{13}] =-z\k_4   \JJ^{14}  $& $[\JJ^{03}, \JJ^{23}] =-z\k_4  \JJ^{24}$&  $[\JJ^{13}, \JJ^{23}] = -z\sqrt{\k_1\k_4}\, \JJ^{23} $\\[0.2cm]
$[\JJ^{04}, \JJ^{14}] =-z \JJ^{14}  $& $[\JJ^{04}, \JJ^{24}] = -z\JJ^{24} $&  $[\JJ^{14}, \JJ^{24}] =0 $\\[0.2cm]
$ [\JJ^{03}, \JJ^{04}] =z \JJ^{03} $& $[\JJ^{13},\JJ^{04}] = 0 $&  $[ \JJ^{23},\JJ^{04}] = 0$\\[0.2cm]
$[\JJ^{03}, \JJ^{14}] = 0  $& $[\JJ^{13}, \JJ^{14}] =z \k_1 \JJ^{03}$& $ [ \JJ^{23},\JJ^{14}] =  z\sqrt{\k_1\k_4}\,  \JJ^{24}  $\\[0.2cm]
 $  [\JJ^{03}, \JJ^{24}] = 0 $&$ [\JJ^{13}, \JJ^{24}] = 0  $&$  [\JJ^{23}, \JJ^{24}] =z\k_2\bigl( \k_1 \JJ^{03} - \sqrt{\k_1\k_4}\,  \JJ^{14} \bigr) $  \\[0.3cm]
$\bigl[ \mathfrak {h}^{(3)}_\asttk, \mathfrak{h}^{(3)}_\asttk] \subset \mathfrak{h}^{(3)}_\asttk $ &   
$\bigl[ \mathfrak{h}^{(3)}_\asttk, \mathfrak{t}^{(3)}_\asttk] \subset \mathfrak{t}^{(3)}_\asttk     $ &   
$\bigl[\mathfrak{t}^{(3)}_\asttk,\mathfrak {t}^{(3)}_\asttk \bigr] \subset \k_3 \k_4\mathfrak{h}^{(3)}_\asttk  $    \\[6pt]
\hline
\\[-0.2cm]
 \multicolumn{3}{l}{$\bullet$   Noncommutative  CK space of 3-hyperplanes    $\mathbb S^{(4)}_{z,\k}\equiv \mathfrak{h}^{(4)}_\asttk=\langle \JJ^{04},\JJ^{14},\JJ^{24},\JJ^{34}\rangle$}\\[0.3cm]
\multicolumn{3}{l}{$ [\JJ^{i4}, \JJ^{04}] = z \JJ^{i4}\qquad  [\JJ^{i4}, \JJ^{j4}]=0\qquad i,j=1,2,3$}\\[0.3cm]
\multicolumn{3}{l}{  $\bigl[ \mathfrak {h}^{(4)}_\asttk, \mathfrak{h}^{(4)}_\asttk] \subset \mathfrak{h}^{(4)}_\asttk   \quad\   \ \,
 \bigl[ \mathfrak{h}^{(4)}_\asttk, \mathfrak{t}^{(4)}_\asttk] \subset \mathfrak{t}^{(4)}_\asttk +\k_1\k_4\mathfrak{h}^{(4)}_\asttk   \quad\   \  
 \bigl[\mathfrak{t}^{(4)}_\asttk,\mathfrak {t}^{(4)}_\asttk \bigr] \subset \k_4\mathfrak{h}^{(4)}_\asttk  + \k_1\k_4 \mathfrak{t}^{(4)}_\asttk$ }   \\[8pt]
\hline  

\hline

\end{tabular}
\end{center}
}
 \end{table}
 

 Now we analyse the structure and properties of such  noncommutative  CK spaces which do strongly depend on the contraction/curvature  parameters. The four noncommutative spaces  close on a  Lie subalgebra $\mathfrak{h}^{(m)}_\asttk$, in agreement with the 
  coisotropy condition (\ref{coisok}), and the   noncommutative spaces of lines    and 2-planes   are  both reductive and symmetric as it was also the case for ${\mathfrak{so}}(5)^\ast$ (see   (\ref{symmetryc})). 
    Furthermore, the explicit presence of the curvature parameters allows us to highlight some properties for the  contracted noncommutative spaces  straightforwardly. In particular, if we set $\k_4=0$ in the noncommutative space of points we find that
\be 
\k_4=0\!:\quad  \bigl[ \mathfrak {h}^{(1)}_\asttk, \mathfrak{h}^{(1)}_\asttk] \subset \mathfrak{h}^{(1)}_\asttk \, , \qquad 
 \bigl[ \mathfrak{h}^{(1)}_\asttk, \mathfrak{t}^{(1)}_\asttk] \subset \mathfrak{t}^{(1)}_\asttk    \, ,  \qquad 
 \bigl[\mathfrak{t}^{(1)}_\asttk,\mathfrak {t}^{(1)}_\asttk \bigr] \subset \k_1\mathfrak{h}^{(1)}_\asttk  .
\nonumber
\ee
Likewise, taking $\k_1=0$ in the noncommutative space of 3-hyperplanes we obtain that
   \be 
\k_1=0\!:\quad  \bigl[ \mathfrak {h}^{(4)}_\asttk, \mathfrak{h}^{(4)}_\asttk] \subset \mathfrak{h}^{(4)}_\asttk   \, , \qquad 
 \bigl[ \mathfrak{h}^{(4)}_\asttk, \mathfrak{t}^{(4)}_\asttk] \subset \mathfrak{t}^{(4)}_\asttk  \,  ,   \qquad 
 \bigl[\mathfrak{t}^{(4)}_\asttk,\mathfrak {t}^{(4)}_\asttk \bigr] \subset \k_4\mathfrak{h}^{(4)}_\asttk    .
\nonumber
\ee
Thus both contracted noncommutative spaces are reductive and symmetric (to be compared with (\ref{mq})).
And a remarkable common property for the four noncommutative spaces is that when $\k_m=0$, the subspace $\mathfrak {t}^{(m)}_\asttk $ becomes an abelian subalgebra $(m=1,2,3,4)$:
   \be 
\k_m=0\!:\quad    \bigl[ \mathfrak {h}^{(m)}_\asttk, \mathfrak{h}^{(m)}_\asttk] \subset \mathfrak{h}^{(m)}_\asttk  \, , \qquad 
 \bigl[ \mathfrak{h}^{(m)}_\asttk, \mathfrak{t}^{(m)}_\asttk] \subset \mathfrak{t}^{(m)}_\asttk  \, ,   \qquad 
 \bigl[\mathfrak{t}^{(m)}_\asttk,\mathfrak {t}^{(m)}_\asttk \bigr] =0  .
\nonumber
\ee
Such relations can be applied,  
for instance, to the inhomogeneous Poincar\'e and Euclidean algebras    with $\k_1=0$ (\ref{ia})  for the noncommutative space of points, to the Newton--Hooke type algebras   with $\k_2=0$ (\ref{ic})  for the noncommutative space of lines, to the twice inhomogeneous algebras (Galilei)  with $\k_1=\k_2=0$ (\ref{ib}) for 
 the noncommutative spaces of points and lines,  and so on up to reach the flag algebra  (\ref{id}) for the four noncommutative spaces.

The dual   polarity  ${d}_z$ (\ref{odzb}) also holds for   $\mathfrak{so}(5)_\k^\ast$  interchanging the noncommutative CK spaces as in (\ref{odzc}) and 
  the curvature parameters in the form $\k_1\leftrightarrow \k_4$ and $\k_2\leftrightarrow \k_3$.   Moreover, if we  consider the   $z$-maps  $\theta^{(m)}_z$  (\ref{invzb})  in the dual CK algebra $\mathfrak{so}_\k(5)^\ast$,  we again find that only $\theta^{(2)}_z$ and $\theta^{(3)}_z$ are always involutive automorphisms of the commutation rules (\ref{oa}) and (\ref{ob})  (as for $\mathfrak{so}(5)^\ast$). However both $\theta^{(1)}_z$ and $\theta^{(4)}_z$ become involutions whenever the product $\k_1\k_4=0$. Consequently, when at least either $\k_1=0$  or $\k_4=0$, the four maps  $\theta^{(m)}_z$  (\ref{invzb})  span a ${\mathbb {Z}}_2^{\otimes 4}$-grading for 
$\mathfrak{so}_\k(5)^\ast$ and the four noncommutative CK spaces (\ref{oobb})  are all reductive and symmetric; recall that these 45 cases correspond to the inhomogeneous algebras $\mathfrak{iso}(p,q)$  with $p+q=4$ with curvature coefficients $(0,\k_2,\k_3,\k_4)$  (\ref{ia})  or  $(\k_1,\k_2,\k_3,0)$  (\ref{iaa}) and their further contractions.

Finally, as we advanced at the end of Section~\ref{s12}, it is worth stressing that the structure of the first-order noncommutative CK space of points $\mathbb S^{(1)}_{z,\k}$, shown in Table~\ref{table2}, is   shared by the 63 CK real Lie bialgebras
with $\k_1\k_4\ge 0$  of Table~\ref{table1} since no $\k_m$ appears within $\mathbb S^{(1)}_{z,\k}$, and  that this  is formally similar to the $\kappa$-Minkowski spacetime (\ref{tss}).  Furthermore, the commutation relations of $\mathbb S^{(1)}_{z,\k}$  are kept linear under   full quantization  for   the 27 CK bialgebras with parameters $(0,\k_2,\k_3,\k_4)$, while  higher-order terms in the quantum coordinates are expected for the CK bialgebras with $\k_1\ne 0$. Similar properties hold for the 
 first-order noncommutative CK space of 3-hyperplanes $\mathbb S^{(4)}_{z,\k}$ which   remains as a linear full noncommutative space for  the 27 CK bialgebras with parameters $( \k_1,\k_2,\k_3,0)$.  
  Nevertheless, if one looks at the {\em four}  first-order noncommutative CK spaces  in Table~\ref{table2} {\em altogether}, then one finds that the four contraction parameters appear explicitly. Thus it turns out that the  set of   four noncommutative spaces $\mathbb S^{(m)}_{z,\k}$ is {\em different} for each specific CK bialgebra except for the nine cases with $\k_1=\k_4=0$, for which all  the terms involving any $\k_m$ vanish.  Consequently,  this observation suggests the necessity of  constructing other noncommutative spaces beyond the usual  noncommutative spacetime for a given quantum deformation. To the best of our knowledge, there are very scarce results in this research direction which concern   noncommutative spaces of lines~\cite{Lines2014, BGH2019worldlinesplb}.

A physical (kinematical) analysis on the  
 noncommutative CK spaces  of points and lines will be addressed in Section~\ref{s5}.

 
\subsection{Drinfel'd Double Structures for   Cayley--Klein Algebras}
\label{s34}

Let us consider the classical $r$-matrix  $\tilde r_D$ (\ref{rdd}) coming from the Drinfel'd double structure of $\mathfrak{so}(5)$. We apply the  composition of the contraction maps   (\ref{maa}) and  (\ref{na}) in the form
 \be
r_D=\bigl( \Phi^{-1}\otimes \Phi^{-1} \bigr)\circ  \Psi^{-1}(\tilde r_D),\qquad \forall\, \k_m\ne 0,
\nonumber
\ee
obtaining that
\bea
&& r_D=z \left(J_{14}\wedge J_{01} + J_{24}\wedge J_{02}
+ J_{34}\wedge J_{03} + \sqrt{\k_1\k_4}\, J_{23}\wedge J_{12}+\frac 1{ \sqrt{\k_2\k_3}} \,J_{13}\wedge J_{04} \right) \nonumber\\[2pt]
&&\quad\  = r+\frac z{ \sqrt{\k_2\k_3}} \, J_{13}\wedge J_{04} ,\qquad \forall\, \k_m\ne 0,
\label{pa}
\eea
which is a superposition of the   classical CK $r$-matrix  $r$  (\ref{nb}) with a Reshetikhin {\em twist} $J_{13}\wedge J_{04}$ formed by the two commuting primitive generators. We remark that, by construction, $r_D$  is a classical $r$-matrix coming from the  Drinfel'd double structure      for the simple Lie algebras contained in the CK family $\mathfrak{so}_\k(5)$. Moreover, $r_D$   leads to the same Schouten bracket as  for $r$ (\ref{nby}) (there are no twist contributions) so that  it  is a solution of the  modified classical Yang--Baxter equation  (\ref{th}).

If we now require  $r_D$  (\ref{pa}) to define a {\em real} Lie bialgebra, $( \mathfrak{so}_\k(5), \delta_D(r_D))$, we have to impose the restriction 
corresponding to $r$ (\ref{constraint}) together with  the new one determined by the twist:
\be
\k_1\k_4> 0\quad \mbox{and}\quad  \k_2\k_3>0 ,
\label{constraints}
\ee
which  lead to {\em four} possible cases shown in Table~\ref{table3}, where we have named them according with their    kinematical interpretation that  we shall show in     Section~\ref{s5}.


\begin{table}[t!] 
{\small
\caption{\small Simple Lie  algebras   with a real $r$-matrix $r_D$ (\ref{pa}) coming from  a Drinfel'd double   structure according to    the sign of the graded contraction parameters $\k=(\k_1,\k_2,\k_3,\k_4)$  and    bilinear form $\bfI_\k$ (\ref{mi}),  along with their  contractions to  non-simple Lie  algebras  endowed with a real Lie bialgebra $( \mathfrak{so}_\k(5), \delta_D(r_D))$.}
\label{table3}
 \begin{center}
 \begin{tabular}{l l l l l l l l}
\hline  

\hline

\\[-0.2cm]
\multicolumn{8}{c}{\textbf{Simple Lie  algebras   with a Drinfel'd double real  structure}}\\[0.2cm]
\hline
\\[-0.2cm] 
(I)&Spherical &$\mathfrak{so}(5)$  & $\k = (+,+,+,+) $&\multicolumn{4}{l}{ \quad  $\bfI_\k= (+,+,+,+,+)$ } \\[4pt]
(II)&    De Sitter &$\mathfrak{so}(4,1)$  & $\k =(+,-,-,+)$&\multicolumn{4}{l}{ \quad $\bfI_\k= (+,+,-,+,+)$} \\[4pt]
(III)&   Anti-de Sitter  & $\mathfrak{so}(3,2)$   & $\k =(-,+,+,-)$ & \multicolumn{4}{l}{\quad $ \bfI_\k= (+,-,-,-,+) $ } \\[4pt]
(IV)&  Anti-de Sitter  & $\mathfrak{so}(3,2)$   & $\k =(-,-,-,-)$ & \multicolumn{4}{l}{\quad $\bfI_\k= (+,-,+,-,+) $ } \\[6pt]
\hline
\\[-0.2cm]
\multicolumn{8}{c}{\textbf{Non-simple Lie  algebras   with a real Lie bialgebra   via contraction}}\\[0.2cm]
\hline
\\[-0.2cm] 
(Ia)&  Euclidean &$\mathfrak{iso}(4)$  & $\k = (0,+,+,+) $   & &   & \\[4pt]
(Ia$^\prime$)&  Para-Euclidean &$\mathfrak{i'so}(4)$&$\k = (+,+,+,0) $ & & &\\[4pt]
(IIa)&   Poincar\'e &$\mathfrak{iso}(3,1)$  & $\k =(0,-,-,+)$   & &  &  & \\[4pt]
 (IIa$^\prime$)&  &$\mathfrak{i'so}(3,1)$ &$\k = (+,-,-,0) $ &  & &\\[4pt]
(IIIa)&  Poincar\'e & $\mathfrak{iso}(3,1)$   & $\k =(0,+,+,-)$ & &  &  & \\[4pt]
 (IIIa$^\prime$)&   Para-Poincar\'e &$\mathfrak{i'so}(3,1)$ &  $\k =(-,+,+,0)$ & & &\\[4pt]
(IVa)&     & $\mathfrak{iso}(2,2)$   & $\k =(0,-,-,-)$ & &  &  &   \\[4pt]
 (IVa$^\prime$)&  &$\mathfrak{i'so}(2,2)$ & $\k =(-,-,-,0)$ & & &\\[4pt]
(Ib) &  Carroll & $\mathfrak{ii'so}(3)$   & $\k =(0,+,+,0)$ & \multicolumn{4}{l}{  (Ib)$\,\equiv\,$(Ib$^\prime$)$\,\equiv\,$(IIIb)$\,\equiv\,$(IIIb$^\prime$) }\\[4pt]
(IIb) &    & $\mathfrak{ii'so}(2,1)$   & $\k =(0,-,-,0)$ &  \multicolumn{4}{l}{  (IIb)$\,\equiv\,$(IIb$^\prime$)$\,\equiv\,$(IVb)$\,\equiv\,$(IVb$^\prime$)  }  \\[6pt]
\hline  

\hline

\end{tabular}
\end{center}
 }
 \end{table}
 

In order to obtain the possible graded contractions of  $r_D$ (\ref{pa}), we firstly point out that this diverges  under the limits $\k_2\to 0$ and $\k_3\to 0$, so that the restriction $\k_2\k_3>0$ must be kept. And, secondly,  both contractions $\k_1\to 0$ and  $\k_4\to 0$ are well defined and consistent with the condition (\ref{constraint}). This, in turn, means that there are {\em ten} possible contractions for $r_D$ which provide   an $r$-matrix generating a real Lie bialgebra    for non-simple Lie algebras (see Table~\ref{table1}). 
 These are  also displayed in Table~\ref{table3} (with the kinematical terminology for the cases that will appear in   Section~\ref{s5}), where  the notation indicates the sequence of contractions  of the real Lie bialgebra  $( \mathfrak{so}_\k(5), \delta_D(r_D))$,  with the ``prime" corresponding to $\k_4=0$.
 For instance:
 \be
\begin{array}{ll}
  {\rm (I)}\,  \xrightarrow{\k_1=\,0}\,  {\rm (Ia)}\, \xrightarrow{\k_4=\, 0}\,  \mbox{(Ib$^\prime$)}
 & \quad   \mbox{Spherical} \  \to \  \mbox{Euclidean} \ \to\  \mbox{Carroll} 
    \\[2pt] 
   \mathfrak{so}(5)  \longrightarrow\,  \mathfrak{iso}(4)  \longrightarrow\, \mathfrak{ii'so}(3) & \quad 
 (+,+,+,+)  \to (0,+, +,+)  \to  {  (0,+,+,0) }       \\[6pt]
   {\rm (III)}\ \xrightarrow{\k_4=\, 0}\  {\rm (IIIa}^\prime)\ \xrightarrow{\k_1=\, 0}\ \mbox {(IIIb)} 
& \quad  \mbox{Anti-de Sitter} \ \to\  \mbox{Para-Poincar\'e} \ \to \  \mbox{Carroll}  
  \\[2pt]  
 \mathfrak{so}(3,2)\longrightarrow\,  \mathfrak{i^\prime so}(3,1)\longrightarrow\, \mathfrak{ii'so}(3)  & \quad 
    (-,+,+,-) \to  (-,+, +,0) \to (0,+,+,0) 
\end{array}
\nonumber
\ee
Recall  the Lie algebra isomorphisms provided by $\mathcal{D}$ (\ref{od}): 
$\mathfrak{i^\prime so}(4) \simeq \mathfrak{i so}(4)$, $ \mathfrak{i^\prime so}(3,1) \simeq \mathfrak{i so}(3,1)$ and
$ \mathfrak{i^\prime so}(2,2) \simeq \mathfrak{i so}(2,2)$.
  Hence any contraction sequence ends on either the Carroll bialgebra (cases (I) and (III)) or on the $\mathfrak{ii'so}(2,1)$ one (cases (II) and (IV)).

 The effect of the twist $J_{13}\wedge J_{04}$ in $r_D$ (\ref{pa})  with respect to the CK $r$-matrix (\ref{nb})  can be highlighted by associating it with  a second deformation parameter $\vartheta$  in a similar form to (\ref{rdde}), that is,
 \be
  r_D=  r+ \vartheta  J_{13}\wedge J_{04} ,
  \label{pf}
 \ee
 such that the $r$-matrix coming from a Drinfel'd double structure corresponds to the one-parame\-tric deformation with 
 \be
 \vartheta\equiv \frac z{ \sqrt{\k_2\k_3}} ,\qquad  \k_2\k_3>0.
 \label{pg}
 \ee
The cocommutator $\delta_D$, obtained with (\ref{tf}), is just the CK cocommutator $\delta$ (\ref{cocok}) plus new terms coming from the twist which are denoted $\delta_\vartheta$. Hence $\delta_D=\delta+ \delta_\vartheta$ with $\delta_\vartheta$ given by
  \bea
&& \delta_\vartheta(J_{04})=0,\qquad  \delta_\vartheta(J_{13})=0,\nonumber\\[2pt]
 &&  \delta_\vartheta(J_{12})=  \vartheta\,\k_2  J_{23}\wedge J_{04}, \qquad
\delta(J_{23})= \vartheta\,\k_3   J_{04}\wedge J_{12} ,\nonumber\\[2pt]
&&    \delta(J_{01})=\vartheta( J_{04}\wedge J_{03}+\k_1  J_{13}\wedge J_{14} ),\nonumber \\[2pt]
&&  \delta(J_{02})=\vartheta\, \k_1\k_2  J_{13}\wedge J_{24} , \label{cocokd}  \\[2pt]
 &&  \delta(J_{03})=\vartheta \, \k_2\k_3 (J_{01}\wedge J_{04}+\k_1  J_{13}\wedge J_{34} ), \nonumber\\[2pt]
&&  \delta(J_{14})=\vartheta \,\k_2\k_3(  J_{04}\wedge J_{34}+ \k_4  J_{01}\wedge J_{13} ),
\nonumber\\[2pt]
&&   \delta(J_{24})=\vartheta \, \k_3\k_4 J_{02}\wedge J_{13} , \nonumber\\[2pt]
&&  \delta(J_{34})=\vartheta( J_{14}\wedge J_{04}+\k_4   J_{03}\wedge J_{13} ).\nonumber
\eea
Consequently, $(\mathfrak{so}_{\k_2,\k_3}(3),\delta_D)$ with generators   $\langle
J_{12},J_{13},J_{23}\rangle$   does  not remain  as a Lie sub-bialgebra   of $( \mathfrak{so}_\k(5), \delta_D (r_D))$. However,  if this is enlarged with the primitive generator $J_{04}$  then it provides the Lie sub-bialgebra    $(\mathfrak{so}_{\k_2,\k_3}(3)\oplus   \mathfrak{so}_{\k_{04}}(2),\delta_D)$.   

The polarity $\mathcal{D}_z$ (\ref{odz}) and the involutions $\Theta^{(2)}_z$ and $\Theta^{(3)}_z$  (\ref{invz}) also hold for the 
     two-parametric  deformation determined by (\ref{pf}) provided that   $\vartheta$ is unchanged. Nevertheless,  in the proper   Drinfel'd double case with a single deformation parameter $z$,  with the identification (\ref{pg}), the above maps do not remain  since $z\to -z$.
 
 As far as the first-order noncommutative spaces  associated to  $( \mathfrak{so}_\k(5), \delta_D(r_D) )$ is concerned,  it is straightforward to  prove that the 
 coisotropy condition (\ref{coisotropy})  is only fulfilled for the subalgebras  $ \mathfrak{h}_\k^{(1)}$ and $ \mathfrak{h}_\k^{(4)}$ (see (\ref{cocokd})):
   \be
\delta_D( \mathfrak{h}_\k^{(l)})\subset \mathfrak{h}_\k^{(l)}\wedge\mathfrak{so}_\k(5) ,  \quad l=1,4 ,\qquad
\delta_D( \mathfrak{h}_\k^{(k)})\not\subset \mathfrak{h}_\k^{(k)}\wedge\mathfrak{so}_\k(5) ,  \quad k=2,3 ,
\nonumber
\ee
for {\em any} of the 14  Lie algebras displayed in Table~\ref{table3}. Therefore, only the {\em  twisted noncommutative CK spaces} of points and 3-hyperplanes  $\mathbb S^{(l)}_{z,\vartheta,\k}$ $(l=1,4)$ can consistently be  constructed. In particular, from (\ref{cocokd}) and applying the quantum duality  (\ref{td}) with  pairing  \eqref{pairing}, we directly obtain the defining commutation relations for the
 twisted noncommutative CK space  of points $\mathbb S^{(1)}_{z,\vartheta,\k}$:
\bea
&& [\JJ^{01}, \JJ^{04}] = z \JJ^{01}+ \vartheta \k_2\k_3  \JJ^{03}  , \qquad   
  [\JJ^{02}, \JJ^{04}] = z \JJ^{02}   , \qquad   
 [\JJ^{03}, \JJ^{04}] = z \JJ^{03}- \vartheta   \JJ^{01}  , \nonumber\\[2pt]   
&&  [\JJ^{0i}, \JJ^{0j}]=0,\qquad i,j=1,2,3 .
\label{ncdd}
  \eea
In the same way, $\mathbb S^{(4)}_{z,\vartheta,\k}$ can also be obtained. Clearly $\mathbb S^{(1)}_{z,\vartheta,\k}$ (\ref{ncdd})  is not isomorphic to $\mathbb S^{(1)}_{z,\k}$ given in Table~\ref{table2}. Moreover, since 
$\k_2\k_3>0$, this factor can be scaled to $+1$ within the commutators (\ref{ncdd}) via the scalings
\be
 \JJ^{03} \to \sqrt{\k_2\k_3}\, \JJ^{03} ,\qquad  \vartheta\to \sqrt{\k_2\k_3}\, \vartheta,
 \nonumber
 \ee
which shows that $\mathbb S^{(1)}_{z,\vartheta,\k}$ is the  {\em common} first-order twisted noncommutative CK space of points for the 14 
 Lie bialgebras shown in Table~\ref{table3}; obviously, higher-order terms in the quantum coordinates may arise for each specific case.

Finally, we stress that it is not ensured at all  that   a given contracted Drinfel'd double  $r$-matrix $r_D$ gives rise to a Drinfel'd double structure for a non-semisimple Lie algebra 
 and, in fact, this problem should be studied case by case.   Nevertheless,   we can answer negatively  to this question for the contracted $r$-matrices of Table~\ref{table3}.  It was established in~\cite{PoinDD},  from the results given in~\cite{Figueroa-OFarrill2018}, that there does not exist any Drinfel'd double structure for Poincar\'e, Euclidean and Carroll algebras at this dimension.   In contrast, as we commented at the end of Section~\ref{s13},  in lower dimensions such structures do exist and the classification of Drinfel'd doubles has recently been performed  for the (2+1)D Poincar\'e~\cite{PoinDD} and 3D Euclidean algebras~\cite{EuclDD}.
Moreover, to the best of our knowledge,   the classification of Drinfel'd doubles for the (anti-)de Sitter   algebras has only been carried out in (2+1) dimensions~\cite{BHM2013cqg}.
 In (3+1)  dimensions there is no such classification for  the simple Lie algebras $\mathfrak{so}(p,q)$ and  there has only been  constructed the Drinfel'd double  structure here considered for $\mathfrak{so}(5)$, reviewed in Section~\ref{s24},   and from it a Drinfel'd double for the   anti-de Sitter algebra $\mathfrak{so}(3,2)$~\cite{BHN2015towards31}, that we advance which is just the case (III) in Table~\ref{table3}.
Therefore we have  obtained  two new $r$-matrices coming from Drinfel'd doubles, one for   the  de Sitter $\mathfrak{so}(4,1)$ and  another for the  anti-de Sitter $\mathfrak{so}(3,2)$     (cases (II) and (IV)), although our results do not convey a complete classification.

The  physical (kinematical) interpretation of the     CK $r$-matrices  $r$ (\ref{nb}) and  $r_D$  (\ref{pa}) along with  their associated first-order noncommutative spaces    will be described in   detail in    Section~\ref{s5}.

  
\section{Kinematical  Algebras and   Homogeneous Spaces}
\label{s4}

 As we have already mentioned in the previous Section, the kinematical algebras introduced in~\cite{BLL} arise as particular cases of graded contractions of  $\mathfrak{so}(5)$~\cite{CKGraded,MontignyKinematical} so that they  appear within the CK family $\mathfrak{so}_\k(5)$ for some specific values of the contraction parameters $\k=(\k_1,\k_2,\k_3,\k_4)$ (except for the static algebra which does not belong to the CK family). These kinematical algebras have recently  been  derived from  deformation theory in~\cite{Figueroa-OFarrill2018,Figueroa-OFarrill2018higher}; in this respect, recall that Lie algebra deformations~\cite{Nijenhuis} can be regarded as the opposite processes to Lie algebra contractions~\cite{Montigny1,IW,Segal,Saletan}.   
   
   In order to deal  with  kinematical algebras let  us introduce a    physical basis   denoting by $P_0$, $\>P=(P_1,P_2,P_3)$, $\>K=(K_1,K_2,K_3)$ and $\>J=(J_1,J_2,J_3)$ the generators of time translations, spatial translations, boosts and rotations, respectively.  These ten generators are isometries of a (3+1)D spacetime of constant curvature. The 11 kinematical algebras~\cite{BLL} are contained within a   three-parametric Lie algebra, here denoted $\kk$, with commutation relations given by
 \be
[J_i,J_j]=\epsilon_{ijk}J_k ,  \qquad [J_i,P_j]=\epsilon_{ijk}P_k ,  \qquad
[J_i,K_j]=\epsilon_{ijk}K_k , \qquad [J_i,P_0]=0,
\label{qa}
\ee
 \be
\begin{array}{lll}
\displaystyle{
  [K_i,P_0]=\la P_i  } , &\qquad \displaystyle {[K_i,P_j]=\frac 1{c^2}\, \delta_{ij}  P_0} ,    &\qquad \displaystyle{[K_i,K_j]=-\la  \frac 1{c^2}\,\epsilon_{ijk} J_k} , 
\\[4pt][P_0,P_i]=-\ka    K_i , &\qquad \displaystyle {  [P_i,P_j]=\ka  \frac 1{c^2}\, \epsilon_{ijk}J_k }, &   
\end{array}
\label{qb}
\ee
where   from now on the indices $i,j,k=1,2,3$ and sum over repeated indices will be understood.
 Recall also that the commutators (\ref{qa}) are a consequence of 3-space isotropy~\cite{BLL} and they are shared by any Lie algebra in $\kk$, while the Lie brackets (\ref{qb}) distinguish the specific kinematical algebra according to the values of the real parameters $\ka$, $c$ and $\la$.

The family $\kk$ has  two   Casimir operators:  a quadratic  one, coming from the Killing--Cartan form, which is given by 
\be
{\cal C}
= \frac 1{c^2}  P_0^2-\la \>P^2 +\ka \left(   \>K^2-\la \frac 1{c^2}\, \>J^2\right) ,
\label{cas}
\ee
 and a fourth-order Casimir~\cite{casimirs} 
 (with the exception of the static algebra~\cite{BLL} corresponding to set $\ka=\la=0$ and $c\to\infty$ in (\ref{qb}), so all of these brackets vanish).

Moreover, $\kk$ is  endowed with the parity $\mathcal {P}$ and the time-reversal $\mathcal{T}$   involutive automorphisms  defined by~\cite{BLL}
    \be
\begin{array}{l}
 \mathcal{P}( P_0,\>P,\>K,\>J)=  (P_0,-\>P,-\>K,\>J),\\[4pt]
 \mathcal{T}( P_0,\>P,\>K,\>J)=  (-P_0,\>P,-\>K,\>J),\\[4pt]
\!\!\!\!\!  \mathcal{PT}( P_0,\>P,\>K,\>J)=  (-P_0,-\>P,\>K,\>J).
\label{qf}
   \end{array}
\ee
Each of them provides a type of contraction:  the composition $ \mathcal{PT}$ corresponds to the (flat) spacetime contraction $(\ka\to 0)$, the parity  $\mathcal{P}$ to the speed-space contraction $(c\to \infty)$, and the time-reversal $ \mathcal{T}$ to the speed-time contraction  $(\la\to 0)$ (see~\cite{CK3d} for the (2+1)D kinematical algebras and contractions within the CK family   $\mathfrak{so}_\k(4)$ and their Drinfel'd--Jimbo quantum deformation). In other words, the quantities $\ka$, $1/c^2$ and $\la$ behave as graded contraction parameters, each of them corresponding to the ${\mathbb {Z}}_2$-grading of  $\kk$ determined by 
$ \mathcal{PT}$, $\mathcal{P}$ and $ \mathcal{T}$, respectively.

From the Lie group $\KK$ of  $\kk$ we   construct the  (3+1)D spacetime and the 6D space of  lines as the coset spaces
\be
\begin{array}{ll}
 \>{ST}^{3+1}=  \KK/ \Hst, \quad& \Hst=\langle \>K,\>J \rangle , \\[2pt]
 {\>L}^{6}=  \KK/ \Hline,\quad &  \Hline= \langle P_0\rangle \otimes \langle\>J \rangle = \langle P_0\rangle \otimes \, {\rm SO}(3) ,
 \end{array}
 \label{qqf}
\ee
such that $\Hst$ and $\Hline$   are the isotropy subgroups of an event and  a  line, respectively.
Thus these   are symmetric homogeneous spaces associated, in this order, with the composition  $ \mathcal{PT}$ and parity $\mathcal{P}$ involutions.  Similarly to the discussion  on the curvature of the CK spaces (\ref{space1})--(\ref{space4}) in  Section~\ref{s31}, we remark that the   (3+1)D spacetime $\>{S
 T}^{3+1}$ 
 is a rank-one homogeneous space such that  all their sectional curvatures $K$ are equal and constant. However, the 
  6D space of  lines $\>{L}^{6}$ is of rank-two  and only the
sectional curvatures $K$  of   any 2-plane direction spanned by any two tangent vectors
 $(P_i, P_j)$,  $(K_i, K_j)$  and $(P_i,K_i)$ $(i,j=1,2,3)$ are equal among themselves and constant, being the remaining ones, $(P_i,K_j)$ with $i\ne j$,  generically non-constant (or zero).

 Furthermore, when $\KK$ is a non-simple Lie group, the metric on either    space~\eqref{qqf} could be degenerate and, in this case, an   invariant foliation  arises so that an additional metric  defined  on each leaf of the foliation is  necessary to  determine completely the metric structure of the space~\cite{HS1997phasespaces}. Moreover, it is important to take into account that, in principle, the (3+1)D spacetime $\>{S
 T}^{3+1}$ does not necessarily    coincides  with the CK space of points $ \mathbb S_\k^{(1)}$  (\ref{space1}) (in most cases it does), and likewise with the
 6D space of  lines  $\>{L}^{6}$ with respect to the   CK space of lines $ \mathbb S_\k^{(2)}$  \eqref{space2}. Nevertheless,  they can always be identified with another CK   space, as for instance  \eqref{other} for the space of lines. 
    This fact will depend on the kinematical assignation 
 of the geometrical  CK generators that we shall study next in Section~\ref{s5}.

 In what follows we describe the 11 kinematical algebras although we shall only focus on the homogeneous spaces  (\ref{qqf}) for   
 nine of them: the Lorentzian, Newtonian and Carrollian cases. Additionally, we shall    show how  the three classical Riemannian algebras (and their homogenous spaces) can also be recovered from the  family  $\kk$. These nine kinematical algebras/spaces plus the three 
 Riemannian ones are those which will appear in Section~\ref{s5} and  they  are  summarized in Table~\ref{table4}.


\subsection{Lorentzian  Algebras}
\label{s41}

If we set the parameter $\la=1$ and consider $c$ finite, we find that    $\kk$ covers the three Lorentzian algebras $\lo_\ka$ of relativistic (3+1)D  spacetimes such that the Lie brackets \eqref{qb} now read 
\be
 \begin{array}{lll}
\displaystyle{
  [K_i,P_0]=  P_i  } , &\qquad \displaystyle {[K_i,P_j]=\frac 1{c^2}\, \delta_{ij}  P_0} ,    &\qquad \displaystyle{[K_i,K_j]=-  \frac 1{c^2}\,\epsilon_{ijk} J_k} , 
\\[4pt][P_0,P_i]=-\ka    K_i , &\qquad \displaystyle {  [P_i,P_j]=\ka  \frac 1{c^2}\, \epsilon_{ijk}J_k }, &   
\end{array}
\label{ra}
\ee
where  $c$ is the speed of light and $\ka$ is the cosmological constant. Then we obtain  the  de Sitter (dS)  $\lo_+= \mathfrak{so}(4,1)$,  anti-de Sitter  (AdS)  $\lo_-=\mathfrak{so}(3,2)$ and Poincar\'e  $\lo_0= \mathfrak{iso}(3,1)$ algebras.  The quadratic Casimir (\ref{cas}) for  $\lo_\ka$ reads
 \be
{\cal C}
= \frac 1{c^2}  P_0^2- \>P^2 +\ka \left(   \>K^2- \frac 1{c^2}\, \>J^2\right) ,
\label{rb}
\ee
  and the fourth-order Casimir, related to the Pauli--Lubanski 4-vector, can be found in~\cite{casimirs}.

     The Lorentz subalgebra corresponds to $\mathfrak{so}(3,1)=\langle \>K,\>J\rangle$ which is the Lie algebra of the isotropy subgroup $\Hst = {\rm SO}(3,1)$ (\ref{qqf}).    The constant sectional  curvature of the (3+1)D spacetime is $K=-\ka$.   Notice that the cosmological constant can be expressed in terms of a  {\em time} universe radius $\tau$ through  $\ka=\pm 1/\tau^2$, so that to 
  take   $\ka=0$ corresponds to the limit $\tau\to \infty$, which is just the spacetime  contraction providing the flat Minkowskian spacetime $\>{M}^{3+1}$ from the (3+1)D (A)dS spacetimes. 
  The isotropy subgroup of a line is $\Hline={\rm SO}_{-\ka}(2)\otimes {\rm SO}(3)$ and the homogeneous space of (time-like) lines (\ref{qqf})    is, in the three cases, of negative constant sectional curvature $K=-1/c^2$~\cite{HS1997phasespaces}. Recall that the notation ${\rm SO}_{-\ka}(2)$ means that 
  ${\rm SO}_{+}(2)= {\rm SO}(2)$,  ${\rm SO}_{-}(2)= {\rm SO}(1,1)$ and ${\rm SO}_{0}(2)= \mathbb R$.


\begin{landscape}

 \begin{table}[htp] 
{\small
\caption{\small Kinematical algebras with commutation relations (\ref{qa}) and (\ref{qb}) together with their corresponding symmetrical homogeneous (3+1)D spacetimes and 6D spaces of lines  (\ref{qqf})  of sectional curvature $K$ according to the values of the graded contraction parameters $(\ka,c,\la)$. The same results for the three Riemannian cases are similarly displayed.}
\label{table4}
  \begin{center}
\begin{tabular}{l l l}
\hline  

\hline

\\[-0.20cm]
\multicolumn{3}{c}{\textbf{Lorentzian  algebras and homogeneous spaces}} \\[0.2cm]
\hline
\\[-0.2cm] 
 $\bullet$  De Sitter&  $\bullet$ Poincar\'e &  $\bullet$  Anti-de Sitter  \\[0.2cm]
$\lo_+=\mathfrak{so}(4,1)$: $\ka>0$, $c$ finite, $\la=1$ & $\lo_0=\mathfrak{iso}(3,1)$: $\ka=0$, $c$ finite, $\la=1$ & $\lo_-=\mathfrak{so}(3,2)$: $\ka<0$, $c$ finite, $\la=1$  \\[0.2cm]
  $\>{dS}^{3+1}={\rm SO}(4,1)/{\rm SO}(3,1)$,\ $K=-\ka<0$ &   $\>{M}^{3+1}={\rm ISO}(3,1)/{\rm SO}(3,1)$,\ $K=0$ &  $\>{AdS}^{3+1}={\rm SO}(3,2)/{\rm SO}(3,1)$,\ $K=-\ka>0$   \\[0.2cm]
  $\>{LdS}^{6}={\rm SO}(4,1)/\bigr( {\rm SO}(1,1)\otimes {\rm SO}(3) \bigl)$,\ $K=-\frac 1{c^2}$ &   $\>{LM}^{6}={\rm ISO}(3,1)/\bigr( \mathbb R\otimes {\rm SO}(3) \bigl)$,\ $K=-\frac 1{c^2}$&  $\>{LAdS}^{6}={\rm SO}(3,2)/\bigr( {\rm SO}(2)\otimes {\rm SO}(3) \bigl)$,\ $K=-\frac 1{c^2}$  \\[0.2cm]
\hline
\\[-0.13cm]
\multicolumn{3}{c}{\textbf{Newtonian  algebras and homogeneous spaces}} \\[0.2cm]
\hline
\\[-0.2cm] 
 $\bullet$ Expanding  Newton--Hooke&  $\bullet$ Galilei &  $\bullet$ Oscillating Newton--Hooke \\[0.2cm]
$\nh_+= \mathfrak{i}_6\bigr(  \mathfrak{so}(1,1) \oplus   \mathfrak{so}(3) \bigl)$:  $\ka>0$, $c=\infty$, $\la=1$& $\nh_0=\mathfrak{iiso}(3)$: $\ka=0$, $c=\infty$, $\la=1$& $\nh_-= \mathfrak{i}_6\bigr(  \mathfrak{so}(2) \oplus   \mathfrak{so}(3) \bigl)$:  $\ka<0$, $c=\infty$, $\la=1$  \\[0.2cm]
  $\>{N}_+^{3+1}={\rm N}_+/{\rm ISO}(3)$,\ $K=-\ka<0$ &   $\>{N}_0^{3+1}\equiv \>{G}^{3+1}={\rm IISO}(3)/{\rm ISO}(3)$,\ $K=0$ &  $\>{N}_-^{3+1}={\rm N}_-/{\rm ISO}(3)$,\ $K=-\ka>0$   \\[0.2cm]
  $\>{LN}_+^{6}={\rm N}_+/\bigr( {\rm SO}(1,1)\otimes {\rm SO}(3) \bigl)$,\ $K=0$ &   $\>{LG}^{6}={\rm IISO}(3)/\bigr( \mathbb R\otimes {\rm SO}(3) \bigl)$,\ $K=0$&  $\>{LN}_-^{6}={\rm N}_-/\bigr( {\rm SO}(2)\otimes {\rm SO}(3) \bigl)$,\ $K=0$  \\[0.2cm]
\hline
\\[-0.13cm]
\multicolumn{3}{c}{\textbf{Carrollian  algebras and homogeneous spaces}} \\[0.2cm]
\hline
\\[-0.2cm] 
 $\bullet$ Para-Euclidean&  $\bullet$  Carroll&  $\bullet$ Para-Poincar\'e \\[0.2cm]
$\ca_+= \mathfrak{i'so}(4)$: $\ka>0$, $c=1$, $\la=0$ & $\ca_0= \mathfrak{ii'so}(3) $: $\ka=0$, $c=1$, $\la=0$ & $\ca_- = \mathfrak{i'so}(3,1) $: $\ka<0$, $c=1$, $\la=0$  \\[0.2cm]
 $\>{C}_+^{3+1}={\rm I'SO(4)}/{\rm ISO}(3)$,\ $K=\ka>0$ &   $\>{C}_0^{3+1}\equiv \>{C}^{3+1}={\rm II'SO}(3)/{\rm ISO}(3)$,\ $K=0$ &  $\>{C}_-^{3+1}={\rm I'SO(3,1)}/{\rm ISO}(3)$,\ $K=\ka<0$   \\[0.2cm]
$\>{LC}_+^{6}={\rm I'SO(4)}/\bigr( \mathbb R\otimes {\rm SO}(3) \bigl)$,\     $K=\ka>0$ &   $\>{LC}^{6}={\rm II'SO}(3)/\bigr( \mathbb R\otimes {\rm SO}(3) \bigl)$,\ $ K=0 $ & $\>{LC}_-^{6}={\rm I'SO(3,1)}/\bigr( \mathbb R\otimes {\rm SO}(3) \bigl)$,\ $K=\ka<0$ \\[0.2cm]
\hline
\\[-0.13cm]
\multicolumn{3}{c}{\textbf{Riemannian  algebras and homogeneous spaces} }\\[0.2cm]
\hline
\\[-0.2cm] 
 $\bullet$ Hyperbolic &  $\bullet$ Euclidean &  $\bullet$ Spherical \\[0.2cm]
$\mathfrak{so}(4,1)$: $\ka>0$, $c={\rm i}$, $\la=1$ or& $\mathfrak{iso}(4)$: $\ka=0$, $c={\rm i}$, $\la=1$ or&$\mathfrak{so}(5)$: $\ka<0$, $c={\rm i}$, $\la=1$  or\\[0.2cm]
 \phantom{$\mathfrak{so}(4,1)$:} $\ka<0$, $c=1$, $\la=-1$  &  \phantom{$\mathfrak{iso}(4)$:} $\ka=0$, $c=1$, $\la=-1$  &\phantom{$\mathfrak{so}(5)$:}  $\ka>0$, $c=1$, $\la=-1$   \\[0.2cm]
  $\>{H}^{4}={\rm SO}(4,1)/{\rm SO}(4)$,\ $K <0$ &   $\>{E}^{4}={\rm ISO}(4)/{\rm SO}(4)$,\ $K=0$ &  $\>{S}^{4}={\rm SO}(5)/{\rm SO}(4)$,\ $K>0$   \\[0.2cm]
  $\>{LH}^{6}={\rm SO}(4,1)/\bigr( {\rm SO}(1,1)\otimes {\rm SO}(3) \bigl)$,\ $K=+1$ &   $\>{LE}^{6}={\rm ISO}(4)/\bigr( \mathbb R\otimes {\rm SO}(3) \bigl)$,\ $K=+1$&  $\>{LS}^{6}={\rm SO}(5)/\bigr( {\rm SO}(2)\otimes {\rm SO}(3) \bigl)$,\ $K=+1$   \\[6pt]
\hline  

\hline

\end{tabular}
 \end{center}
}
 
 \end{table}

  \end{landscape}



\subsection{Newtonian  Algebras}
\label{s42}
   
   The non-relativistic limit $c\to \infty$  (or speed-space contraction) of  $\lo_\ka$ (\ref{ra}) gives rise to three Newtonian algebras $\nh_\ka$ with Lie brackets
   \be
 \begin{array}{lll}
\displaystyle{
  [K_i,P_0]=  P_i  } , &\qquad \displaystyle {[K_i,P_j]= 0} ,    &\qquad \displaystyle{[K_i,K_j]= 0} , 
\\[4pt][P_0,P_i]=-\ka   K_i , &\qquad \displaystyle {  [P_i,P_j]=0 }, &   
\end{array}
\nonumber
\ee
where $\ka=\pm 1/\tau^2$   and $\tau$ is again the time universe radius. The second-order Casimir \eqref{rb} reduces to 
 \be
{\cal C}
=  - \>P^2 +\ka   \>K^2 
\nonumber
\ee
  and the corresponding fourth-order Casimir can be found in~\cite{casimirs}. This non-relativistic limit  is obtained by setting $\la=1$ and $c\to\infty$ in (\ref{qb}) and (\ref{cas}).  
    In this way we find the  expanding Newton--Hooke (NH) $\nh_+$,  oscillating NH  $\nh_-$  and the 
 Galilei $\nh_0\equiv \mathfrak{iiso}(3)$ algebras,   which have the following structure (see (\ref{ic}) and \eqref{ib}, respectively):
  \be
\begin{array}{llll}
\nh_+\equiv  \mathfrak{i}_6\bigr(  \mathfrak{so}(1,1) \oplus   \mathfrak{so}(3) \bigl)\\[2pt] 
\quad\ \,  \equiv
 \mathbb{R}^6\dsum\bigr(  \mathfrak{so}(1,1) \oplus   \mathfrak{so}(3) \bigl):&
  \mathbb{R}^6=\langle \>P,\>K\rangle ,&  \mathfrak{so}(1,1)=\langle   P_0\rangle,&  \mathfrak{so}(3)= \langle  \>J \rangle.\\[2pt] 
   \nh_-\equiv  \mathfrak{i}_6\bigr(  \mathfrak{so}(2) \oplus   \mathfrak{so}(3) \bigl)
\\[2pt] 
\quad\ \, \equiv \mathbb{R}^6\dsum\bigr(  \mathfrak{so}(2) \oplus   \mathfrak{so}(3) \bigl):&
  \mathbb{R}^6=\langle \>P,\>K\rangle ,& \mathfrak{so}(2)=\langle   P_0\rangle,& \mathfrak{so}(3)= \langle  \>J \rangle.
     \end{array}
\nonumber
\ee
    \be
\begin{array}{llll}
\nh_0\equiv    \mathfrak{iiso}(3) \equiv \mathbb{R}^4\dsum\bigr(    \mathbb{R}^3 \dsum  \mathfrak{so}(3) \bigl):&
   \mathbb{R}^4=\langle P_0, \>P\rangle ,&    \mathbb{R}^3=\langle  \>K \rangle,&  \mathfrak{so}(3)= \langle  \>J \rangle.
   \end{array}
 \nonumber
\ee
  
The isotropy subgroup $\Hst$  (\ref{qqf}) is now the 3D Euclidean subgroup ${\rm ISO}(3)$ spanned by rotations and (commuting) Newtonian boosts  and the 
(3+1)D spacetime has the same sectional  curvature as in the Lorentizan spacetimes:   $K=-\ka$. The metric is degenerate and corresponds to an ``absolute-time", so that  there exists an invariant foliation under the  action   of the  Newtonian group ${\rm N}_\ka$, whose leaves are defined by a constant time, which is determined by a  3D non-degenerate 
  Euclidean spatial metric  restricted   to each leaf of the foliation~\cite{HS1997phasespaces,SnyderG}.    The isotropy subgroup of a line is again $\Hline={\rm SO}_{-\ka}(2)\otimes {\rm SO}(3)$, but  the homogeneous space of  lines (\ref{qqf}) is flat, i.e.~$K=0$~\cite{HS1997phasespaces}.
 

 \subsection{Carrollian  Algebras}
\label{s43}

We set $\la=0$ and $c=1$ in the commutators (\ref{qb}) yielding the three Carrollian algebras $\ca_\ka$ with Lie brackets and second-order Casimir given by
 \be
 \begin{array}{lll}
\displaystyle{
  [K_i,P_0]=0  } , &\qquad \displaystyle {[K_i,P_j]=  \delta_{ij}  P_0} ,    &\qquad \displaystyle{[K_i,K_j]= 0} , 
\\[4pt][P_0,P_i]=-\ka    K_i , &\qquad \displaystyle {  [P_i,P_j]=\ka \epsilon_{ijk}J_k }, &   
\end{array}
\label{rf}
\ee
\be
{\cal C}
=  P_0^2 +\ka \>K^2 .
\nonumber
\ee
Notice that now the parameter $\ka$ has dimensions of ${\rm length}^{-2}$ instead of ${\rm time}^{-2}$  and the Carrollian boosts have dimensions of speed instead of ${\rm speed}^{-1}$ (which were the cases in    the Lorentizan and Newtonian algebras).

The algebra  $\ca_\ka$ comprises the   Para-Euclidean algebra $\ca_+\equiv\mathfrak{i'so}(4)$ (isomorphic to the Euclidean   $\mathfrak{iso}(4)$), the
Para-Poincar\'e algebra $\ca_-\equiv\mathfrak{i'so}(3,1)$ (isomorphic to the Poincar\'e   $\mathfrak{iso}(3,1)$) and the proper Carroll algebra 
$\ca_0\equiv\mathfrak{ii'so}(3)\equiv \mathfrak{i'iso}(3) $, which have the following structure~\cite{SnyderG} (see (\ref{iaa}), (\ref{ibb}) and (\ref{ibbc})):
  \be
\begin{array}{lll}
\ca_+\equiv \mathfrak{i'so}(4) =  \mathbb{R}'^4 \dsum  \mathfrak{so}(4)\!:&  
  \mathbb{R}'^4=\langle P_0,\>K\rangle ,&   \mathfrak{so}(4)=\langle   \>P,\>J\rangle .
\\[4pt] 
\ca_- \equiv \mathfrak{i'so}(3,1) =  \mathbb{R}'^4 \dsum  \mathfrak{so}(3,1)\!:&  
  \mathbb{R}'^4=\langle P_0,\>K\rangle ,&   \mathfrak{so}(3,1)=\langle   \>P,\>J\rangle .
\\[4pt] 
\ca_0\equiv \mathfrak{i'iso}(3) =  \mathbb{R}'^4 \dsum \bigr( \mathbb{R}^3 \dsum  \mathfrak{so}(3)  \bigl):&  
   \mathbb{R}'^4=\langle P_0, \>K\rangle ,&     \mathbb{R}^3=\langle  \>P \rangle, \quad\ \, \mathfrak{so}(3)= \langle  \>J \rangle,\\[4pt] 
\ca_0\equiv  \mathfrak{ii'so}(3) =  \mathbb{R}^4 \dsum \bigr( \mathbb{R}'^3 \dsum  \mathfrak{so}(3)  \bigl):&  
   \mathbb{R}^4=\langle P_0, \>P\rangle ,&     \mathbb{R}'^3=\langle  \>K \rangle, \quad  \mathfrak{so}(3)= \langle  \>J \rangle .
   \end{array}
\label{rh}\nonumber
\ee
  The isotropy subgroup $\Hst$  (\ref{qqf}) is again the 3D Euclidean subgroup ${\rm ISO}(3)$ spanned by rotations and (commuting) Carrollian boosts,   but
     the  (3+1)D spacetime has    sectional  curvature $K=+\ka$ (instead of $K=-\ka$   as in the Lorentizan and Newtonian spacetimes). The metric   is degenerate   corresponding  to an ``absolute-space" and there exists 
an invariant foliation under   the action of the Carrollian group  characterized by a  1D  time metric    restricted to each leaf of the foliation~\cite{SnyderG}.    The isotropy subgroup of a line is   $\Hline=\mathbb{R}\otimes {\rm SO}(3)$ and the homogeneous space of  lines (\ref{qqf})  has the same curvature as the   spacetimes so   equal to $+\ka$.
  

 \subsection{The Two Remaining Kinematical  Algebras}
\label{s44}

For the sake of completeness, we also mention that the Para-Galilei algebra~\cite{BLL} arises for $\la=0$ and $c\to \infty$, that is,  the commutators (\ref{qb}) reduce to
 \be
\begin{array}{lll}
\displaystyle{
  [K_i,P_0]=0 } , &\qquad \displaystyle {[K_i,P_j]= 0} ,    &\qquad \displaystyle{[K_i,K_j]=0} , 
\\[4pt][P_0,P_i]=-\ka   K_i , &\qquad \displaystyle {  [P_i,P_j]=0 }, &   
\end{array}
\nonumber
\ee
for any value of $\ka\ne 0$ (apply the map $P_0\to \pm P_0/\ka)$, while the second-order Casimir  \eqref{cas}  simply reads
${\cal C}=  \ka   \>K^2$.
And the   static algebra~\cite{BLL} corresponds to the most contracted algebra within the kinematical family for $\ka=\la=0$ and $c\to \infty$,
\be
\begin{array}{lll}
\displaystyle{
  [K_i,P_0]=0 } , &\qquad \displaystyle {[K_i,P_j]= 0} ,    &\qquad \displaystyle{[K_i,K_j]=0} , 
\\[4pt][P_0,P_i]=0 , &\qquad \displaystyle {  [P_i,P_j]=0 },&   
\end{array}
\label{frh}
\ee
with trivial  second-order Casimir  ${\cal C}=0$. In fact, the static algebra is the only kinematical one which does not appear within the CK family  $\mathfrak{so}_\k(5)$~\cite{CKMontigny},  but it can be obtained from the   general solution of the grading equations for $\mathfrak{so}(5)$~\cite{CKGraded,MontignyKinematical}.  Observe that the static algebra is not a quasisimple Lie algebra in the sense that
it does not have the same number of Casimir invariants as the simple Lie algebra  $\mathfrak{so}(5)$.

When one compares the commutation relations of the static algebra  (\ref{frh}) with those for the Carroll one  (\ref{rf}) with $\ka=0$, one finds that the Carroll algebra can be regarded as a  centrally extended algebra, with non-trivial central extension  $P_0$,  from the  static algebra, and there cannot be added any other central extension to the  Carroll algebra~\cite{BLL} (see~\cite{extensions} for the central extensions of the CK algebras in any dimension). In this respect, we remark that in~\cite{Figueroa-OFarrill2018} the (3+1)D kinematical algebras have been constructed  from the   static algebra through deformation theory  (see~\cite{Figueroa-OFarrill2018higher} for higher dimensions). We also recall that twist deformations for the Para-Galilei,  static  and Carroll  algebras have been obtained in~\cite{Daszkiewicz2019}. 

Neither the Para-Galilei  nor the static algebra  will   appear within the   deformations that  we  shall describe in     Section~\ref{s5} so that they are omitted in Table~\ref{table4}.

  
\subsection{Riemannian  Algebras}
\label{s45}

   Additionally, but not kinematically, we can set $\la=1$ and the speed of light equal to the imaginary unit $c={\rm i}$ in (\ref{qb}) finding the commutators
   \be
\begin{array}{lll}
\displaystyle{
  [K_i,P_0]= P_i  } , &\qquad \displaystyle {[K_i,P_j]=- \delta_{ij}  P_0} ,    &\qquad \displaystyle{[K_i,K_j]= \epsilon_{ijk} J_k} , 
\\[4pt][P_0,P_i]=-\ka    K_i , &\qquad \displaystyle {  [P_i,P_j]=-\ka    \epsilon_{ijk}J_k } ,&   
\end{array}
\label{ri}
\ee
with second-order Casimir  (\ref{cas}) given by
\be
{\cal C}
=- P_0^2- \>P^2 +\ka \left(   \>K^2+ \>J^2\right) .
\nonumber
\ee
  In this way, we obtain $\mathfrak{so}(5)$ for $\ka<0$, $\mathfrak{iso}(4)$ for $\ka=0$ and $\mathfrak{so}(4,1)$ for $\ka>0$. The generator $P_0$ now behaves as another space translation, while the generators $\>K$ are no longer boosts  but rotations. The isotropy subgroup $\Hst$ (\ref{qqf}) is       ${\rm SO}(4)=\langle \>K,\>J\rangle$, such that we recover  the three classical 4D Riemannian spaces of constant sectional curvature $K=-\Lambda$: spherical ($K>0$), Euclidean ($K=0$) and  hyperbolic ($K<0$) spaces. 
  The  isotropy subgroup of a line is   $\Hline={\rm SO}_{-\ka}(2)\otimes {\rm SO}(3)$   and the  corresponding 6D space of lines  has positive curvature $K=+1$ for any value of $\ka$~\cite{HS1997phasespaces}.
  
  Alternatively, we can set $\la=-1$ and $c=1$ in (\ref{qb}) obtaining that
     \be
\begin{array}{lll}
\displaystyle{
  [K_i,P_0]= -P_i  } , &\qquad \displaystyle {[K_i,P_j]=\delta_{ij}  P_0} ,    &\qquad \displaystyle{[K_i,K_j]= \epsilon_{ijk} J_k} , 
\\[4pt][P_0,P_i]=-\ka    K_i , &\qquad \displaystyle {  [P_i,P_j]=\ka   \epsilon_{ijk}J_k } , &   
\end{array}
\nonumber
\ee
\be
{\cal C}
=  P_0^2+\>P^2 +\ka \left(   \>K^2+ \>J^2\right) ,
\nonumber
\ee
which are equivalent to the Lie brackets (\ref{ri}) by means of the maps $P_0\to  -P_0$ and $\ka\to -\ka$.
  Therefore we again obtain  the same Riemannian  algebras (and homogeneous spaces), but now $\mathfrak{so}(5)$ for $\ka>0$, $\mathfrak{iso}(4)$ for $\ka=0$ and $\mathfrak{so}(4,1)$ for $\ka<0$.

 
\section{Kinematical Lie Bialgebras and Noncommutative Spaces}
\label{s5}

Our aim now is to   interpret in the kinematical framework the Lie bialgebras coming from  the classical CK $r$-matrix $r$ (\ref{nb}) and  the   Drinfel'd doubles further provided by $r_D$ (\ref{pa}) together with the corresponding first-order noncommutative spaces of points and lines displayed in Table~\ref{table2} and the twisted one (\ref{ncdd}). With this in mind, we shall perform different identifications between the ``geometrical" generators $J_{ab}$ of  $\mathfrak{so}_\k(5)$  \eqref{mf} and the kinematical ones of $\kk$ (\ref{qa}) and (\ref{qb}), which will convey physical correspondences between the contraction/curvature CK parameters $\k$ and $\Lambda$, $c$, $\lambda$.

According to~\cite{CK4d}, the main idea is to start with the main primitive generator $J_{04}$ and to identify it   either  with  a spatial translation $P_i$ or with  the time translation $P_0$.  Since the product $zJ_{04}$ must be dimensionless we shall obtain the so-called~\cite{CK3d}   
 ``space-like" deformations  $J_{04}\equiv P_i$, with the deformation parameter $z$ being a fundamental {\em length scale}, and the  ``time-like" deformations $J_{04}\equiv P_0$, with $z$ being a fundamental {\em time scale}. 
In particular,   we shall study firstly three classes of kinematical deformations, called A, B  and C, such that their properties are determined by the two primitive (undeformed) generators $(J_{04},J_{13})$ corresponding to  $(P_3,K_2)$, $(P_2,J_2)$ and  $(P_0,J_2)$, respectively.
We remark that in these three classes the   time translation generator
  $P_0\equiv J_{0l}$ for $l=1,2,4$, and the remaining case $P_0\equiv J_{03}$  would provide results which are equivalent, under certain Lie algebra automorphisms, to those    already contained in the class~A, so that we omit it. Therefore,  the classes A and B will   give rise to  space-like deformations, while the class C will    lead to  time-like ones. Additionally, we shall construct an AdS Lie bialgebra for which $z$ is   dimensionless with primitive generators $(J_2,P_0)$,   and it will correspond to the new Drinfel'd double of case (IV) in Table~\ref{table3}; we shall call it class D. We point out that such four classes of kinematical deformations will contain   the four Drinfel'd doubles for the simple Lie algebras of  Table~\ref{table3}.

  The main results that will be obtained along     Sections~\ref{s51}--\ref{s54} concerning the kinematical $r$-matrices are presented in Table~\ref{table5}. From them,  their  corresponding    first-order noncommutative spacetimes and spaces of lines will be computed being summarized  in Table~\ref{table6}. Comments on these results will be performed in Sections~\ref{s55} and~\ref{s56}.
  

\begin{table}[t!] 
{\small
\caption{\small Four classes of real classical $r$-matrices for the kinematical and Riemannian algebras with commutation relations (\ref{qa}) and (\ref{qb}). For each class, it is displayed the dimensions of the quantum deformation parameter $z$ and the primitive generators (determined by  $(J_{04},J_{13})$), the $r$-matrix, the specific Lie algebras according to the values of the graded contraction parameters $(\ka,c,\la)$ as in Table~\ref{table4}, the CK parameters $\k$, and the   Drinfel'd double $r$-matrix $r_D$   together with  the corresponding case    given in Table~\ref{table3}.}
\label{table5}
 \begin{center}
\begin{tabular}{l l l l l  l l l }
\hline  

\hline

\\[-0.2cm]
\multicolumn{1}{l}{  {\#}} & \multicolumn{1}{c}{  $  z $ } & 
\multicolumn{3}{l}{ {Kinematical real $ {r}$-matrices and  Lie algebras}}&$ $& \multicolumn{1}{l}{ {Drinfel'd double} $ {r_D}$} \\[0.2cm]
\hline
\\[-0.2cm] 
A&Length &\multicolumn{4}{l}{  $r =z\bigl(K_3\wedge P_0+J_1\wedge P_2- J_2\wedge P_1 +\sqrt{-\ka}\, J_3\wedge K_1 \bigr)$}  & $r_{D}=r  +zK_2\wedge P_3 $  \\[4pt]
   &$(P_3,K_2) $    &AdS &$\mathfrak{so}(3,2)$ &$\ka<0,\ c, \ \la=1$&  $\k=(+,-,+,+)$  & No  \\[4pt]
  &     &Poincar\'e &$\mathfrak{iso}(3,1)$& $\ka=0,\ c, \ \la=1$&  $\k=(0,-,+,+)$  & No \\[4pt]
 &     &Oscillating~NH &$\mathfrak{n}_-$ &$\ka<0,\ c=\infty ,\ \la=1$&  $\k=(+,0,+,+)$  &  No \\[4pt]
  &     &Galilei &$\mathfrak{iiso}(3)$& $\ka=0,\ c=\infty ,\ \la=1$&  $\k=(0,0,+,+)$  & No \\[4pt]
 &     &Spherical &$\mathfrak{so}(5)$ &$\ka<0,\ c={\rm i} ,\ \la=1$&  $\k=(+,+,+,+)$  &  (I) \\[4pt]
  &     &Euclidean& $\mathfrak{iso}(4)$ &$\ka=0,\ c={\rm i}, \ \la=1$&  $\k=(0,+,+,+)$  &  (Ia)  \\[6pt]
 \hline
\\[-0.2cm]
B&Length &\multicolumn{4}{l}{  $r =z\bigl(K_2\wedge P_0+J_3\wedge P_1- J_1\wedge P_3 +\sqrt{\ka} \, K_3\wedge K_1 \bigr)$}  &$r_{D}=r - z J_2\wedge P_2 $   \\[4pt]
   & $(P_2,J_2) $     &dS &$\mathfrak{so}(4,1)$ &$\ka>0,\ c=1, \ \la=1$&  $\k=(+,-,-,+)$  &   (II)   \\[4pt]
  &     &Poincar\'e &$\mathfrak{iso}(3,1)$ &$\ka=0,\ c=1, \ \la=1$&  $\k=(0,-,-,+)$  &   (IIa)   \\[6pt]
 \hline
\\[-0.2cm]
C&Time &\multicolumn{4}{l}{  $r =z\bigl( K_1\wedge P_1+K_2\wedge P_2 + K_3\wedge P_3+\sqrt{-\la\,\ka} \, J_1\wedge J_3   \bigr) $}  &  $r_{D}=r - z J_2\wedge P_0 $   \\[4pt]
   & $(P_0,J_2) $     &AdS &$\mathfrak{so}(3,2)$ &$\ka<0,\ c=1, \ \la=1$&  $\k=(-,+,+,-)$  &  (III)    \\[4pt]
  &     &Poincar\'e &$\mathfrak{iso}(3,1)$ &$\ka=0,\ c=1, \ \la=1$&  $\k=(0,+,+,-)$  & (IIIa)     \\[4pt]
   &     &Para-Euclidean &$\mathfrak{i'so}(4)$ &$\ka>0,\ c=1, \ \la=0$&  $\k=(+,+,+,0)$  &  (Ia$^\prime$)  \\[4pt]
    &     &Para-Poincar\'e &$\mathfrak{i'so}(3,1)$ &$\ka<0,\ c=1, \ \la=0$&  $\k=(-,+,+,0)$  &  (IIIa$^\prime$)   \\[4pt]
   &     &Carroll &$\mathfrak{ii'so}(3)$ &$\ka=0,\ c=1, \ \la=0$&  $\k=(0,+,+,0)$  & (Ib) $\equiv$ (IIIb)    \\[4pt]
 &     &Spherical &$\mathfrak{so}(5)$ &$\ka>0,\ c=1 ,\ \la=-1$&  $\k=(+,+,+,+)$  &  (I) \\[4pt]
  &     &Euclidean& $\mathfrak{iso}(4)$ &$\ka=0,\ c=1, \ \la=-1$&  $\k=(0,+,+,+)$  &  (Ia)   \\[6pt]
 \hline
\\[-0.2cm]
D&None &\multicolumn{4}{l}{  $r=z\bigl(K_1\wedge K_3+K_2\wedge P_2 + P_1\wedge P_3+  J_3\wedge J_1\bigr) $}  &  $r_{D}=r + z P_0 \wedge J_2$   \\[4pt]
   & $(J_2,P_0) $     &AdS &$\mathfrak{so}(3,2)$ &$\ka=-1,\ c=1, \ \la=1$&  $\k=(-,-,-,-)$  &  (IV)    \\[6pt]
\hline  

\hline
\end{tabular}
\end{center}
}
 \end{table}
 


\begin{table}[htp]
{ \small
\caption{ \small The   first-order noncommutative  spacetimes $\>{ST}^{3+1}_z=\langle  \hat x^0, \hat x^1, \hat x^2, \hat x^3\rangle$ and spaces of  lines $\>{L}^{6}_z=\langle     \hat x^1, \hat x^2, \hat x^3,   \hat \xi^1, \hat \xi^2, \hat \xi^3 \rangle$     for the   classes A, B and C of kinematical and Riemannian Lie bialgebras   shown in Table~\ref{table5} with the notation of Table~\ref{table4}; the case D has no associated noncommutative  space.    First-order twisted noncommutative spacetimes coming from  the Drinfel'd double structures and their contractions presented in Table~\ref{table3} are also written in terms of the twist deformation parameter $\vartheta$ such that $\vartheta\equiv z$ corresponds to the proper (or contracted)  Drinfel'd double.}
\label{table6}
 \begin{center}
\begin{tabular}{l l l l l}
\hline  

\hline

\\[-0.15cm]
\!\!\! \!\!\!  A&\multicolumn{4}{l}{\!\!\! \!\!\!   $\bullet$   Noncommutative   spacetimes   $\Lambda\le 0$, $\lambda=1$: $\>{AdS}^{3+1}_z$, $\>{M}^{3+1}_z$, $\>{N}^{3+1}_{-,z}$, $\>{G}^{3+1}_z$, $\>{S}^{4}_z$\,, $\>{E}^{4}_z$ }\\[0.25cm]
&\quad  $ [\hat x^{0},\hat x^{3}] = z \hat x^{0} $&$  
[\hat x^{1},\hat x^{3}] = z \hat x^{1} $&$ [\hat x^{2},\hat x^{3}] = z \hat x^{2} $&$\quad
[\hat x^{0},\hat x^{1}] =[\hat x^{0},\hat x^{2}] = [\hat x^{1},\hat x^{2}] = 0$   \\[9pt]
 &\multicolumn{4}{l}{ \!\!\! \!\!\!  $\bullet$   Twisted noncommutative   spaces of points    $\Lambda\le 0$, $\lambda=1$, $c={\rm i}$:  $\>{S}^{4}_{z,\vartheta}$\,, $\>{E}^{4}_{z,\vartheta}$ }\\[0.25cm]
&   $ \quad   [\hat x^{0},\hat x^{3}] = z \hat x^{0} +\vartheta  \hat x^{2}  $&$ 
[\hat x^{1},\hat x^{3}] = z \hat x^{1}  $&$  [\hat x^{2},\hat x^{3}] = z \hat x^{2}-  \vartheta   \hat x^{0}   $&$   \quad
[\hat x^{0},\hat x^{1}] =[\hat x^{0},\hat x^{2}] = [\hat x^{1},\hat x^{2}] = 0$    \\[9pt]
& \multicolumn{4}{l}{\!\!\! \!\!\!   $\bullet$   Noncommutative   space of lines     $\Lambda< 0$, $\lambda=1$:    $\>{LAdS}^{6}_z$\,,  $\>{LN}^{6}_{-,z}$\,,  $\>{LS}^{6}_z$ }\\[0.25cm]
&\quad$ [\hat x^{1},\hat x^{2}] =0$& $ [\hat x^{1},\hat x^{3}] =z \hat x^{1} $&  $ [\hat x^{2},\hat x^{3}] =z \hat x^{2}$&\\[0.2cm]
&\quad$ [{\hat \xi}^{1},\hat \xi^{2}] =z\sqrt{-\Lambda}\, \hat \xi^{1}$& $ [\hat \xi^{1},\hat \xi^{3}] =- z \Lambda\hat x^{1} $&  $ [\hat \xi^{2},\hat \xi^{3}] =-z \Lambda \hat x^{2}$ &\\[0.2cm]
&\multicolumn{4}{l}{\quad$ [\hat x^{1}, \hat\xi^{1}] =z (\sqrt{-\Lambda}\, \hat x^{2} - \hat\xi^{3} )\qquad\quad  [\hat x^{2}, \hat\xi^{1}] =  -z\sqrt{-\Lambda}\,  \hat x^{1} \qquad\quad [\hat x^{3}, \hat\xi^{1}] = 0$}\\[0.2cm]
&\quad$ [\hat x^{1}, \hat \xi^{2}] = 0  $& $[\hat x^{2}, \hat \xi^{2}] =-z  \hat \xi^{3} $&  $[\hat x^{3}, \hat \xi^{2}] = 0$&\\[0.2cm]
&\quad$[\hat x^{1}, \hat \xi^{3}] = 0   $& $[\hat x^{2}, \hat \xi^{3}] = 0 $&  $[\hat x^{3}, \hat \xi^{3}] =-z \hat \xi^{3} $  & \\[9pt]
& \multicolumn{4}{l}{\!\!\! \!\!\!   $\bullet$   Noncommutative   space of lines     $\Lambda= 0$, $\lambda=1$:    $\>{LM}^{6}_z$\,,  $\>{LG}^{6}_{z}$\,,  $\>{LE}^{6}_z$ }\\[0.25cm]
&\quad$ [\hat x^{1},\hat x^{2}] =0$& $ [\hat x^{1},\hat x^{3}] =z \hat x^{1} $&  $ [\hat x^{2},\hat x^{3}] =z \hat x^{2}$&\\[0.2cm]
&\quad$ [{\hat \xi}^{i},\hat \xi^{j}]  =0$& $ [\hat x^{i},\hat \xi^{j}] =- z \delta_{ij}  \hat \xi^{3} $&  $ $  &  \\[8pt]
\hline
\\[-0.15cm]
\!\!\! \!\!\!   B&\multicolumn{4}{l}{\!\!\! \!\!\!  $\bullet$   Noncommutative   spacetimes     $\Lambda\ge  0$,    $c=1$, $ \la=1$: $\>{dS}^{3+1}_z$, $\>{M}^{3+1}_z$  }\\[0.25cm]
&\quad  $ [\hat x^{0},\hat x^{2}] = z \hat x^{0} $&$  
[\hat x^{1},\hat x^{2}] = z \hat x^{1} $&$ [\hat x^{3},\hat x^{2}] = z \hat x^{3}$&$  \quad
[\hat x^{0},\hat x^{1}] =[\hat x^{0},\hat x^{3}] = [\hat x^{1},\hat x^{3}] = 0$  \\[8pt]
 &\multicolumn{4}{l}{\!\!\! \!\!\!  $\bullet$   Twisted noncommutative   spacetimes     $\Lambda\ge  0$,   $c=1$, $ \la=1$:   $\>{dS}^{3+1}_{z,\vartheta}$, $\>{M}^{3+1}_{z,\vartheta}$  }
\\[0.25cm]
&   \quad $    [\hat x^{0},\hat x^{2}] = z \hat x^{0} $ &$ 
[\hat x^{1},\hat x^{2}] = z \hat x^{1}+\vartheta  \hat x^{3}   $ &$  [\hat x^{3},\hat x^{2}] = z \hat x^{3}-  \vartheta   \hat x^{1}  $ &$ \quad
[\hat x^{0},\hat x^{1}] =[\hat x^{0},\hat x^{3}] = [\hat x^{1},\hat x^{3}] = 0$    \\[8pt]
& \multicolumn{4}{l}{\!\!\! \!\!\!   $\bullet$   Noncommutative   space of lines      $\Lambda= 0$, $c=1$, $ \la=1$:    $\>{LM}^{6}_z$ }\\[0.25cm]
&\quad$ [\hat x^{1},\hat x^{2}] =z \hat x^{1} $& $ [\hat x^{1},\hat x^{3}] =0$&  $ [\hat x^{3},\hat x^{2}] =z \hat x^{3}$&\\[0.2cm]
&\quad$ [{\hat \xi}^{i},\hat \xi^{j}]  =0$& $ [\hat x^{i},\hat \xi^{j}] =- z \delta_{ij}  \hat \xi^{2} $&  $ $  &  \\[8pt]
\hline
\\[-0.15cm]
\!\!\! \!\!\!   C&\multicolumn{4}{l}{\!\!\! \!\!\!  $\bullet$   Noncommutative   spacetimes     $-\la \ka\ge  0$, $c=1$: $\>{AdS}^{3+1}_z$, $\>{M}^{3+1}_z$, $\>{C}^{3+1}_{+,z}$, $\>{C}^{3+1}_{-,z}$,  $\>{C}^{3+1}_z$, $\>{S}^{4}_z$\,, $\>{E}^{4}_z$  }\\[0.25cm]
&\quad  $ [\hat x^{1},\hat x^{0}] = z \hat x^{1} $&$  
[\hat x^{2},\hat x^{0}] = z \hat x^{2} $&$ [\hat x^{3},\hat x^{0}] = z \hat x^{3} $&$ \quad
[\hat x^{1},\hat x^{2}] =[\hat x^{1},\hat x^{3}] = [\hat x^{2},\hat x^{3}] = 0$    \\[8pt]
 &\multicolumn{4}{l}{\!\!\! \!\!\!  $\bullet$   Twisted noncommutative   spacetimes     $-\la \ka\ge  0$,   $c=1$:   $\>{AdS}^{3+1}_{z,\vartheta}$, $\>{M}^{3+1}_{z,\vartheta}$, $\>{C}^{3+1}_{+,{z,\vartheta}}$, $\>{C}^{3+1}_{-,{z,\vartheta}}$,  $\>{C}^{3+1}_{z,\vartheta}$, $\>{S}^{4}_{z,\vartheta}$, $\>{E}^{4}_{z,\vartheta}$  }
\\[0.25cm]
& \quad  $    [\hat x^{1},\hat x^{0}] = z \hat x^{1} +\vartheta  \hat x^{3}   $&$ 
[\hat x^{2},\hat x^{0}] = z \hat x^{2}  $&$  [\hat x^{3},\hat x^{0}] = z \hat x^{3}-  \vartheta   \hat x^{1}   $&$ \quad
[\hat x^{0},\hat x^{1}] =[\hat x^{0},\hat x^{3}] = [\hat x^{1},\hat x^{3}] = 0$   \\[8pt]
& \multicolumn{4}{l}{\!\!\! \!\!\!  $\bullet$   Noncommutative   space of lines      $-\la \ka\ge  0$, $c=1$:     $\>{LAdS}^{6}_z$\,, $\>{LM}^{6}_z$\,, $\>{LC}^{6}_{+,z}$\,, $\>{LC}^{6}_{-,z}$\,,  $\>{LC}^{6}_z$\,, $\>{LS}^{6}_z$\,, $\>{LE}^{6}_z$ }\\[0.25cm]
&\quad$ [\hat x^{i},\hat x^{j}] =0$& $ [\hat \xi^{i},\hat \xi^{j}] =0 $&  $ [\hat x^{i},\hat \xi^{j}] = 0$ \\[8pt]
\hline  

\hline

\end{tabular}
\end{center}
}
 \end{table}


 
\subsection{Class A: Space-like Deformations with Primitive Generators $(P_3,K_2)$}
\label{s51}

We consider the following kinematical assignation~\cite{CK4d}
\be
P_0=J_{01}, \quad\ \  \>P=(J_{02},J_{03},J_{04}),\quad\  \ \>K=(J_{12},J_{13},J_{14}),\quad\  \  \>J=(J_{34},-J_{24},J_{23}),
\label{wa}
\ee
which in the  array form used in Section~\ref{s21}  gives
\be
\noindent
\begin{tabular}{cccc}
$J_{01} $&$ J_{02} $& $  J_{03} $&$  J_{04} $ \\ 
  &$  J_{12} $& $  J_{13} $&$  J_{14} $ \\ 
 & & $  J_{23} $&$  J_{24} $ \\ 
&  & &$  J_{34} $ 
\end{tabular}
\quad  \equiv  \quad 
\begin{tabular}{cccc}
$P_0 $&$ P_1 $& $  P_2 $&$  P_3 $ \\ 
  &$  K_1 $& $  K_2 $&$  K_3 $ \\ 
 & & $  J_{3} $&$  -J_{2} $ \\ 
&  & &$  J_{1} $ 
\end{tabular}
\label{wb}
\ee
which shows that  the spaces  $ \mathbb S_\k^{(1)}$  (\ref{space1}) and  $ \mathbb S_\k^{(2)}$  (\ref{space2}) coincide with the spacetime $\>{ST}^{3+1}$ and the space of (time-like) lines 
$\>{L}^{6}$ (\ref{qqf}), respectively (see  (\ref{dia}) and  (\ref{diab})).
 
When we impose, under the identification (\ref{wa}), that the commutation rules of $\mathfrak{so}_\k(5)$  (\ref{mf}) fulfil the common Lie brackets of any kinematical algebra (\ref{qa}) we find that $\k_3=\k_4=+1$. Next the remaining specific kinematical commutation relations  (\ref{qb}) imply that $\k_1=-\ka$, $\k_2=-1/c^2$ and $\la=1$. In this way we find  a set of six kinematical algebras in $\mathfrak{so}_\k(5)$ with graded contraction parameters
\be
(\k_1,\k_2,\k_3,\k_4)=\bigl(-\ka,-1/c^2,+1,+1 \bigr),\quad\  \la=1 ,
\label{wc}
\ee
and     bilinear form $\bfI_\k$ (\ref{mi})  given by
\be
\bfI_\k=\biggl(     +1,-\ka,\frac{\ka}{c^2},\frac{\ka}{c^2},\frac{\ka}{c^2} \biggr).
\nonumber
\ee
These are the three Lorentzian  ($c$ finite) and the three Newtonian ($c=\infty$)  algebras described in Sections~\ref{s41} and~\ref{s42}. 
Moreover, the three Riemannian algebras of Section~\ref{s45} also appear for  $c=\rm{i}$, that is, $\k_2=+1$. 
Thus this class A cover nine of the Lie algebras shown in Table~\ref{table4}. Recall that in the Lorentzian cases the sectional curvature of the (3+1)D spacetime $\>{ST}^{3+1}$ is minus the cosmological constant $\k_1=-\ka$, while the 6D space of (time-like) lines $\>{L}^{6}$ is of negative curvature $\k_2=-1/c^2$. In the Newtonian cases,   $\>{L}^{6}$ is a flat space with $\k_2=0$ $(c=\infty)$. 
The kinematical automorphisms (\ref{qf}) are related to the CK ones $\Theta^{(m)}$ (\ref{inv}) through
\be
 \mathcal{P} = \Theta^{(2)} ,\qquad  \mathcal{T}=\Theta^{(1)}\Theta^{(2)} ,\qquad \mathcal{PT}=\Theta^{(1)} .
\nonumber
\ee
 
Now we apply the geometrical-kinematical identification (\ref{wb}) to the       CK $r$-matrix $r$  \eqref{nb} and to the Drinfel'd double one $r_D$  (\ref{pa})  obtaining the following  kinematical $r$-matrices
\be
r=z\bigl(K_3\wedge P_0+J_1\wedge P_2- J_2\wedge P_1 +\sqrt{-\ka}\, J_3\wedge K_1 \bigr) ,\quad\   r_D=r +z\sqrt{-c^2} \, K_2\wedge P_3 ,
\label{wf}
\ee
with primitive generators $P_3\equiv J_{04}$ and $K_2\equiv J_{13}$,  so  that $z$ has dimensions of a {\em length} with dimensionless product $zP_3$. Next, the constraint (\ref{constraint}) $\k_1\k_4=-\ka \ge 0$ excludes three cases in order to deal with real bialgebras:  dS, expanding NH and hyperbolic algebras, all of them with $\ka>0$ (first column in Table~\ref{table4}). And 
the Drinfel'd double $r$-matrix $r_D$ subjected to the additional condition  (\ref{constraints})    $\k_2\k_3=-1/c^2>0$ only holds for $c={\rm i}$, that is, for the spherical and Euclidean algebras  so recovering the cases (I) and (Ia) in Table~\ref{table3}. Thus these  results finally comprise {\em six} real Lie bialgebras shown  in Table~\ref{table5}.
 
 The cocommutator $\delta$  for $r$ (\ref{wf}) can then be deduced by applying  (\ref{tf}), or by   introducing directly the kinematical assignations (\ref{wb}) and (\ref{wc}) into  the CK cocommutator \eqref{cocok}. It can be checked that   the isotropy subalgebras of an event $\mathfrak{h}_{\rm st}$ and a line $\mathfrak{h}_{\rm line}$  given by (see (\ref{qqf}))
 \be
 \mathfrak{h}_{\rm st}=   \langle \>K,\>J \rangle ,\qquad 
 \mathfrak{h}_{\rm line}= \langle P_0\rangle \oplus \langle\>J \rangle ,
\label{wfx1}
 \ee
 both satisfy the coisotropy condition (\ref{coisotropy}).
 
 Now we proceed to obtain the  corresponding first-order noncommutative spacetimes $\>{ST}^{3+1}_z$ and spaces of (time-like) lines $\>{L}^{6}_z$ associated with the homogeneous spaces \eqref{qqf}. For this purpose,  we introduce the quantum coordinates $(\hat x^0,\hat x^i,\hat \xi^i ,\hat\theta^i)$ dual, in this order, to the generators $(P_0,P_i,K_i,J_i)$  $(i=1,2,3)$   via the  canonical pairing (\ref{tdd}), so with non-zero entries:
 \be
\langle  \hat x^0,P_0\rangle= \langle  \hat x^i,P_i\rangle= \langle  \hat \xi^i,K_i\rangle=\langle  \hat \theta^i,J_i\rangle=1 .
\label{wf1}
\ee
Hence  the first-order noncommutative spaces are defined as the annihilators of the vector subspaces $ \mathfrak{h}_{\rm st}$ and $ \mathfrak{h}_{\rm line}$ (\ref{wfx1}):
\be
\>{ST}^{3+1}_z=\langle  \hat x^0, \hat x^1, \hat x^2, \hat x^3\rangle,\qquad \>{L}^{6}_z=\langle     \hat x^1, \hat x^2, \hat x^3,   \hat \xi^1, \hat \xi^2, \hat \xi^3 \rangle .
\label{wf2}
\ee
Notice that for the Riemannian cases the  quantum time coordinate $\hat x^0$  corresponds to a spatial one, while the 
noncommutative rapidities $\hat \xi^i$ become  quantum angular coordinates. The corresponding defining commutation relations for these noncommutative spaces can be deduced either from the dual of the cocommutator 
$\delta$ for $r$ (\ref{wf}), or by introducing the following identification between the  quantum CK coordinates $\hat x^{ab}$ and the kinematical ones (\ref{wf1}) in    $ \mathbb S_{z,\k}^{(1)}$ and   $ \mathbb S_{z,\k}^{(2)}$  given in Table~\ref{table2}:
\be
\noindent
\begin{tabular}{cccc}
$\JJ^{01} $&$ \JJ^{02} $& $  \JJ^{03} $&$  \JJ^{04} $ \\[2pt] 
  &$  \JJ^{12} $& $  \JJ^{13} $&$  \JJ^{14} $ \\[2pt] 
 & & $  \JJ^{23} $&$  \JJ^{24} $ \\[2pt] 
&  & &$  \JJ^{34} $ 
\end{tabular}
\quad  \equiv  \quad 
\begin{tabular}{cccc}
$\hat x^0 $&$ \hat x^1 $& $  \hat x^2 $&$  \hat x^3 $ \\[2pt] 
  &$  \hat \xi^1 $& $   \hat \xi^2 $&$   \hat \xi^3 $ \\[2pt] 
 & & $   \hat \theta^{3} $&$  -\hat \theta^{2} $ \\[2pt] 
&  & &$  \hat \theta^{1} $ 
\end{tabular}
\nonumber
\ee
(which is the dual counterpart of (\ref{wb})), together with  (\ref{wc}).  Likewise, the  first-order twisted noncommutative   spaces of points  can be obtained from  $\mathbb S^{(1)}_{z,\vartheta,\k}$ (\ref{ncdd}) by taking into account that it only covers the spherical and Euclidean spaces with $c={\rm i}$ ($\k_2\k_3=+1$) so that     $\hat x^0 $ is another quantum spatial coordinate; recall that the proper Drinfel'd double structure corresponds to set $\vartheta\equiv z$  (\ref{pg}).  All of these noncommutative   spaces  are explicitly presented in Table~\ref{table6}.

 
\subsection{Class B: Space-like Deformations  with    Primitive Generators $(P_2,J_2)$}
\label{s52}

We perform the identification~\cite{CK4d}
\be
P_0=J_{02}, \quad\ \  \>P=(J_{01},J_{04},J_{03}),\quad\ \  \>K=(J_{12},J_{24},J_{23}),\quad\ \  \>J=(-J_{34},-J_{13},J_{14}),
\label{wg}
\ee
that is,
\be
\noindent
\begin{tabular}{cccc}
$J_{01} $&$ J_{02} $& $  J_{03} $&$  J_{04} $ \\ 
  &$  J_{12} $& $  J_{13} $&$  J_{14} $ \\ 
 & & $  J_{23} $&$  J_{24} $ \\ 
&  & &$  J_{34} $ 
\end{tabular}
\quad  \equiv  \quad 
\begin{tabular}{cccc}
$P_1 $&$ P_0 $& $  P_3 $&$  P_2 $ \\ 
  &$  K_1 $& $ - J_2 $&$  J_3 $ \\ 
 & & $ K_{3} $&$  K_{2} $ \\ 
&  & &$ - J_{1} $ 
\end{tabular}
\label{wh}
\ee 
 The fulfilment of the  kinematical commutators  (\ref{qa}) from the CK ones  (\ref{mf})   requires to  fix $\k_2=\k_3=-1$ and $\k_4=+1$. And   the  remaining commutation relations  (\ref{qb}) 
 lead to $\la=1$, $c=1$ and $\k_1=\ka$. Hence we obtain      the  three Lorentzian   algebras of  Section~\ref{s41}  within $\mathfrak{so}_\k(5)$ with  contraction parameters
\be
(\k_1,\k_2,\k_3,\k_4)=\bigl(\ka,-1 ,-1,+1 \bigr),\quad\  \la=1 , \quad \  c=1 ,
\label{wi}
\ee
and     bilinear form $\bfI_\k$ (\ref{mi})  given by
\be
\bfI_\k=\bigl(     +1,\ka, -\ka , \ka  , \ka \bigr).
\nonumber
\ee
In terms of $\Theta^{(m)}$ (\ref{inv}), the kinematical automorphisms (\ref{qf}) read
\be
 \mathcal{P} = \Theta^{(1)} \Theta^{(2)} \Theta^{(3)} ,\qquad  \mathcal{T}=\Theta^{(2)}\Theta^{(3)} ,\qquad \mathcal{PT}=\Theta^{(1)} .
\nonumber
\ee
  With the assignations (\ref{wg})  and (\ref{wi}) we find that  the space   $ \mathbb S_\k^{(1)}$  (\ref{space1}) is related   to the spacetime $\>{ST}^{3+1}$ (since $\mathcal{PT}=\Theta^{(1)}$), but the former has curvature $K=\k_1$, while the latter has $K=-\k_1 =-\ka$.  The space of (time-like) lines 
 $\>{L}^{6}$ (\ref{qqf}) cannot be    identified with   $ \mathbb S_\k^{(2)}$  (\ref{space2}) (now $\mathcal{P}\ne \Theta^{(2)}$), but it can be so  with the  rank-2 CK   space associated with the composition of involutions $ \mathcal{P} = \Theta^{(1)} \Theta^{(2)} \Theta^{(3)} $ (see the comments at the end of Section~\ref{s31}).

The     $r$-matrices (\ref{nb}) and   (\ref{pa})   turn out to be
\be
r=z\bigl( K_2\wedge P_0+J_3\wedge P_1- J_1\wedge P_3 +\sqrt{\ka} \, K_3\wedge K_1   \bigr) ,\qquad  r_D=r -z J_2\wedge P_2 .
\label{wl}
\ee
The primitive generators are $P_2\equiv J_{04}$ and $J_2\equiv -J_{13}$,  and      $z$ has dimensions of a {\em  length} since the product $zP_2$ is dimensionless; notice that (\ref{wl}) is written in units with $c=1$.
The constraint  $\k_1\k_4=\ka \ge 0$ excludes  the AdS algebra. Moreover, since   $\k_2\k_3=+1$, the $r$-matrix $r_D$ is well defined for the dS and Poincar\'e cases, which correspond to the cases (II) and (IIa)   in Table~\ref{table3}, as shown in Table~\ref{table5}.  The cocommutator for $r$ 
 (\ref{wl}) can be obtained straightforwardly showing that the coisotropy condition  (\ref{coisotropy}) is satisfied 
  for  $\mathfrak{h}_{\rm st}$ (\ref{wfx1}) in both cases,  but only for $\mathfrak{h}_{\rm line}$ 
  for the Poincar\'e bialgebra,    thus precluding the construction of the noncommutative  dS space of (time-like) lines $\>{L}^{6}_z$ (\ref{wf2}).

 We stress that to set $c=1$ in (\ref{wi}) implies that the $r$-matrices (\ref{wl}) are not well defined neither for    the non-relativistic algebras with $c\to\infty$, nor for the Riemannian ones with $c={\rm i}$ (remind that $\lambda=1$). In fact, the speed of light can be introduced explicitly in  $r$ and $r_D$ (\ref{wl}) providing the commutation rules (\ref{qb})  by means of the   scalings
 \be
\tilde {\>P}= \frac 1c\, \>P ,\qquad \tilde {\>K}= \frac 1c\, \>K ,\qquad \tilde z= c \, z ,
\label{wwmm}
\ee
 (that preserve the product $zP_2= \tilde z\tilde P_2$)  yielding the classical $r$-matrices
\be
\tilde r=\tilde z\bigl( \tilde K_2\wedge P_0+J_3\wedge \tilde P_1- J_1\wedge \tilde P_3 +c \sqrt{\ka} \, \tilde K_3\wedge \tilde K_1   \bigr) ,\qquad 
\tilde r_D= \tilde r -\tilde z J_2\wedge \tilde P_2 ,
\nonumber
\ee
  showing the above exclusions.      Note also that it is possible to transform the deformation parameter as  $ \tilde z= c^2 \, z$  (without preserving $zP_2$)  obtaining that 
\be
\tilde r=\frac{\tilde z}{c}\bigl( \tilde K_2\wedge P_0+J_3\wedge \tilde P_1- J_1\wedge \tilde P_3  \bigr) + \tilde z \sqrt{\ka} \, \tilde K_3\wedge \tilde K_1  ,\qquad 
\tilde r_D= \tilde r -\frac{\tilde z}{c} J_2\wedge \tilde P_2 ,
\nonumber
\ee
which is not real for $c={\rm i}$ but both of them  have a well defined limit $c\to \infty$  reducing  to  a Reshetikhin twist $ \tilde r\equiv\tilde r_D=   \tilde z \sqrt{\ka} \, \tilde K_3\wedge \tilde K_1  $. 
  
  Next we construct the  two first-order noncommutative spacetimes  $\>{ST}^{3+1}_z$  and the noncommutative Minkowskian space of 
     (time-like) lines $\>{L}^{6}_z$ (\ref{wf2})
  by means of the dual of the cocommutator for $r$ (\ref{wl}) or, alternatively, by introducing the  kinematical identification
  \be
\noindent
\begin{tabular}{cccc}
$\JJ^{01} $&$ \JJ^{02} $& $  \JJ^{03} $&$  \JJ^{04} $ \\[2pt] 
  &$  \JJ^{12} $& $  \JJ^{13} $&$  \JJ^{14} $ \\[2pt] 
 & & $  \JJ^{23} $&$  \JJ^{24} $ \\[2pt] 
&  & &$  \JJ^{34} $ 
\end{tabular}
\quad  \equiv  \quad 
\begin{tabular}{cccc}
$\hat x^1 $&$ \hat x^0 $& $  \hat x^3 $&$  \hat x^2 $ \\[2pt] 
  &$  \hat \xi^1 $& $   -\hat \theta^{2}  $&$   \hat \theta^{3}  $ \\[2pt] 
 & & $    \hat \xi^3    $&$ \hat \xi^2  $ \\[2pt] 
&  & &$ - \hat \theta^{1} $ 
\end{tabular}
\nonumber
\ee
   dual to (\ref{wh}), together with the contraction parameters (\ref{wi}) in  the commutation relations of the dual CK algebra (\ref{oa}) and 
   (\ref{ob}). 
 Similarly,    the  first-order twisted noncommutative dS and Minkowskian  spacetimes can be deduced  from  $\mathbb S^{(1)}_{z,\vartheta,\k}$ (\ref{ncdd})   ($\k_2\k_3=+1$).  All of these   structures are presented in Table~\ref{table6}.

 
\subsection{Class C: Time-like Deformations with      Primitive Generators $(P_0,J_2)$}
\label{s53}

We consider the kinematical assignation~\cite{CK4d}
\be
P_0=J_{04}, \quad \ \  \>P=(J_{01},J_{02},J_{03}),\quad\ \  \>K=(J_{14},J_{24},J_{34}),\quad\ \ \>J=(J_{23},-J_{13},J_{12}),
\nonumber
\ee
that is,
\be
\noindent
\begin{tabular}{cccc}
$J_{01} $&$ J_{02} $& $  J_{03} $&$  J_{04} $ \\ 
  &$  J_{12} $& $  J_{13} $&$  J_{14} $ \\ 
 & & $  J_{23} $&$  J_{24} $ \\ 
&  & &$  J_{34} $ 
\end{tabular}
\quad  \equiv  \quad 
\begin{tabular}{cccc}
$P_1 $&$ P_2 $& $  P_3 $&$  P_0 $ \\ 
  &$  J_3 $& $  -J_2 $&$ K_{1} $ \\ 
 & & $ J_{1} $&$  K_{2} $ \\ 
&  & &$ K_{3} $ 
\end{tabular}
\label{wo}
\ee 
Starting from the CK algebra (\ref{mf}), we find that
the   Lie brackets  (\ref{qa})   imply that   $\k_2= \k_3=+1$,  while the commutators  (\ref{qb}) 
give rise to $\k_1=\ka$, $\k_4=-\la$ and  $c=1$. Thus the contraction parameters and  the   bilinear form $\bfI_\k$ (\ref{mi})  are  given by
\be
(\k_1,\k_2,\k_3,\k_4)=\bigl(\ka,+1 ,+1,-\la \bigr),\quad \   c=1 , 
\label{wp}
\ee
\be
\bfI_\k=\bigl(     +1, \ka, \ka , \ka  , -\la \ka \bigr).
\nonumber
\ee
Therefore we find nine algebras in the CK family $\mathfrak{so}_\k(5)$: the three Lorentzian algebras of Section~\ref{s41} for $\la=1$, the three Carrollian algebras of Section~\ref{s43} for $\la=0$, and the   three Riemannian ones  of Section~\ref{s45} for $\la=-1$ (see Table~\ref{table4}).
 The kinematical automorphisms (\ref{qf}) turn out to be
\be
 \mathcal{P} = \Theta^{(1)}  \Theta^{(4)} ,\qquad  \mathcal{T}= \Theta^{(4)} ,\qquad \mathcal{PT}=\Theta^{(1)} .
\nonumber
\ee
In this case,  the spacetime $\>{ST}^{3+1}$  is again related   to the CK space   $ \mathbb S_\k^{(1)}$  (\ref{space1}) (since $\mathcal{PT}=\Theta^{(1)}$), while the space of   lines $\>{L}^{6}$ cannot be associated with $ \mathbb S_\k^{(2)}$,   but it can with the rank-2  symmetric CK space   (\ref{other}) with automorphism $ \mathcal{P} = \Theta^{(1)}  \Theta^{(4)} $.

The  CK   $r$-matrices (\ref{nb})  and    (\ref{pa})  now become
\be
r=z\bigl( K_1\wedge P_1+K_2\wedge P_2 + K_3\wedge P_3+\sqrt{-\la \ka} \, J_1\wedge J_3   \bigr) , \quad\ \   r_D=r -z J_2\wedge P_0 ,
\label{ws}
\ee
which lead to primitive generators   $P_0\equiv J_{04}$ and $J_2\equiv -J_{13}$,  thus   with   $z$ having dimensions of a time (recall that $c=1$) provided that   the product $zP_0$ is dimensionless. 
The constraint  $\k_1\k_4=-\la \ka \ge 0$ excludes  the dS algebra   ($\la=1$,  $\ka>0$) and the hyperbolic one  ($\la=-1$, $\ka<0$). 
Since   $\k_2\k_3=+1$,  any real $r$-matrix $r$ always provides a real  $r_D$ in this class. Consequently, 
the resulting {\em seven} real Lie bialgebras  given in Table~\ref{table5} also appear in Table \ref{table3} with the simple algebra AdS corresponding to the case (III). 
 The spherical and Euclidean algebras  (cases (I) and (Ia))  are again  recovered as in class A, but here for   different values for the parameters $(\ka,c,\la)$.   Once  the cocommutator for $r$ 
 (\ref{ws}) has been computed  it can be checked that the coisotropy condition  (\ref{coisotropy}) is fulfilled  
  for both subalgebras  $\mathfrak{h}_{\rm st}$ and $\mathfrak{h}_{\rm line}$ (\ref{wfx1}) allowing   the   construction of the two noncommutative  spaces  (\ref{wf2}).

It is worth stressing that this class of time-like deformations cover the so-called kappa-deformations such that the deformation parameters $z$ and $\kappa$ are related through $z\sim 1/\kappa$. Hence the $r$-matrix  for the Poincar\'e algebra of case (IIIa)  in Table~\ref{table3} underlies the well known $\kappa$-Poincar\'e
deformation~\cite{LRNT1991,GKMMK1992,LNR1992fieldtheory,Maslanka1993,MR1994}, and that for the AdS algebra of case (III) provides the $\kappa$-AdS algebra~\cite{BHN2015towards31, BHMN2017kappa3+1}.

  As far as the non-relativistic limit $c\to \infty$ is concerned, we remark that the condition $c=1$, in principle, precludes it. To be precise, 
     if we apply the same scalings  (\ref{wwmm})   to the Lie generators keeping $z$ unchanged and so the product $zP_0$ as well,   then $c$ appears explicitly in the commutators  (\ref{qb}) and    the $r$-matrices (\ref{ws}) now read 
 \be
\tilde r=  z \,c^2( \tilde K_1\wedge \tilde P_1+ \tilde K_2\wedge \tilde P_2+\tilde K_3\wedge \tilde P_3) +z  \sqrt{-\la \ka} \, J_1\wedge J_3   , \qquad 
\tilde r_D= \tilde r -  z J_2\wedge   P_0,
\nonumber
\ee
which diverge  under the limit $c\to \infty$. Nevertheless, we can introduce the scalings  \eqref{wwmm} but with transformed deformation parameter $\tilde z= c^2  z$  (not preserving $zP_0$) finding that~\cite{kappaN}  
 \be
\tilde r=\tilde z( \tilde K_1\wedge \tilde P_1+ \tilde K_2\wedge \tilde P_2+\tilde K_3\wedge \tilde P_3) +\frac{\tilde z}{c^2}  \sqrt{-\la \ka} \, J_1\wedge J_3   ,  \qquad 
\tilde r_D= \tilde r -\frac{\tilde z}{c^2}   J_2\wedge   P_0,
\nonumber
\ee
thus allowing one  to apply the limit $c\to \infty$ obtaining that 
    \be
\tilde r\equiv \tilde r_D=\tilde z( \tilde K_1\wedge \tilde P_1+ \tilde K_2\wedge \tilde P_2+\tilde K_3\wedge \tilde P_3)    ,
\nonumber
\ee
which coincides with the $\kappa$-Poincar\'e $r$-matrix. The remarkable point is that    the corresponding cocommutator is trivial, that is, $\delta(X)=0$ for all $X$, so that there is no deformation for the (contracted) Newtonian algebras.  In other words, the scheme of contractions for  the CK $r$-matrix (\ref{nb}) and cocommutator $\delta$ (\ref{cocok}) ensures to always obtain   both a non-trivial $r$-matrix and cocommutator.
In this respect, it is worth stressing that it is possible to apply a Lie bialgebra contraction    in such a manner that the initial $r$-matrix diverges but the initial cocommutator gives a non-trivial result~\cite{LBC} (this   is called fundamental but non-coboundary Lie bialgebra contraction). This contraction process was  applied in~\cite{GKMMK1992,Azcarraga95} in order to obtain the $\kappa$-Galilei algebra by contracting $\kappa$-Poincar\'e, finding a non-coboundary quantum Galilei algebra which explains its absence in our approach. In fact, such a non-coboundary Lie bialgebra contraction has recently been applied in~\cite{kappaN} in order to deduce the $\kappa$-Newtonian algebras containing both $\kappa$-NH algebras  $(\Lambda\ne 0)$ together with the above  $\kappa$-Galilei one  $(\Lambda= 0)$.

   The    first-order noncommutative spacetimes and spaces of lines for the seven Lie bialgebras contained in this class can be obtained 
   by introducing   the identification dual to (\ref{wo}) given by
   \be
\noindent
\begin{tabular}{cccc}
$\JJ^{01} $&$ \JJ^{02} $& $  \JJ^{03} $&$  \JJ^{04} $ \\[2pt] 
  &$  \JJ^{12} $& $  \JJ^{13} $&$  \JJ^{14} $  \\[2pt] 
 & & $  \JJ^{23} $&$  \JJ^{24} $  \\[2pt] 
&  & &$  \JJ^{34} $ 
\end{tabular}
\quad  \equiv  \quad 
\begin{tabular}{cccc}
$\hat x^1 $&$ \hat x^2 $& $  \hat x^3 $&$  \hat x^0 $  \\[2pt] 
  &$  \hat  \theta^{3} $& $  -\hat \theta^{2}   $&$   \hat \xi^1 $  \\[2pt] 
 & & $   \hat \theta^{1} $&$\hat  \xi^2  $ \\[2pt] 
&  & &$  \hat  \xi^3 $ 
\end{tabular}
\nonumber
\ee
together with the contraction parameters (\ref{wp}) in the commutation rules (\ref{oa}) and (\ref{ob}). In the same way, the 
  first-order twisted noncommutative spacetimes are deduced from     $\mathbb S^{(1)}_{z,\vartheta,\k}$ (\ref{ncdd}) for the seven cases since for all of them $\k_2\k_3=+1$. The explicit expressions for all of these  noncommutative spaces can be found  in Table~\ref{table6}.

 
\subsection{Class D: Dimensionless Deformation with    Primitive Generators $(J_2,P_0)$}
\label{s54}

As the last class we study how to obtain the AdS deformation of case (IV) in Table~\ref{table3} in a kinematical basis. With this aim we 
 consider the identification (not considered in~\cite{CK4d}) given by
\be
P_0=J_{13}, \quad\ \ \>P=(J_{14},-J_{12},J_{01}),\quad\ \ \>K=(J_{34},-J_{23},J_{03}),\quad \ \ \>J=(J_{02},J_{04},J_{24}) .
\label{wv}
\ee
Then  the   Lie brackets  (\ref{qa})   gives that   $\k_1= \k_2=\k_3=\k_4=-1$ and the commutators  \eqref{qb} 
lead to set $\ka=-1$, $c=1$ and $ \la=1$. Therefore   we obtain a {\em single} Lie algebra in this class, ${\rm  AdS}\simeq \mathfrak{so}(3,2)$, such that
\be
(\k_1,\k_2,\k_3,\k_4)= (-1,-1 ,-1,-1 \bigr),\quad\   \ka=-1, \quad\  c=1 , \quad\   \la=1 ,
\nonumber
\ee
\be
\bfI_\k=\bigl(     +1, -1, +1 , -1  , +1  \bigr).
\nonumber
\ee
 The kinematical automorphisms (\ref{qf}) read
\be
 \mathcal{P} = \Theta^{(1)} \Theta^{(2)} \Theta^{(3)}  \Theta^{(4)} ,\qquad  \mathcal{T}=\Theta^{(3)}  \Theta^{(4)} ,\qquad \mathcal{PT}= \Theta^{(1)} \Theta^{(2)} .
\nonumber
\ee
And  the  CK   $r$-matrices (\ref{nb})  and    (\ref{pa})  turn out to be
\be
r=z\bigl( K_1\wedge K_3+K_2\wedge P_2 + P_1\wedge P_3+  J_3\wedge J_1   \bigr) , \qquad  r_D=r+ zP_0 \wedge  J_2  .
\label{wz}
\ee
The primitive generators  are $J_2\equiv J_{04}$ and $P_0\equiv J_{13}$, while   $z$ is dimensionless like the product $zJ_2$ (we are working with units with $\ka=-1$ and $c=1$). 
Note that the constraints  $\k_1\k_4=+1$  and $\k_2\k_3=+1$ are automatically satisfied, so that $r_D$ is the kinematical expression of the new Drinfel'd double $r$-matrix of case (IV) in Table~\ref{table3}. By computing the cocommutator for $r$ 
 (\ref{wz}) (or from (\ref{cocok}) with the identification (\ref{wv})), it can be checked that the coisotropy condition  (\ref{coisotropy}) is not  satisfied 
  for any subalgebra   (\ref{wfx1}), so that there do  not exist  noncommutative  spacetime  and space  of lines  (\ref{wf2}) associated with this bialgebra.

 It is rather natural to analyse whether there may exist some possible contraction from this AdS deformation, although by following our approach the answer is negative whenever one requires to keep a non-trivial $r$-matrix and cocommutator.
 Starting from the AdS commutation relations  (\ref{qb}) with $\ka=-1$, $c=1$ and $\la=1$, it is possible to introduce explicitly such parameters by means of the scalings (coming from the    automorphisms  (\ref{qf}))
 \be
 \tilde P_0=\sqrt{-\ka} \sqrt{\la}\, P_0,\qquad   \tilde {\>P}= \sqrt{-\ka}\, \frac 1c\, \>P ,\qquad \tilde {\>K}= \sqrt{\la}\,\frac 1c\, \>K ,\qquad  \tilde {\>J}=\>J,
 \label{wza}
 \ee
 keeping the dimensionless parameter $z$. By introducing (\ref{wza}) in $r$ (\ref{wz}) we obtain that
 \be
\tilde r=-\frac{z\,c^2}{\ka \la}\left(-\ka  \tilde K_1\wedge \tilde K_3+\sqrt{-\ka} \sqrt{\la}\, \tilde K_2\wedge \tilde P_2 + \la \tilde P_1\wedge \tilde P_3-\frac{\ka \la}{c^2} \, \tilde J_3\wedge\tilde  J_1   \right) ,  
\nonumber
\ee
so that this diverges under the contractions $\ka\to 0$, $c\to \infty$ and $\la\to 0$. If we transform the deformation parameter as 
\be
\tilde z=-\frac{z\,c^2}{\ka  \la}
\nonumber
\ee
then the above contractions are well defined but only provide   twisted $r$-matrices   
\bea
&&\lim_{\ka\to 0}\tilde r=\tilde z\, \la\, \tilde P_1\wedge \tilde P_3 ,\qquad  \lim_{\la\to 0}\tilde r=-\tilde z \,\ka\, \tilde K_1\wedge \tilde K_3  ,
\nonumber\\[2pt]
&&\lim_{c\to \infty}\tilde r=\tilde z \left(-\ka \tilde K_1\wedge \tilde K_3+\sqrt{-\ka}\sqrt{\la}\, \tilde K_2\wedge \tilde P_2 + \la \tilde P_1\wedge \tilde P_3   \right) ,
\nonumber
\eea
 whose terms are all formed by commuting generators.

 
\subsection{Quantum Kinematical Algebras}
\label{s55}

 Tables~\ref{table5} and \ref{table6} highlight the main results so far obtained from a global kinematical viewpoint; the former
 covers all the information of kinematical bialgebras, while the latter shows their corresponding first-order noncommutative spacetimes and spaces of lines.  We now make some observations and    also comment on known results as well as on some open problems concerning the kinematical bialgebras and their complete quantum deformation. Similarly, some remarks for the full noncommutative spaces will be addressed in  Section~\ref{s56}.
 
  All the kinematical $r$-matrices presented in Table~\ref{table5}    underlie   quantum kinematical algebras ${\mathcal U}_z(\kk)$
 with  real Lie bialgebras $(\kk,\delta(r))$ of quasitriangular or standard type as it was described in Section~\ref{s11}. If we focus on the Poincar\'e bialgebra we get  three sequence of coboundary contractions:
   \be
\begin{array}{llll}
\mbox{Class A: space-like} &    \mbox{AdS} \ \  \xrightarrow{\ka\,\to\,0}    &\mbox{Poincar\'e} \ 
 \  \xrightarrow{c\,\to\, \infty}    & \mbox{Galilei}  
  \\[2pt]  
\mbox{\footnotesize (No Drinfel'd double)}    &   
    \mathfrak{so}(3,2)       & \mathfrak{i so}(3,1)     & \mathfrak{iiso}(3) \\[4pt]
    
    \mbox{Class B: space-like} & \mbox{dS}\quad  \ \,  \xrightarrow{\ka\,\to\,0}    &\mbox{Poincar\'e}      & 
  \\[2pt]  
\mbox{\footnotesize (With Drinfel'd double)}    &   
   \mathfrak{so}(4,1)       & \mathfrak{i so}(3,1)     & \\[4pt]

        \mbox{Class C: time-like} &   \mbox{AdS} \ \  \xrightarrow{\ka\,\to\,0}    &\mbox{Poincar\'e} \ 
 \  \xrightarrow{\la\,\to\, 0}    & \mbox{Carroll}  
  \\[2pt]  
\mbox{\footnotesize (With Drinfel'd double)}    &   
    \mathfrak{so}(3,2)       & \mathfrak{i so}(3,1)     & \mathfrak{ii^\prime so}(3)     
\end{array} 
\label{f1}
\ee
Recall that the class D only contains an isolated AdS bialgebra. From this approach, the non-relativistic contraction leading to a   Galilei bialgebra can only be performed within the space-like deformations belonging to  the class A, but none   of the three bialgebras in this sequence   can be endowed with a (contracted) Drinfel'd double   structure. The only possibility    to get a space-like   Poincar\'e bialgebra with associated contracted  Drinfel'd double   structure is provided by the class B, but now coming from a   dS bialgebra, instead of the AdS one   of    class A. Notice that the  difference  between the    classical $r$-matrices $r_D$ of   classes A and B can clearly been appreciated in the kinematical basis. In the class A there appears the term   $z   K_2\wedge P_3$, which only holds for the Riemannian bialgebras (with $c={\rm i}$ and $K_2$ becoming a rotation generator), while in the class B the  Drinfel'd double structure requires  to add the term $- z J_2\wedge P_2$ with a proper rotation generator. The sequence  for the class B  corresponds to perform
 $\mbox{(II)}\to  \mbox{(IIa)}$ in Table~\ref{table3}.   

 The class C deserves a special mention since it corresponds to the kappa-deforma\-tion. The sequence \eqref{f1} starts with the $\kappa$-AdS bialgebra~\cite{BHN2015towards31,BHMN2017kappa3+1}, continues with the $\kappa$-Poincar\'e bialgebra~\cite{LRNT1991,GKMMK1992,LNR1992fieldtheory,Maslanka1993,MR1994}, ending in the $\kappa$-Carroll one~\cite{CK4d,kappaN} while keeping a (contracted) Drinfel'd double   structure through all the process  which is denoted 
 $\mbox{(III)}\to  \mbox{(IIIa)}\to \mbox{(IIIb')}$ in Table~\ref{table3}.  

The complete Hopf algebra structure for the quantum inhomogeneous kinematical  algebras and their further contractions can be found in~\cite{CK4d}, which belong to the quantum CK family ${\mathcal U}_z (\mathfrak{so}_\k(5))$ with arbitrary $\k=(0,\k_2,\k_3,\k_4)$. Thus such results comprise the  quantum deformations of the Poincar\'e, Galilei and Euclidean bialgebras of  class A, the Poincar\'e bialgebra of  class B, along with   the $\kappa$-Poincar\'e and $\kappa$-Carroll  ones of class C; observe that the   Euclidean bialgebra of class C is equivalent to that of class A, being just case (Ia) in Table~\ref{table3}. The   $\kappa$-deformation for the curved Carrollian bialgebras (with $\ka\ne 0$) of class C,    $\kappa$-Para-Euclidean and $\kappa$-Para-Poincar\'e, can be deduced from the results given in~\cite{CK4d} by applying the   $z$-polarity $\mathcal{D}_z$ (\ref{odz}) since  this  map interchanges the CK bialgebras   as in (\ref{nc}),  $(0,\k_2,\k_3,\k_4)  \leftrightarrow (\k_4,\k_3,\k_2,0) $, providing their explicit expressions which are given in~\cite{kappaN}. Generalized results on twisted (space- and time-like) Poincar\'e  algebras can be found in~\cite{BP2014sigma}. For  twist deformations of   $\kappa$-Poincar\'e and their contractions to $\kappa$-Galilei algebras  we refer to~\cite{Daszkiewicz2008}, and 
for  twist deformations of the  Carroll algebra see~\cite{Daszkiewicz2019}.

Nevertheless,   quantum deformations for the  simple AdS and dS bialgebras have only been achieved for the  $\kappa$-AdS of class C in~\cite{BHMN2017kappa3+1},    as a Poisson--Hopf algebra, showing the hard difficulties of this task. Consequently, the obtention of the   quantum algebras for AdS of class A, dS of class B and AdS of class D remain as open problems.
 In contrast to this (3+1)D case,   quantum Drinfel'd--Jimbo deformations for the semisimple (A)dS algebras,  $\mathfrak{so}(3,1)$ and $\mathfrak{so}(2,2)$, are well known and their space- and time-like deformations were formerly obtained in~\cite{CK3d}   within a CK framework. Later on, the (2+1)D $\kappa$-(A)dS algebras were  considered in a quantum gravity context in~\cite{ASS2004}, and their  twisted deformations, with underlying Drinfel'd double structures~\cite{BHM2013cqg},  were studied in~\cite{BHMN2014sigma}. 

By taking into account the above comments, it is worth  comparing the (2+1)D case with the (3+1)D one with more detail, since they are quite different. In fact, the former is somewhat ``special" as it is very well known in quantum gravity.
In particular, the six generators  $\{J_{ab}\}$   $(a<b;\,  a,b=0,1,\dots,3)$ span the CK family  $\mathfrak{so}_{\k_1,\k_2,\k_3}(4)$  which turns out to be a Lie subalgebra of $\mathfrak{so}_\k(5)$ so with non-vanishing Lie brackets included in \eqref{mf}. It was proven in~\cite{LBC} that the Drinfel'd--Jimbo $r$-matrix for the family  $\mathfrak{so}_{\k_1,\k_2,\k_3}(4)$ simply reads
\be
r=z(J_{13}\wedge J_{01}+J_{23}\wedge J_{02}),
\label{f2}
\ee
which is $\k$-independent in contrast to (\ref{nb}). Moreover the $r$-matrix (\ref{f2})   gives rise to a cocommutator $\delta$ through  \eqref{tf} 
  determining a real Lie bialgebra for any value of the  three contraction parameters $(\k_1,\k_2,\k_3)$. The pair of main and secondary primitive generators is $(J_{03},J_{12})$. Therefore the family of CK bialgebras $(\mathfrak{so}_{\k_1,\k_2,\k_3}(4),\delta(r))$ contains all the $3^3=27$ possibilities at this dimension~\cite{CK3d}. 
  
  Next we consider the family of   kinematical algebras $\mathfrak{so}_{\ka, c ,\la}(4)$ spanned by the six generators $\{P_0,P_1,P_2,K_1,K_2,J_3\}$ in such a manner that the commutation rules are given by (\ref{qa}) and (\ref{qb}) setting the indices $i,j=1,2$ and fixing $k=3$. Starting with the CK $r$-matrix (\ref{f2}), we look for  space- and time-like   $\mathfrak{so}_{\ka, c ,\la}(4)$ bialgebras    which would be  the (2+1)D counterparts of the classes A, B and C in Table~\ref{table5}. Clearly the class B, with commuting primitive generators $(P_2,J_2)$, has no (2+1)D counterpart  since at this dimension there does not exist  a spatial  generator $P_i$ commuting with $J_3$. Hence we are led to  the two  classes A and C for  $\mathfrak{so}_{\ka, c ,\la}(4)$ such that the former requires to set $\la=1$, while the latter obliges to fix $c=1$.  Each of them covers nine real Lie bialgebras which are  displayed in Table~\ref{table7}.


\vskip-0.5cm

\begin{table}[t!] 
{\footnotesize
\caption{\small Space- and time-like real classical $r$-matrices for the (2+1)D kinematical and 3D Riemannian algebras with commutation relations (\ref{qa}) and (\ref{qb}) with indices $i,j=1,2$ and   $k=3$.}
\label{table7}
 \begin{center}
 \begin{tabular}{l l l l }
 \hline  

\hline

 \\[-0.2cm]
\multicolumn{4}{l}{    {Space-like class A:}  \quad \  \, $\la=1$ \qquad  {Primitive}\ $(P_2,K_1)$ \qquad \ $r=z(K_2\wedge P_0 +J_{3}\wedge P_1) $   }  \\[0.2cm]
$\bullet$ Lorentzian\ ($c$   finite): & {dS}\  $\mathfrak{so}(3,1)$\  $\ka>0$ & Poincar\'e\  $\mathfrak{iso}(2,1)$\  $\ka=0$ & AdS\   $\mathfrak{so}(2,2)$\  $\ka<0$
 \\[0.2cm]
  $\bullet$ Newtonian\ ($c=\infty$): & {Expanding NH}\  $\mathfrak{n}_+$\  $\ka>0$ & Galilei\ \ $\mathfrak{iiso}(2)$\  $\ka=0$ &Oscillating    NH\   $\mathfrak{n}_-$\ $\ka<0$
 \\[0.2cm]
 $\bullet$ Riemannian\ ($c={\rm i}$): & {Hyperbolic}\  $\mathfrak{so}(3,1)$\  $\ka>0$ & Euclidean\  $\mathfrak{iso}(3)$\  $\ka=0$ & Spherical\   $\mathfrak{so}(4)$\ $\ka<0$
 \\[0.2cm]
  \hline
\\[-0.2cm]
\multicolumn{4}{l}{    {Time-like class C:}  \qquad   $c=1$ \qquad \, {Primitive}\ $(P_0,J_3)$ \qquad\, $r=z(K_1\wedge P_1 +K_2 \wedge P_2)  $   }  \\[0.2cm]
 $\bullet$ Lorentzian\ ($\la=1$    ): & {dS}\  $\mathfrak{so}(3,1)$\  $\ka>0$ & Poincar\'e\ $\mathfrak{iso}(2,1)$\  $\ka=0$ & AdS\  $\mathfrak{so}(2,2)$\ \ $\ka<0$
 \\[0.2cm]
 $\bullet$ Carrollian\ ($\la=0$): & {Para-Euclidean}\,$\mathfrak{i'so}(3)$\,$\ka>0$ & Carroll\  $\mathfrak{ii'so}(2)$\  $\ka=0$ &Para-Poincar\'e\,$\mathfrak{i'so}(2,1)$\,$\ka<0$
 \\[0.2cm]
 $\bullet$ Riemannian\ ($\la= -1$):&{Hyperbolic}\ \ $\mathfrak{so}(3,1)$\ \ $\ka<0$ & Euclidean\ \ $\mathfrak{iso}(3)$\ \ $\ka=0$ & Spherical\ \  $\mathfrak{so}(4)$\ \ $\ka>0$ \\[0.25cm]
\hline  

\hline
\end{tabular}
\end{center}
}
 \end{table}


The  (2+1)D  NH algebras are $\mathfrak{n}_+= \mathfrak{i}_4 (  \mathfrak{so}(1,1) \oplus   \mathfrak{so}(2)  )$ and   $\mathfrak{n}_-= \mathfrak{i}_4 (  \mathfrak{so}(2) \oplus   \mathfrak{so}(2)  )$. Obviously, there exists a   second  class of space-like deformations with primitive generators $(P_1,K_2)$, but this leads to equivalent results already contained within the class A~\cite{CK3d}. The time-like class C, corresponding to the kappa-deformation, does  not only show the known fact that the (2+1)D  $\kappa$-Poincar\'e $r$-matrix is shared by its curved neighbours (see~\cite{BHMN2014sigma} and references therein), but  also that it holds for the three Carrollian and Riemannian deformations.
  The strong differences between the (2+1)D and (3+1)D deformations become evident when comparing the   expressions of Table~\ref{table7} with those given in Table~\ref{table5}.

  Finally, we would like to point out that the CK approach to kinematical deformations may appear to be  rather restrictive since this starts with the specific Drinfel'd--Jimbo  $r$-matrix  (\ref{ec}) for $ \mathfrak{so}(5) $ and, from it, the   CK $r$-matrix (\ref{nb}) is introduced, which together with $r_D$ (\ref{pa}) become the cornerstone  of this work. However, we stress that this is not the case  provided that one searches for deformations with a fundamental scale determined by the quantum deformation parameter.
         In particular, the problem of finding  time-like  classical $r$-matrices for (3+1)D (A)dS algebras was   addressed in~\cite{BrunoAdS}. There, it was initially considered the           Lorentzian algebras, with commutation relations  (\ref{qa}) and \eqref{qb} (set $\la =1$),  together with the most generic classical $r$-matrix depending on 45 deformation parameters.  Then it was required to   keep underformed the time translation generator $P_0$ and another commuting generator which was chosen $J_3$, that is, $\delta(P_0)=\delta(J_3)=0$ (recall that $P_0$ only commutes with  generators of rotations). Under these conditions, the solution of the modified classical Yang--Baxter equation  \eqref{th}  gave rise  to two two-parametric classical $r$-matrices. One of them was formed  by the superposition of the $\kappa$-AdS $r$-matrix with the twist $P_0\wedge J_3$, which turns out to be  just $r_D$ of the  class C in Table~\ref{table5} by identifying the two deformation parameters in~\cite{BrunoAdS} and interchanging the indices $3\leftrightarrow 2$  in the generators (so $J_3\leftrightarrow J_2$) through the appropriate  Lie algebra automorphism. Likewise, the second solution can be identified with $r_D$ of the  class D by    identifying again the two deformation parameters and applying the permutation of indices $3\to 2\to 1\to   3$   with another algebra automorphism. We remark that no analysis on real Lie bialgebras and Drinfel'd doubles was carried out in~\cite{BrunoAdS}.
Moreover, although it was claimed that the second solution ($r_D$ of class D)  was determined by a dimensionful deformation parameter this is not  exactly correct if one requires a dimensionless classical $r$-matrix. In fact, from a dimensionless deformation parameter (like $z$ in class D) it can be introduced a dimensionful one by trivially multiplying it by a global factor.

 
\subsection{Noncommutative Spacetimes and Spaces of Lines}
\label{s56}

The first-order noncommutative spaces  for the kinematical bialgebras of Table~\ref{table5} 
 are shown in Table~\ref{table6}.  There are several different situations among the four classes, ranging from the class D, where there is   no  noncommutative space (and so omitted), 
  to the classes A and C, for which there exist both noncommutative spacetimes and spaces of lines for all the bialgebras. In this sense, the latter classes can be regarded as the prototypes   for space- and time-like  noncommutative spaces. However, when Drinfel'd double structures are taken into account, the classes B and C become the relevant ones, also providing twisted noncommutative  spacetimes.

Although these results do not convey, in general, the full noncommutative spaces, for which all   orders in the quantum coordinates must be considered, in some cases they do. Concerning the  (3+1)D  noncommutative spacetimes,  which are those commonly studied in the literature, the first-oder noncommutative spacetimes in Table~\ref{table6}  turns out to be the complete   ones for all the cases associated with a flat classical spacetime $ \>{ST}^{3+1}$ \eqref{qqf}, so with vanishing sectional curvature $K$; these are just the four spaces displayed in the middle column of Table~\ref{table4}. Consequently,  Table~\ref{table6} comprises the following full (3+1)D (linear) noncommutative  spacetimes:  the   space-like  Minkowskian and  Galilean ones  together with  the   4D Euclidean space in the class A; another  space-like  Minkowskian  spacetime of class B which is equivalent to that of class A under the automorphism corresponding to the permutation of indices $ 1\to 2 \to 3 \to 1 $; and the (time-like)  $\kappa$-Minkowski and $\kappa$-Carroll spacetimes of class C (the Euclidean case is equivalent to that of class A). Additionally,  complete twisted noncommutative spacetimes coming from  contracted Drinfel'd double structures cover the    4D Euclidean  space of class A, the space-like Minkowskian one of class B,  and the time-like Minkowskian and Carroll spaces of class C (again the Euclidean space here is equivalent to that of class A).
We remark that more general results on    twisted space- and time-like Minkowskian noncommutative spacetimes can be found in~\cite{BP2014sigma}.

To the best of our knowledge, results for   (3+1)D  noncommutative spacetimes related to curved spacetimes $ \>{ST}^{3+1}$ \eqref{qqf} (so with $\ka\ne 0)$ only comprise  the (nonlinear)  $\kappa$-AdS space (and its twisted version)~\cite{kappaAdS}, as well as the $\kappa$-Para-Euclidean and $\kappa$-Para-Poincar\'e~\cite{kappaN} of class C. In such noncommutative spaces there appear higher-order terms in the quantum coordinates governed by the cosmological constant/curvature parameter $\ka$. This fact allows one to distinguish them from the linear (flat) 
$\kappa$-Minkowski and $\kappa$-Carroll spaces with $\ka=0$, but the latter share the same   linear structure.
Hence the construction of the space-like noncommutative AdS (class A) and dS (class B) spacetimes remain as open problems, which could be faced by 
computing   their Poisson--Lie structure by means of the Sklyanin  bracket \eqref{sklyanin} and next studying their quantization (similarly to the   
 $\kappa$-AdS spacetime~\cite{kappaAdS}).

Noncommutative  spaces of  lines have scarcely been  explored and they have only been  constructed 
 for the $\kappa$-Minkowski of class C in~\cite{BGH2019worldlinesplb}   
 and in lower dimensions  for the 4D (A)dS noncommutative spaces of worldlines in~\cite{Lines2014}; recall that    the three classical Lorentzian spaces of worldlines are of    non-zero curvature equal to $-1/c^2$, while both NH and Galilean spaces of lines are flat.   Although the first-order noncommutative Minkowskian space of lines  of class C  has vanishing commutators, we stress that     the brackets defining its full quantum space  are not trivial at all and, in fact, it can be endowed with a symplectic structure everywhere but in the origin. By contrast, observe   that the structure of the first-order noncommutative spaces of lines of class A is not
 trivial (see Table~\ref{table6}).  From this viewpoint,    noncommutative spaces of lines deserve a deeper study and moreover it would be necessary   to construct more noncommutative spaces of lines  which when read altogether with their corresponding noncommutative spacetimes could allow for a deeper insight into the structure of each precise quantum deformation.

 
\section{Conclusions and Outlook}
\label{s6}

This paper can be seen as a two-fold work with two interlinked parts that we proceed to comment separately.

 In the first part of the work (Sections~\ref{s2} and \ref{s3}), we have considered   the CK formalism for quasiorthogonal Lie algebras and their associated symmetric homogeneous spaces  in order to  next study their Drinfel'd--Jimbo quantum deformations.
The CK approach conveys a built-in scheme of Lie algebra contractions in terms of explicit graded contraction/curvature parameters $\k$, in such a manner that semisimple together with non-semisimple Lie algebras and their homogeneous spaces can be described in a unified setting, which ranges from the semisimple  $\mathfrak{so}(p,q)$ algebras  (providing  curved spaces)
 to the most contracted case in the CK family, the flag algebra (with associated flat spaces). In all the contraction sequence the same number of Casimir invariants (two, in our case) is preserved which, in turn, implies that these CK algebras share many structural properties as we have shown along the paper. As a novelty, we stress that we have not only considered the usual space of points (i.e.~spacetimes), but also the symmetric homogeneous CK spaces of lines, 2-planes and 3-hyperplanes. In this global framework,
    Drinfel'd--Jimbo  CK bialgebras have been obtained from the one corresponding to  $\mathfrak{so}(5)$  in a rotational basis,    by always requiring the condition  of getting a real Lie bialgebra, which finally led to the 63 real Lie bialgebras shown in Table~\ref{table1}. From these results,   their dual quantum counterparts have also been deduced giving rise to their corresponding 
      first-order noncommutative spaces of points, lines, etc.,  for which the coisotropy condition has been imposed, thus ensuring to always obtain a noncommutative space as a subalgebra of the dual Lie bialgebra; the final results have been summarized in Table~\ref{table2}. Furthermore, $r$-matrices coming from Drinfel'd double  structures have   been studied in detail as well. In particular,  starting with the one corresponding to the real compact form  $\mathfrak{so}(5)$ in the rotational CK basis, three classical $r$-matrices for the  $\mathfrak{so}(p,q)$ algebras  together with ten contracted  $r$-matrices  have  explicitly been achieved and displayed in Table~\ref{table3}. New results correspond to the dS $\mathfrak{so}(4,1)$ algebra of case (II) and the AdS $\mathfrak{so}(3,2)$ one of case (IV), along with  the contractions from the four classical $r$-matrices $r_D$ for the simple Lie algebras. 
 We remark that such $r$-matrices, coming from Drinfel'd doubles have  provided, in a natural way,  first-order  twisted noncommutative CK spaces of points and of 3-hyperplanes for the 14 real Lie bialgebras given in Table~\ref{table3}. 
 
 Concerning this first part of the paper, there are, at least, two research lines which we plan   to face  in the future:

    \begin{enumerate} 
   
   \item  To construct new dual   homogeneous CK spaces with isotropy subalgebras corresponding to the first-order noncommutative CK spaces  $ \sz^{(m)} \equiv\mathfrak{h}^{(m)}_\asttk $ $(m=1,\dots, 4)$ (\ref{oobb})   in a similar form to that followed in~\cite{PHS}, but moreover considering their symmetric character according to the $z$-involutions    $\theta^{(m)}_z$  (\ref{invzb}), which in some cases would provide a ${\mathbb {Z}}_2^{\otimes 4}$-grading in this dual framework. And also
    to study their mathematical/physical properties.
 
\item  To   perform  a similar construction to the one here developed for the  Drinfel'd--Jimbo quantum deformations of quasiorthogonal CK algebras for   other  families of  CK algebras~\cite{Santander2013}. Among them, we remark   the  quasiunitary CK algebras~\cite{unitaryCK,GromovSU} (starting with the    $\mathfrak{su}(p,q)$ algebras) since they are naturally related to the physical quantum space of states for   any quantum system~\cite{Santander2002,CKsu2}.

  \end{enumerate} 

 In the second part of the work we have focused on the kinematical algebras together with their associated symmetric homogeneous spacetimes and spaces of lines  displayed in Table~\ref{table4}. Then we have   applied the previous CK approach in order to deduce their corresponding classical $r$-matrices, $r$ and $r_D$, given in Table~\ref{table5}, thus providing kinematical bialgebras, and from them we have constructed the first-order noncommutative spacetimes and spaces of lines shown in Table~\ref{table6}.
    A detailed physical discussion on known results and open problems concerning their full quantum algebra deformation and complete quantum spaces has already been carried out in Sections~\ref{s55} and~\ref{s56}, respectively. Therefore, to end with, we summarize the main conclusions and open lines of research on this issue:
    
       \begin{enumerate} 
   
   \item   In this paper we have only considered {\em coboundary} Lie bialgebra contractions, that is, those Lie bialgebras coming from a contracted classical $r$-matrix. However, there also exist {\em fundamental} Lie bialgebra contractions, under the which the $r$-matrix diverges but the cocommutator $\delta$ is well defined~\cite{LBC} (so ensuring the existence of well defined coproduct $\Delta_z$). Hence a systematic study of all the possible fundamental but non-coboundary Lie bialgebra contractions starting with the four classical $r$-matrices for $\mathfrak{so}(3,2)$ and $\mathfrak{so}(4,1)$  in Table~\ref{table5} is still lacking. These could give rise  to new quantum deformations for non-simple kinematical algebras as it was the case   for the $\kappa$-Newtonian ones already obtained in~\cite{kappaN}
 
\item   The quantum algebra deformations for the simple algebra   AdS $\mathfrak{so}(3,2)$  of the classes A  and D,  
 and  for dS $\mathfrak{so}(4,1)$ of the class B are still unknown. Such structures would be moreover useful in order to obtain the corresponding contracted quantum algebras for the kinematical algebras with $\ka\ne 0$ covering  the NH,  Para-Euclidean and Para-Poincar\'e algebras. In this contraction process, both coboundary and fundamental non-coboundary   Lie bialgebra contractions may be applied.

\item From Table~\ref{table6} it directly follows that   quantum deformations for different kinematical algebras share the same underlying first-order noncommutative   spacetime structure. When dealing with the curved cases with $\ka\ne 0$, differences among them could arise  when higher-orders in the quantum coordinates are taking into account (as it happened for  the $\kappa$-spacetimes of class C obtained  in~\cite{kappaAdS,kappaN}). Nevertheless,  the linear 
noncommutative spacetime structure remains the same for the flat cases with $\ka=0$ and this fact holds for the Minkowkian, Galilean and Carroll spaces. Consequently, the construction of the ``accompanying" noncommutative spaces of lines may be of interest in order to distinguish them. And, furthermore, new physical consequences could be extracted from such new structures. In this respect, we would like to emphasize that  the noncommutative space of worldlines already constructed for the $\kappa$-Poincar\'e algebra in~\cite{BGH2019worldlinesplb} constitutes a prototype   example in this direction.

  \end{enumerate}

 Work on the above lines is currently in progress.

\section*{Acknowledgments}

 \small

This work has been partially supported by   Agencia Estatal de Investigaci\'on (Spain)  under grant  PID2019-106802GB-I00/AEI/10.13039/501100011033,  and by Junta de Castilla y Le\'on (Spain) under grants BU229P18 and BU091G19. 
 The authors would like to acknowledge the contribution of the   European Cooperation in Science and Technology COST Action CA18108.



\end{document}